\documentclass[11pt]{article}
\pdfoutput=1

\usepackage{stolenstyle}

\usepackage{amsmath,epsf,amssymb,latexsym,amsthm,setspace,bbm,array,pifont,enumerate}

\DeclareFontFamily{U}{rsf}{}
\DeclareFontShape{U}{rsf}{m}{n}{
  <5> <6> rsfs5 <7> <8> <9> rsfs7 <10-> rsfs10}{}
\DeclareMathAlphabet\Scr{U}{rsf}{m}{n}

\def\CO#1#2{{[#1,#2]}}

\def\iden{{\mathbbm 1}}

\def\C{{\mathbb C}}

\def\P{{\mathbb P}}
\def\R{{\mathbb R}}
\def\Z{{\mathbb Z}}

\def\Gr{\operatorname{Gr}}

\def\Hom{\operatorname{Hom}}

\def\im{\operatorname{im}}

\def\tr{\operatorname{tr}}

\def\SL{\operatorname{SL}}

\def\PSL{\operatorname{PSL}}
\def\GO{\operatorname{O{}}}
\def\SU{\operatorname{SU}}
\def\GU{\operatorname{U{}}}

\def\GE{\operatorname{E}}

\def\so{\operatorname{\mathfrak{so}}}

\def\su{\operatorname{\mathfrak{su}}}
\def\Lu{\operatorname{\mathfrak{u}}}
\def\Le{\operatorname{\mathfrak{e}}}
\def\Lg{\operatorname{\mathfrak{g}}}
\def\LLh{\operatorname{\mathfrak{h}}}

\def\p{\partial}
\def\pb{\bar{\partial}}

\def\ff#1#2{{\textstyle\frac{#1}{#2}}}

\def\cA{{\cal A}}
\def\cB{{\cal B}}
\def\cC{{\cal C}}
\def\cD{{\cal D}}
\def\cE{{\cal E}}
\def\cF{{\cal F}}
\def\cG{{\cal G}}

\def\cL{{\cal L}}

\def\cP{{\cal P}}

\def\cR{{\cal R}}
\def\cS{{\cal S}}
\def\cT{{\cal T}}

\def\cV{{\cal V}}

\def\cX{{\cal X}}

\def\ep{{\epsilon}}

\newcommand\taub{\overline{\tau}}

\newcommand\Thetab{\overline{\Theta}}

\newcommand\Omegab{\overline{\Omega}}

\newcommand\bb{\overline{b}}

\newcommand\zb{\overline{z}}

\newcommand\pt{\widetilde{p}}

\newcommand\Ch{\widehat{C}}

\newcommand\Fb{\overline{F}}

\newcommand\Xt{\widetilde{X}}
\newcommand\Yt{\widetilde{Y}}

\theoremstyle{definition}

\usepackage{upgreek}

\usepackage{tikz}

\usetikzlibrary[shapes.geometric]
\usetikzlibrary{positioning} 
\usetikzlibrary{calc,intersections,through,backgrounds}
\usetikzlibrary{cd}

\tikzset{>=stealth}
\tikzset{every picture/.style={very thick}}

\def\bA{{{\boldsymbol{A}}}}

\def\bF{{{\boldsymbol{F}}}}
\def\bL{{{\boldsymbol{L}}}}

\def\be{{{\boldsymbol{e}}}}

\def\bsigma{{{\boldsymbol{\sigma}}}}

\def\bv{{{\boldsymbol{v}}}}
\def\bell{{\boldsymbol{\ell}}}

\def\bLambda{{{\boldsymbol{\Lambda}}}}
\def\balpha{{\boldsymbol{\alpha}}}
\def\bbeta{{\boldsymbol{\beta}}}

\def\bsigma{{\boldsymbol{\sigma}}}
\def\bkappa{{\boldsymbol{\kappa}}}

\def\bUplambda{{\boldsymbol{\Uplambda}}}

\def\bTheta{{\boldsymbol{\Theta}}}

\def\ba{{{\textbf{a}}}}
\def\bb{{{\textbf{b}}}}

\def\bp{{\boldsymbol{p}}}
\def\bpi{{\boldsymbol{\pi}}}
\def\bpit{{\widetilde{\bpi}}}

\def\sleft{\text{\tiny{L}}}
\def\sright{\text{\tiny{R}}}

\def\preprint{ ~~ }

\title{Topology change and heterotic flux vacua}
\author[a] {Dan Isra\"el,}
\author[b] {Ilarion V.~Melnikov,}
\author[c] {Ruben Minasian,}
\author[a] {Yann Proto\,}
\affiliation[a] {Sorbonne Universit\'e, CNRS, Laboratoire de Physique Th\'eorique et Hautes {\'E}energies, LPTHE, F-75005 Paris, France}
\affiliation[b] {Department of Physics and Astronomy,
James Madison University,
Harrisonburg, VA 22807, USA}
\affiliation[c] {Institut de Physique Th{\'e}orique,  
Universit{\'e} Paris-Saclay, CNRS, CEA, F-9119, Gif-sur-Yvette, France
}

\emailAdd{israel@lpthe.jussieu.fr}
\emailAdd{melnikix@jmu.edu}
\emailAdd{ruben.minasian@ipht.fr}
\emailAdd{yproto@lpthe.jussieu.fr}

\abstract{We investigate the interrelation between topology and Narain T-duality of heterotic flux vacua.  We present evidence that all $5$ and $4$-dimensional Minkowski space heterotic flux backgrounds with $8$ supercharges have a locus in the moduli space with a T-dual description in terms of a compactification on the product of a K3 surface with a circle or a torus.  A test of this equivalence is provided by calculating the new supersymmetric index on both sides of the duality. We examine the implications of these dualities for CHL-like orbifolds that reduce the rank of the gauge group, as well as those that lead to minimal supersymmetry in $4$ dimensions.  We also discuss properties of flux vacua that preserve minimal supersymmetry in $4$ dimensions that cannot be related to conventional compactifications by Narain T-duality.   Along the way we point out a number of properties of these vacua, including the role played by non-trivial flat gerbes, the appearance of rational worldsheet CFTs in decompactification limits, and the role of attractive K3 surfaces in backgrounds with minimal supersymmetry.  Finally, we discuss the dual pairs from the perspective of M-theory/heterotic duality.}

\begin{document}

\maketitle

\section{Introduction} \label{s:Introduction}
A generic string vacuum, when realized by some compactification geometry, is a flux vacuum with non-zero background values for the field strengths of various $p$-form fields that appear in the massless spectrum of a ten-dimensional string theory.  Even in the presence of supersymmetry the study of such generic configurations is challenging.  For instance, it is typically not easy to argue that a supergravity solution with fluxes extends to a solution of the string equations of motion.

In this context the heterotic string is particularly appealing, since the fluxes in question involve the heterotic $3$-form $H$ and the curvature of the gauge bundle, both of which can be understood directly in the RNS heterotic worldsheet.  Indeed, the possibility of heterotic flux vacua was appreciated early on in the study of string compactification~\cite{Hull:1986kz,Strominger:1986uh}, but the efforts to overcome challenges in working with a non-K\"ahler target space in spacetime or a generic (0,2) supersymmetric non-linear sigma model on the worldsheet have only met with partial success in subsequent years.

Although we are far from understanding the stringy geometry that underlies a generic heterotic flux vacuum, there is a particularly nice class of such vacua that preserve at least $4$ supercharges on $\R^{1,3}$~\cite{Goldstein:2002pg,Fu:2006vj,Becker:2006et}.  The target space $X$ is a holomorphic principal $T^2$ bundle over a smooth K3 surface $M$.  When the bundle is topologically non-trivial, $X$ is a topologically non-K\"ahler $\SU(3)$ structure manifold.  The $H$-flux measures the non-closure of the Hermitian form, and its quantization requires $X$ to have string scale cycles associated to the fiber directions.

The Ansatz for these geometries was initially motivated from a dual M-theory construction~\cite{Dasgupta:1999ss} and developed further by a number of authors, as discussed in~\cite{Goldstein:2002pg}, but is it required by the heterotic string?
Certainly not for for vacua preserving $4$ supercharges:  after all, we know that there are supersymmetric string-perturbative heterotic compactifications on generic Calabi--Yau manifolds.   On the other hand, the preservation of $8$ supercharges in spacetime leads to strong constraints on the worldsheet superconformal theory~\cite{Banks:1988yz}.
These constraints were applied to heterotic non-linear sigma models in~\cite{Melnikov:2010pq,Melnikov:2012cv}, where it was argued that if a compactification has a geometric description and preserves $8$ supercharges, then the geometry should indeed be a principal $T^2$ bundle over a K3 surface, and the fibration must be anti-self-dual.  That is, the two curvature $2$-forms that characterize the principal $T^2$ fibration must lie in $H^2_-(M,\R)$.  The analysis of~\cite{Melnikov:2010pq,Melnikov:2012cv} missed one important point:  it is possible to obtain a smooth $X$ as a principal $T^2$ fibration over a singular K3 surface $M$~\cite{Fino:2019mvp}.  Given that generalization, it is reasonable that these geometries exhaust the possibilities of heterotic flux vacua with $8$ supercharges.

Precisely for the class of flux geometries where $M$ is smooth and the compactification to $4$ dimensions preserves $8$ supercharges there are gauged linear sigma models with low energy limit believed to be the corresponding two-dimensional heterotic CFT~\cite{Adams:2006kb}.  When combined with supersymmetric localization, this linear sigma model technology can be used to calculate various protected quantities~\cite{Adams:2009zg,Israel:2015aea,Israel:2016xfu,Angelantonj:2016gkz}.   Thus, we can be reasonably confident that these flux geometries are indeed vacua of the perturbative heterotic string, and we have tools to study a number of their properties beyond the supergravity limit.  This is a crucial point because the flux geometries have string-scale cycles, and the supergravity analysis is at best a leading term in a formal $\alpha'$ expansion.

At first glance it appears that the heterotic flux vacua provide a significant generalization of the classic compactification on the K\"ahler geometry $X_0 = M\times T^2$.  This is certainly the case from the point of view of heterotic supergravity: the geometries are topologically distinct, and the flux vacua are topologically non-K\"ahler.  In addition, while the vacua based on $X_0$ have a large volume limit where supergravity is reliable and lifts to both $8$ and $6$ dimensions, the
same is not true of more generic $X$:  these have string-scale cycles and appear to only have a lift to $8$ dimensions (at least when $M$ is smooth), where the base of the fibration $M$ is taken to be large.

Nevertheless, over the years there have been a number of works suggesting that there are relations between the K\"ahler and non-K\"ahler vacua.  For example, in~\cite{Becker:2007ea,Sethi:2007bw} it was argued via duality with M-theory compactification on K3$\times$K3 that certain heterotic flux backgrounds could have a dual description as a compactification on $X_0$.  It has also been pointed out that heterotic T-duality on the torus fiber can relate flux compactifications to K\"ahler compactifications~\cite{Evslin:2008zm,Andriot:2009fp,Martelli:2010jx,Andriot:2011iw,Israel:2013hna}.

In the present work we focus on heterotic flux vacua with an $8$-dimensional lift.  We point out that in general $X$ is not simply connected, and the non-trivial fundamental group allows for more general choices of gauge bundle than usually considered.  In particular, the bundle can be topologically distinct from any gauge bundle pulled back from $M$, and this allows for CHL-like constructions in flux compactifications.  

However, we also argue that a heterotic flux vacuum will have a locus in the moduli space where the geometry can be T-dualized to a K\"ahler space.  So, while on one hand we enlarge the class of heterotic flux vacua, we also show that all such compactifications can be mapped to the more familiar class of $\text{K3}\times T^2$ compactifications.   The map includes motion to an appropriate locus in the moduli space and T-duality, and as usual, certain features are much easier to understand in one duality frame than another.  We illustrate this point by mapping global symmetries of the worldsheet theory between the dual descriptions and considering the corresponding orbifold theories.  A further discussion of heterotic flux vacua with $8$ supercharges based on the geometries introduced in~\cite{Fino:2019mvp} and which are not covered by this analysis is given in section \ref{s:outlook}.

\subsection{An extended example}
Let us illustrate our main findings in an example---a class of $5$ dimensional heterotic flux compactifications.  

Consider the $(\GE_8\times\GE_8)\rtimes\Z_2$ heterotic string compactified to $9$ dimensions on a circle of radius $r$ with a generic choice of Wilson line $\ba$ valued in the Cartan subalgebra $\LLh \subset \Le_8\oplus \Le_8$.  The resulting $9$-dimensional theory has gauge algebra $\Lu(1)^{\oplus 17}$, which we decompose as $\Lu(1)_{\text{phys}} \oplus \Lu(1)^{\oplus 16}_{\text{gauge}}$, where the first factor is associated to the vector multiplet coming from the translation symmetry of the circle.\footnote{There is also an additional gauge boson that arises from the supersymmetric (in our conventions right-moving) sector of the string.  This gauge boson resides in the spacetime gravity multiplet.  In what follows when we discuss the gauge algebra we will leave it out from our considerations.}  The Narain moduli space is then given as a quotient of a Grassmannian $\Gr(1,17)/\GO(\Gamma_{1,17})$ where $\Gamma_{1,17}$ is the even self-dual lattice embedded in $\R^{1,17}$, with $\GO(\Gamma_{1,17})$ the set of lattice isomorphisms, and we take the metric on $\R^{1,17}$ to be
\begin{align}
\eta = \begin{pmatrix} 0 & 1 & 0 \\ 1 & 0 & 0 \\ 0 & 0 & \iden_{16} \end{pmatrix}~.
\end{align}

We now compactify the theory further on a smooth K3 surface $M$.  The resulting theory is equivalent to a compactification of the $10$-dimensional theory on the $5$-dimensional space $X_{v}$, where $v\in \Gamma_{1,17} \otimes H^2(M,\Z)$ determines the topology of the space, and, from the $9$-dimensional perspective, the choice of instanton used to solve the $9$-dimensional heterotic Bianchi identity.   When $v$ has a non-zero component corresponding to $\Lu(1)_{\text{phys}}$, the ten-dimensional interpretation of the vacuum is a $5$-dimensional heterotic flux compactification.  On the other hand, when the $\Lu(1)_{\text{phys}}$ component of $v$ is zero, we obtain a $5$ dimensional heterotic compactification on a space that is topologically $M\times S^1$.

As a simple choice of compactification we consider $X_{k,l} = X_{v_0}$ labeled by two non-negative integers $k,l$, where
\begin{align}
v_0 = \begin{pmatrix}  k \upomega \\ l \upomega \\ 0 \end{pmatrix}~, 
\end{align}
and $\upomega$ is a primitive anti-self-dual class in $H^2(M,\Z)$.  The heterotic Bianchi identity requires
\begin{align}
kl (-\upomega.\upomega) = 24~,
\end{align}
where we use $.$ to denote the pairing on $H^2(M,\Z)$.  Recall that $M$ is smooth if and only if it does not contain an anti-self-dual $-2$ curve~\cite{Barth:2004ne,Aspinwall:1996mn}.  Thus, to have a smooth $M$ it is necessary that $\upomega$ satisfies $\upomega.\upomega \le -4$, and there are just a few compatible values of $k$ and $l$.

 Flux quantization leads to constraints on the Narain moduli:
\begin{align}
\label{eq:quantizationrA}
r^2  + \frac{1}{2} \ba \cdot \ba = \frac{l}{k}~,
\end{align}
where $\cdot$ denotes the standard Euclidean inner product on $\R^{16}$.\footnote{As we will discuss in more detail below, this condition implies that the worldsheet CFT associated to the circle degrees of freedom is rational.  This structure also generalizes to $4$ dimensional flux compactifications based on $T^2$ principal bundles.}  For each choice of $\upomega$ we then obtain a set of compactifications $X_{k,l}$.  Geometrically we will see that $k \upomega$ determines the curvature of the circle bundle over $M$, while the second component in $v_0$, $l\upomega$, determines a quantized component of the heterotic $H$-flux shifted by the gauge bundle curvature.

For a fixed choice of $\upomega$ the $X_{k,l}$ are related to each other in a simple fashion.  $X_{1,kl}$ is a simply-connected manifold with a freely acting $G = \Z_k$ symmetry, and the resulting quotient is precisely $X_{k,l}$.     We can modify the quotient by combining the $\Z_k$ geometric action with an additional symmetry action on the gauge sector, resulting in a free action $G'$.  The quotient by $G'$ then leads to a modified CHL-like theory, which can also be interpreted as a compactification on $X_{k,l}$ with gauge bundle holonomy supported by the non-trivial $\pi_1(X_{k,l}) = \Z_k$.

On the other hand, each $X_{k,l}$ is a special case of $X_{v}$, and elements of $\GO(\Gamma_{1,17})$ act on $v$ in the fundamental representation, i.e. $g\in \GO(\Gamma_{1,17})$ is represented as an $18\times 18$ integer-valued matrix satisfying
\begin{align}
g^{\text{t}} \eta g = \eta~.
\end{align}
The configurations $X_{v}$ and $X_{g v}$ are T-dual, and by choosing $g$ appropriately, we can relate a flux compactification to a compactification on $M\times S^1$.  For example, given a configuration $X_{1,l}$, as above
we can find $g \in\GO(\Gamma_{1,17})$ such that
\begin{align}
v_1 = g v_0 = \begin{pmatrix} 0 \\ \upomega \\ -\bkappa \upomega\end{pmatrix} ~,
\end{align}
where $\bkappa$ is a lattice vector $\bkappa \in \Gamma_8 + \Gamma_8$  obeying $\bkappa\cdot\bkappa = 2 l$.

The T-duality relates the circle moduli as follows.  The compactification $X_{v_0}(r,\ba)$ is T-dual to $X_{v}(r',\ba')$, with
\begin{align}
r '& =\frac{r}{2l + \ba\cdot\bkappa}~,&
\ba' & = -\frac{\ba +\bkappa}{2l +  \ba\cdot\bkappa}~.
\end{align}
The quantization condition~(\ref{eq:quantizationrA}) is mapped in the T-dual description to
\begin{align}
\ba' \cdot \bkappa = -1~.
\end{align}
Thus, we traded the non-trivial fibration of a circle with its quantization condition for a trivial fibration but a non-trivial abelian gauge bundle with a quantization condition on the projection of the Wilson line along the abelian instanton.

The compactifications on $X_{v_0}$ and $X_{v_1}$ are isomorphic at the level of worldsheet CFT.
Denoting the corresponding CFTs by $\cC[X_{v_0}]$ and $\cC[X_{v_1}]$, and the isomorphism by $\cT : \cC[X_{v_0}]\rightarrow \cC[X_{v_1}]$,  we can then map the action of any symmetry of one theory to the dual formulation.  More precisely, given a symmetry group $G_0$ of $\cC[X_{v_0}]$ with elements $g \in G_0$, we also have an isomorphic action on $\cC[X_{v}]$ by a group $G$ with elements $\cT g \cT^{-1}$.  It follows that we also have an isomorphism of orbifold CFTs:
\begin{align}
\cC[X_{v_0}] /G_0 \simeq \cC[X_{v}]/G~.
\end{align}
As an example of a quotient action, we can consider $G_0 = \Z_2$, with generator $g_{\text{chl}}$, which we can write as a product of two actions:  $g_{\text{shift}}$, which acts as the shift orbifold of $S^1$, and $g_{\text{flip}}$, which exchanges the two $\Le_8$ factors.   The latter is a symmetry of $\cC[X_{v_0}]$ if the Wilson line satisfies
\begin{align}
\begin{pmatrix} 0 & \iden_8 \\ \iden_8 & 0 \end{pmatrix}  \ba = \ba .
\end{align}
Just as in~\cite{Chaudhuri:1995bf}, the resulting CHL orbifold $\cC[X_{v_0}]/G_0$ has a straightforward geometric interpretation:  it is a compactification on $X_{2,l}$ with a holonomy turned on for the $\Z_2$ factor in the $(\GE_8\times\GE_8 )\rtimes \Z_2$ gauge group of the heterotic string.  By our arguments this heterotic flux CHL vacuum has a T-dual description in terms of a CHL orbifold $\cC[X_{v}]/G$.

On one hand, the example shows that the heterotic flux vacua based on $X_{v_0}$ and their orbifolds do not add new components to the moduli space of $5$--dimensional compactifications: there is always an equivalent product geometry $X_{v} \simeq M\times S^1$.  However, as usual with duality, there may be features that are simpler to understand in terms of the description based on $X_{v_0}$.    For instance, it is very easy to see in the $X_{v_0}$ description that setting $\ba = 0$ results in unbroken $\Le_8\oplus \Le_8$ gauge symmetry.  This is less obvious in the dual description $X_v$ with $r = \frac{1}{2 \sqrt{l}}$ and $\ba = -\frac{\bkappa}{2l}$.  As another example, we can consider the equivalence between the two CHL actions.  As we just discussed, the action on $X_{v_0}$ by $g_{\text{chl}} = g_{\text{shift}} g_{\text{flip}}$ has a straightforward geometric interpretation.  Its T-dual is given by
\begin{align}
g'_{\text{chl}} = \cT g_{\text{chl}} \cT^{-1} = (\cT g_{\text{shift}}\cT^{-1}) (\cT g_{\text{flip}}\cT^{-1})~.
\end{align}
We will see below that the dual action $\cT g_{\text{shift}} \cT^{-1}$ does not have a simple geometric interpretation.

\subsection{Non-abelian gauge sector}
Our example only involved a choice of abelian gauge background over $M$ in the $9$-dimensional gauge group.   More generally, we can consider a decomposition for each $\Le_8$ factor into
\begin{align}
(\Le_8)_{1,2} \supset \Lg_{1,2} \oplus \mathfrak{C}_{1,2}~,
\end{align}
where $\Lg_{1,2}$ is a semi-simple subalgebra and $\mathfrak{C}_{1,2}$ is its commutant subalgebra.  We can then turn on irreducible Hermitian Yang-Mills (HYM) connections that fill out the $\Lg_{1,2}$ factors and at the same time take a generic Wilson line valued in the Cartan subalgebra of the commutants $\LLh=\LLh_1\oplus\LLh_2$.  Supposing $\LLh$ to have rank $r$, the topology of the flux vacuum is then determined by the instanton numbers $k_1$ and $k_2$ for the HYM connections, as well as the vector $v$ of the form
\begin{align}
v & = \begin{pmatrix} \upomega \\ \upnu \\ \bUplambda \end{pmatrix}~,
\end{align}
where $\bUplambda \in (\Gamma_8+\Gamma_8)\otimes H^2(M,\Z)$ is constrained to lie in the orthogonal complement to $\Gamma_{\Lg_1} + \Gamma_{\Lg_2}$ and describes the topology of an additional abelian instanton configuration. 
Given such a flux vacuum, we can consider the orbit of the T-duality group $G_{\text{T}} \subset \GO(\Gamma_{1,17})$, where $G_{\text{T}}$ is the subgroup that preserves the $\Gamma_{\Lg_1} +\Gamma_{\Lg_2}$ sublattice.    

Because the action of T-duality is now restricted, it is not necessarily the case that the T-duality orbit of a heterotic flux vacuum contains a product geometry $M\times S^1$.  For example, setting $k_1=k_2 = 10$ and choosing $\Lg_1 = \Lg_2 = \Le_8$, we must take
\begin{align}
v & = \begin{pmatrix} k \upomega \\ l \upomega \\ 0\end{pmatrix}~,
\end{align}
with
\begin{align}
kl (-\upomega.\upomega) = 4~.
\end{align}
With a smooth K3 the only solution is $\upomega.\upomega = -4$, and $k=l=1$.  Since in this case $G_{\text{T}}$ must preserve the zero entries in $v$, there is no product geometry in the T-duality orbit.

We can get a better perspective by considering the quantization of the moduli.
Since the Wilson line must also be valued in the commutant of the holonomy of the HYM connection in this case $\ba = 0$, and thus the radius is fixed to its self-dual value:  $r^2 = 1$.  Since the circle is now at the self-dual radius, the $9$--dimensional theory has an $\su(2) \oplus \Le_8\oplus\Le_8$ gauge symmetry, and the worldsheet theory for the $5$-dimensional target space is better thought of as a fibered WZW model, as described in~\cite{Melnikov:2012cv}.  From this perspective we recognize the compactification as a $5$-dimensional lift of a familiar  $4$-dimensional vacuum with $2$ vector multiplets and $129$ neutral hypermultiplets discussed in~\cite{Kachru:1995wm}.  In this case there is no element in $G_{\text{T}}$ that can lead to a dual product geometry, simply because a compactification on $M\times S^1$ will necessarily have a $\Lu(1)$ spacetime gauge symmetry. 

On the other hand, each $\Le_8$ HYM connection has a moduli space of  quaternionic dimension $30 k -248 = 52$, and there is a locus in the moduli space where the connection is contained in an $\su(2)$ subalgebra with instanton number $10$.  At this locus in the moduli space it is possible to T-dualize the theory by using the map discussed above, leading to a dual configuration with geometry $M\times S^1$ and a gauge instanton with holonomy $\SU(2) \times \SU(2) \times \SU(2)$, with the first factor carrying instanton number $4$, and the latter each having instanton number $10$.

\subsection{\texorpdfstring{$4$}{4}-dimensional flux vacua}
While the $5$-dimensional example is simple and already illustrates some of our main points, compactification to $4$ dimensions offers new possibilities.  For example, there are quotients $\cC[X_{v}]/\Gamma$ that preserve exactly $4$ supercharges in $4$ dimensions.  The simplest of these has a geometric action that combines an Enriques involution on the base K3 $M$ with a reflection symmetry on the torus fiber.  We will show that the T-dual geometry is in this case the Enriques Calabi--Yau $3$-fold equipped with an appropriate gauge bundle.\footnote{Unlike in~\cite{Ferrara:1995yx}, where this manifold first appeared in the context of string duality, here we consider it as the target space for a heterotic compactification.}  We will also discuss a class of such quotients based on a torsional linear sigma model~\cite{Israel:2023itj}, and show how the T-duality can be understood as a relationship between torsional and conventional linear sigma models.

As already observed in~\cite{Goldstein:2002pg,Becker:2006et}, it is also possible to construct $\SU(3)$ structure heterotic flux vacua as holomorphic $T^2$ principal bundles over $M$.  These are considerably more subtle to analyze, and their status as vacua of the perturbative heterotic string is on less solid ground~\cite{Becker:2009df,Melnikov:2014ywa}, but, on the other hand, it is possible to think of them as compactifications of the heterotic $8$-dimensional theory, and therefore, to the extent that these vacua exist, we can also consider the action of T-duality on the corresponding configurations.  As expected, their T-duality orbits cannot include a $M\times T^2$ geometry.  We explore these orbits and use the description to characterize these compactifications.   Our main finding is that the base K3 must have Picard number $\ge 18$.  Such surfaces are quite special, and there is a mathematical conjecture~\cite{Oda:1980}, proven for Picard number $20$ and $19$~\cite{Morrison:1984}, that such K3 surfaces can be classified through a generalization of the Shioda--Inose construction.  Thus, although they cannot be T-dualized to K\"ahler geometries, these N=1 heterotic flux vacua may well admit a classification.

\subsection{Organization}

In the previous section we gave a sketch of some of the key results of our study.  In what follows we will expand upon this by providing the details and necessary background.  Since our analysis relies on T-duality in a Narain compactification, we will begin in section~\ref{s:lattice} with an overview of our conventions for $\Gamma_{1,17}$ and the relevant CFT structures, leaving a few technical aspects to appendices.  In section~\ref{s:fluxvacua} we will review the geometry of heterotic flux vacua, emphasizing the quantization conditions on the moduli.  Sections~\ref{s:5dimflux8supercharge} and \ref{s:4dimflux8supercharge} are devoted to flux vacua that preserve $8$ supercharges and their orbifolds:  we describe the ``unwinding T-dualities'' that relate flux vacua to compactifications on $M\times T^2$ and follow the implications these have for orbifold theories.  At the end of section~\ref{s:4dimflux8supercharge}, we use linear sigma model technology to give a non-trivial test of the unwinding isomorphisms. In section~\ref{s:su3structure} we turn to the four-dimensional flux vacua with $\SU(3)$ structure.  We then conclude with an outlook, where we comment on type II and M-theory dual descriptions, as well as possible extensions to flux vacua of~\cite{Fino:2019mvp}.

\subsection*{Acknowledgements}
IVM's work is partially supported by the Humboldt Research Award and the Jean d'Alembert Program at the University of Paris--Saclay, as well as the Educational Leave program at James Madison University.  
Part of this work was carried out while IVM and RM~were visiting  the Albert Einstein Institute (Max Planck Institute for Gravitational Physics) as well as LPTHE at Sorbonne Universit\'e, and they are grateful to both AEI and LPTHE for hospitality.  We would like to thank P.~Cheng, M.~Gra\~na, and A.~Sarti for useful conversations.

\section{Elements of \texorpdfstring{$\GO(\Gamma_{d,d+16})$}{O(d,d+16)}} \label{s:lattice}
We begin by setting up conventions for the heterotic Narain compactification to $d$ dimensions.\footnote{This is a well-known story.  A classic review is~\cite{Giveon:1994fu}.  A recent overview can be found in introductory sections of~\cite{Font:2020rsk}.  Our lattice conventions, apart from a signature change, largely follow~\cite{Cheng:2022nso} and~\cite{Aspinwall:1996mn}.}   

\subsection{Lattice set up}
\label{subsec:latticesetup}
Consider $\R^{d,d+16}$ equipped with the Minkowski metric $\eta$ given as above:
\begin{align}
\label{eq:eta}
\eta = \begin{pmatrix} 0 & \iden_{d} & 0 \\ \iden_{d} & 0 & 0 \\ 0 & 0 & \iden_{16} \end{pmatrix}~.
\end{align}
We denote the corresponding inner product by $\cdot$, so that  $v_1 \cdot v_2 = v^{\text{t}}_1 \eta v_2$ for any two vectors $v_{1,2} \in \R^{d,d+16}$.  Next we choose a fiducial embedding of the lattice\footnote{We choose this form for the lattice because our interest is in the $\GE_8\times\GE_8$ heterotic string.  The relationship to $\Gamma_{d,d}+\Gamma_{16}$~\cite{Ginsparg:1986bx} is nicely reviewed as the ``HE$\rightarrow$HO'' map in~\cite{Font:2020rsk}.} 
\begin{align}
\Gamma_{d,d+16} = \Gamma_{d,d} + \Gamma_8 + \Gamma_8 \subset \R^{d,d+16}
\end{align} 
so that the lattice inner product is given by ``$\cdot$''.   We denote the generators of $\Gamma_{d,d}$ by $\be_I$ and $\be^{\ast I}$,  with $I=1,\ldots, d$.  These are null vectors satisfying $\be_I\cdot\be^{\ast J} = \delta_I^J$.  For each $\Gamma_8 \subset \R^{8}$ we choose the generators to be simple roots $\balpha_i$ with inner product $\balpha_i \cdot \balpha_j$ normalized so that roots have length squared $2$, and we denote the two mutually orthogonal sets of roots by $\balpha_i$ and $\balpha'_i$.  We will sometimes abuse the notation and combine the simple roots into a single set $\balpha_1,\ldots,\balpha_{16}$, with the first $8$ corresponding to the first $\Gamma_8$ factor, and the last $8$ to the second $\Gamma_8$ factor.  With that notation, every lattice point $\bp \in \Gamma_{d,d+16}$ is uniquely written as 
\begin{align}
\bp = \text{w}^I \be_I + \text{n}_I \be^{\ast I} + \bL~,
\end{align}
where 
\begin{align}
\bL = \bell + \bell' =  \textstyle\sum_i \ell^i \balpha_i +  \textstyle\sum_i \ell'^i \balpha'_i~,
\end{align}
and $\text{w}$, $\text{n}$, $\ell^i$ $\ell'^i$ are integer coefficients.  It will also be convenient for us to fix a Cartan--Killing basis for $\R^{16}$ with orthonormal basis vectors $\bv_a$,  $a=1,\ldots,16$, with respect to which 
\begin{align}
\label{eq:CKbasis}
\bL = \sum_a \ell^a \bv_a~.
\end{align}
It follows that for any $\bp \in \Gamma_{d,d+16}$
\begin{align}
\bp\cdot \bp = 2\text{n}_I\text{w}^I + \bL \cdot \bL  \in 2\Z~.
\end{align}

Specializing to $d=1$, we denote the heterotic moduli as above by a pair $(r, \ba)$, where $\ba\in\R^{16}$ is the set of Wilson line parameters.  A choice of $(r,\ba)$ determines a point in the Grassmannian $\Gr(1,17)$ through the orthogonal basis for $\R^{1,17}$ consisting of vectors  $\bpit$ , $\bpi$, and $\bpi^\circ_a$, $a=1,\ldots,16$ written with respect to the fiducial basis as
\begin{align}
\bpit  & = \be -\left(r^2 + \ff{1}{2} \ba \cdot \ba \right) \be^\ast - \ba~,\nonumber\\
\bpi   & = \bpit + 2r^2 \be^\ast~, \nonumber\\
\bpi^\circ_a & = \bv_a + (\bv_a \cdot \ba) \be^\ast~.
\end{align}
These have non-zero inner products
\begin{align}
\bpit \cdot \bpit &=-2r^2~,&
\bpi \cdot \bpi & = 2r^2~,&
\bpi^\circ_a \cdot \bpi^\circ_b & = \delta_{ab}~.
\end{align}
More generally, we introduce the torus metric and $B$-field $\cG_{IJ}$ and $\cB_{IJ}$, so that
\begin{align}
\label{eq:NarainCoset}
\bpit_I & = \be_I - \left(\cG_{IJ} - \cB_{IJ} +\ff{1}{2} \ba_I \cdot \ba_J\right) \be^{\ast J} - \ba_I~, \nonumber\\
\bpi_I & = \bpit_I + 2 \cG_{IJ} \be^{\ast J}~,\nonumber\\
\bpi_a^{\circ} & = \bv_a + (\bv_a \cdot \ba_I) \be^{\ast I}~
\end{align}
have non-zero inner products
\begin{align}
\bpit_I \cdot \bpit_J & = -2 \cG_{IJ}~,&
\bpi_I \cdot \bpi_J & = 2 \cG_{IJ}~,&
\bpi^\circ_a \cdot \bpi^\circ_b & = \delta_{ab}~.
\end{align}
The parametrization of the Narain  moduli space is chosen to be consistent with the NLSM conventions of~\cite{Melnikov:2012cv}, so that we can meaningfully relate this structure to the quantization conditions on the moduli obtained in that work.  

\subsection{Vertex operators}
The heterotic worldsheet theory consists of the Narain CFT with $d+16$ left-moving chiral bosons $X^I_{\sleft}(z)$, $\cX_{\sleft}^a(z)$, and $d$ right-moving chiral bosons $X^I_{\sright}(\zb)$.  For a geometric description we think of $X_{\sleft}^I$ and $X_{\sright}^I$ as the holomorphic and antiholomorphic components of the compact bosons describing the torus, while the $\cX_{\sleft}^{a}$ can be thought of as representing the heterotic worldsheet (Weyl) fermions. 

In order to build the full heterotic string we should also include the right-moving fermion $\psi^I_{\sright}(\zb)$---these are the superpartners of $X^I_{\sright}$, as well as the Minkowski degrees of freedom for $\R^{1,9-d}$ and the $bc$--$\beta\gamma$ ghost system.  These degrees of freedom (and the accompanying right-moving GSO projection) will play a spectator role in our discussion, so we will for the most part focus on the Narain CFT.

At a generic point in the moduli space the Narain CFT has a $\Lu(1)^{\oplus d+16}_{\sleft} \oplus \Lu(1)^{\oplus d}_{\sright}$ Kac--Moody symmetry, and the primary operators with respect to this symmetry are the vertex operators $\cV_{\bp}$ labeled by lattice points in $\Gamma_{d,d+16}$.  These have weights $h_{\sleft}(\bp)$, $h_{\sright}(\bp)$ that depend on the moduli.  With our basis they are expressed as follows.  First, the operator's spin is determined by $\bp$ alone:
\begin{align}
s(\bp) = h_{\sleft}(\bp) - h_{\sright}(\bp) = \frac{\bp \cdot \bp}{2} = \text{n}_I\text{w}^I + \ff{1}{2} \bL\cdot \bL~.
\end{align}
The right-moving weight is given by
\begin{align}
h_{\sright}(\bp) & = \frac{1}{4} \cG^{IJ} (\bpit_I \cdot \bp) (\bpit_J \cdot\bp)~, &
\bpit_I\cdot \bp = \text{n}_I- (\cG_{IJ} - \cB_{IJ} +\ff{1}{2} \ba_I \cdot \ba_J)\text{w}^J - \ba_I \cdot \bL~,
\end{align}
where $\cG^{IJ}$ denotes the inverse torus metric.  In the special case of $d=1$ this reduces to 
\begin{align}
\label{eq:rightweight}
h_{\sright}(\bp) = \frac{1}{4r^2} \left(\bpit \cdot \bp\right)^2 =\frac{1}{4 r^2} \left( \text{n} - \left(r^2 + \ff{1}{2} \ba \cdot\ba\right) \text{w} - \ba \cdot \bL\right)^2~.
\end{align}

\subsection{Dualities and symmetries}
Consider the action of a group $G$ on the vertex operators $\cV_{\bp}$ of the following form:  for any element $g$ the action is
\begin{align}
g \circ \cV_{\bp} = U(g,\bp) \cV_{\varphi_{g}(\bp)}~,
\end{align}
where $\varphi_{g}(\bp)$ is a lattice automorphism, and $U(g,\bp)$ is a phase.  Consistency with group multiplication requires the phases and lattice isomorphisms to obey
\begin{align}
\label{eq:cocycleproduct}
U(g_2,\varphi_{g_1}(\bp)) U(g_1,\bp) &= U(g_2 g_1,\bp)~,&
\varphi_{g_2} (\varphi_{g_1}(\bp)) & = \varphi_{g_2 g_1} (\bp)~.
\end{align}
Thus,  $\varphi$ can be thought of as a map $\varphi : G \to \GO(\Gamma)$, and $G_{\Gamma} = \GO(\Gamma)/\ker \varphi$ is a subgroup of the lattice automorphisms.\footnote{To slightly lighten notation we abbreviate $\Gamma_{d,d+16}$ to $\Gamma$ when we are making general statements and there is not much potential for confusion.}

In order to give a  well-defined map of conformal field theories, the action should also be compatible with the OPE.  This requires $\varphi_{g}$ to be a lattice automorphism, i.e. an invertible linear map that preserves the inner product, and it also leads to constraints on the phases $U(g,\bp)$.  When $G_{\Gamma} = 1$, i.e. $\varphi_{g} = \text{id}$ for all $g \in G$, the requirement is simply that $U$ is a map $U: G\to \Hom(\Gamma,\GU(1))$.  More generally, the constraint on the $U(g,\bp)$ is more subtle and involves a choice of cocycle.\footnote{This is nicely reviewed in some detail in~\cite{Tan:2015nja}.  Our presentation here borrows from a slightly more comprehensive discussion in~\cite{Cheng:2022nso}.}

Recall that when we represent the Narain vertex operators in terms of chiral bosons we need to include additional zero mode operators to ensure that the operators obey the correct commutation relations, and the algebra of these operators must satisfy additional requirements in order to have an associative OPE~\cite{Polchinski:1998rq,Green:1987mn}.  This requires a choice of representative $\varepsilon$ for a class in the group cohomology $H^2(\Gamma,\GU(1))$ that satisfies
\begin{align}
\varepsilon(\bp_2,\bp_1) = e^{i\pi \bp_1\cdot\bp_2} \varepsilon(\bp_1,\bp_2)~.
\end{align}
We give an explicit formula for this representative in appendix~\ref{app:cocyclerep}.  In order to be consistent with the OPE, the phase assignment $U(g,\bp)$ must obey~ for all $\bp_1,\bp_2\in\Gamma$
\begin{align}
\label{eq:cocyclecondition}
\frac{ U(g,\bp_1 + \bp_2)}{U(g,\bp_1) U(g,\bp_2)} = \frac{ \varepsilon(\varphi_g(\bp_1),\varphi_g(\bp_2))}{\varepsilon(\bp_1,\bp_2)}~.
\end{align}
For a fixed $g$ the ratio of any two solutions $U$ and $U'$ to this condition belongs to $\Hom(\Gamma,\GU(1))$. 

The action of $g$ on the Narain CFT is not in general a symmetry because it acts on the moduli.  More precisely, we have seen above how to express the right-moving weight $h_{\sright}(\bp)$ in terms of the parameters $\zeta \in \Gr(d,d+16)$.  Making this dependence explicit by writing $h_{\sright}(\zeta;\bp)$, we then see that for any $g$ we obtain a map on the parameter space $\mu_g :  \Gr(d,d+16) \to \Gr(d,d+16)$ defined by demanding that for all $\bp \in \Gamma$
\begin{align}
h_{\sright}(\mu_g(\zeta);\bp) = \varphi_{g}^\ast(h_{\sright}(\zeta;\bp)) = h_{\sright}(\zeta;\varphi_g(\bp))~.
\end{align}
It is not hard to see that $\mu_g$ defined this way is unique.  An action $g$ is then a symmetry of the Narain CFT with moduli $\zeta$ if and only if $\mu_g(\zeta) = \zeta$.

\subsubsection*{T-duality for $\Gamma_{1,17}$}
Having reviewed the general structure, we specialize to the case of heterotic string on a circle and discuss the T-duality group $\GO(\Gamma_{1,17})$.  The group has the following generators~\cite{Fraiman:2018ebo}, for each of which we provide the lattice automorphism as an action on
\begin{align}
\bp = \text{w} \be + \text{n} \be^\ast + \bL=  \text{w} \be + \text{n} \be^\ast  +\textstyle\sum_{i} \ell^i \balpha_i + \textstyle\sum_{i} \ell'^i \balpha'_i~,
\end{align} 
a phase $U(g,\bp)$, and the map of the moduli.\footnote{The details of the phase computations are given in appendix~\ref{app:cocyclerep}.}
\begin{enumerate}
\item The factorized duality transformation $g_{\text{i}}$ with
\begin{align}
\varphi_{g_\text{i}} (\bp) &= \text{n} \be + \text{w} \be^\ast + \bL~, &
U(g_{\text{i}},\bp) &= e^{i \pi \text{nw}}~, \nonumber\\
\mu_{g_{\text{i}}}(r,\ba) & = \left( \frac{r}{r^2 + \ff{1}{2} \ba\cdot\ba},\frac{-\ba}{r^2 + \ff{1}{2} \ba\cdot\ba}\right)~.
\end{align}
\item The circle reflection $g_{\text{ref}}$ with
\begin{align}
\varphi_{g_\text{ref}}(\bp) &= -\text{w} \be - \text{n} \be^\ast + \bL~; & 
U(g_{\text{r}},\bp) & = 1~, \nonumber\\
\mu_{g_{\text{ref}}}(r,\ba)& = (r,-\ba)~.
\end{align}
\item Wilson line shifts $g_{\text{s}}[{\boldsymbol{\kappa}}]$ labeled by ${\boldsymbol{\kappa}} \in \Gamma_8 + \Gamma_8$ with
\begin{align}
\varphi_{g_\text{s}}[{\boldsymbol{\kappa}}] (\bp) = \text{w} \be + \left(\text{n} + {\boldsymbol{\kappa}}\cdot\bL -\ff{1}{2} {\boldsymbol{\kappa}}\cdot{\boldsymbol{\kappa}}\text{w}\right) \be^{\ast} + \bL - {\boldsymbol{\kappa}} \text{w}~.
\end{align}
The phase here is a little more complicated and involves a choice of simple roots $\balpha_I$, $I=1,\ldots,16$ for $\Gamma_8+\Gamma_8$.  We leave the details to the appendix and merely quote the result in terms of the bilinear integer-valued map $S_{{\boldsymbol{\kappa}}}(\bL)$ defined therein:
\begin{align}
U(g_{\text{s}}[{\boldsymbol{\kappa}}],\bp) = e^{i \pi (\text{w}+1) S_{{\boldsymbol{\kappa}}}(\bL)}~.
\end{align}
The map on the moduli is just a shift of the Wilson line by the lattice vector:
\begin{align}
\mu_{g_{\text{s}}}[{\boldsymbol{\kappa}}] (r,\ba) = (r,\ba- {\boldsymbol{\kappa}})~.
\end{align}

\item The final set of generators consists of rotations in $\Gamma_8+\Gamma_8$.  As we will not need the detailed form for these actions, we relegate the general discussion to the appendix, and here just mention one particular rotation: the exchange of the two $\Gamma_8$ factors, which acts by
\begin{align}
\label{eq:CHLflip}
\varphi_{g_{\text{flip}}} (\bp) &= \text{w} \be + \text{n} \be^\ast  +\textstyle\sum_{i} \ell'^i \balpha_i + \textstyle\sum_{i} \ell^i \balpha'_i~, &
U(g_{\text{flip}},\bp) = 1~, \nonumber\\[2mm]
\mu_{g_{\text{flip}}}(r,\ba) &= \mu_{g_{\text{flip}}}(r,\text{a},\text{a}') = (r,\text{a}',\text{a})~.
\end{align}

\end{enumerate}
All of these actions can be interpreted as redefinitions on the chiral bosons.  For example, $g_{\text{i}}$ corresponds to $X_{\sright} \to -X_{\sright}$ while keeping the other bosons fixed, while $g_{\text{ref}}$ reflects both $X_{\sleft}$ and $X_{\sright}$.  This has an important consequence for the heterotic conformal field theory, because compatibility with worldsheet supersymmetry requires that the reflection of $X_{\sright}$ must be accompanied by a reflection on its right-moving superpartner fermion.  Notice, however, that the action by $g_{\text{i}} g_{\text{ref}}$ is simply in the left-moving sector, and so does not have this complication.

\subsubsection*{T-duality for $\Gamma_{2,18}$}
A description of the group $\GO(\Gamma_{2,18})$ associated to the heterotic string on $T^2$ will also be useful for studying T-duality orbits of four-dimensional compactifications. We leave the specifics of the cocycle representatives and moduli transformation to the appendix~\ref{app:cocyclerep}, and only list our choice of generators along with their lattice action below. 
\begin{enumerate}
\item Factorized dualities $g_{\text{i},I}$ in the two circle directions with
\begin{align}
\varphi_{g_{\text{i},1}}(\bp)&=\text{n}_1\be_1+\text{w}^2\be_2+\text{w}^1\be^{\ast 1}+\text{n}_2\be^{\ast 2}+\bL~,\nonumber\\
\varphi_{g_{\text{i},2}}(\bp)&=\text{w}^1\be_1+\text{n}_2\be_2+\text{n}_1\be^{\ast 1}+\text{w}^2\be^{\ast 2}+\bL~.
\end{align}
\item Torus isometries $g_{\text{t}}[\cR]$ labelled by a $\SL(2,\Z)$ matrix $\cR$ with
\begin{align}
\varphi_{g_{\text{t}}}[\cR](\bp)&=\cR^I{}_J\text{w}^J\be_I+(\cR^{-1})^J{}_I\text{n}_J\be^{\ast I}+\bL~.
\end{align}
\item Reflections $g_{\text{ref},I}$ in the two circle directions with
\begin{align}
\varphi_{g_{\text{ref},1}}(\bp)&=-\text{w}^1\be_1+\text{w}^2\be_2-\text{n}_1\be^{\ast 1}+\text{n}_2\be^{\ast 2}+\bL~,\nonumber\\
\varphi_{g_{\text{ref},2}}(\bp)&=\text{w}^1\be_1-\text{w}^2\be_2+\text{n}_1\be^{\ast 1}-\text{n}_2\be^{\ast 2}+\bL~.
\end{align}
\item $B$-field shifts $g_{\text{b}}[m]$ labelled by $m\in\Z$ with
\begin{align}
\varphi_{g_{\text{b}}}[m](\bp)&=\text{w}^I\be_I+(\text{n}_I+m\epsilon_{IJ}\text{w}^J)\be^{\ast I}+\bL~.
\end{align}
\item Wilson line shifts $g_{\text{s}}[\bkappa_I]$ labelled by two lattice vectors\footnote{For the transformation $g_{\text{s}}[\bkappa_1,\bkappa_2]$ to belong to $\GO(\Gamma_{2,18})$, the lattice vector should obey $\bkappa_1\cdot\bkappa_2\in2\Z$.} $\bkappa_I\in\Gamma_{8}+\Gamma_{8}$ with
\begin{align}
\varphi_{g_{\text{s}}}[\bkappa_I](\bp)&=\text{w}^I\be_I+(\text{n}_I+\bkappa_I\cdot\bL-\tfrac{1}{2}\bkappa_I\cdot\bkappa_{J}\text{w}^{J})\be^{\ast I}+\bL-\bkappa_I\text{w}^I~.
\end{align}
It will be useful to consider Wilson line shifts $g_{\text{s},I}[\bkappa]$ in a single circle direction, defined by $g_{\text{s},1}[\bkappa]=g_{\text{s}}[\bkappa,0]$ and $g_{\text{s},2}[\bkappa]=g_{\text{s}}[0,\bkappa]$.
\item Rotations in the gauge lattice $g_{\text{g}}[R]$ labelled by a matrix $R\in\GO(\Gamma_{8}+\Gamma_{8})$ with
\begin{align}
\varphi_{g_{\text{g}}}[R](\bp)&=\text{w}^I\be_I+\text{n}_I\be^{\ast I}+R(\bL)~.
\end{align}
\end{enumerate}

\section{A review of heterotic flux vacua} \label{s:fluxvacua}

\subsection{The topology of principal torus bundles}
In this section we review aspects of the topology of the $6$-dimensional geometry $X$ underlying the heterotic flux vacua.   While a significant portion of the discussion is a review of known results, we give a more complete discussion of the topology of $X$ and the quantization conditions on the geometric parameters imposed by the presence of fluxes, both in the case of backgrounds with $8$ and $4$ supercharges.

Let $p:X\to M$ be a $T^2$ principal bundle over a smooth K3 surface $M$.  The topology of the torus bundle is specified by two classes $\upomega^I \in H^2(M,\Z)$,  $I=1,2$.    Denote by $\cT_I$ the Hermitian line bundles over $M$ associated to the principal bundle with $\upomega^I = c_1(\cT_I)$.  Recall that $H^2(M,\Z)$ is an even self-dual lattice $\Gamma_{3,19}$, while $H_1(T^2,\Z) = \Z\times\Z$.   By fixing a basis for each of these lattices, the specification of the $\upomega^I$ amounts to choosing an integral $2\times 22$ matrix, and by bringing this matrix to Smith normal form we find that the $\upomega^I$ can be expressed as
\begin{align}
\label{eq:torusclasses}
\upomega^1 & = m_1 \upomega^1_{\text{p}}~, &
\upomega^2 & = m_1 m_2 \upomega^2_{\text{p}}~,
\end{align}
where $m_1$, $m_2$ are non-negative integers, while $\upomega^I_{\text{p}}$ are linearly independent primitive elements in $\Gamma_{3,19}$.  We can complete the latter to a basis for $H^2(M,\Z)$:
\begin{align}
\label{eq:specialbasis}
H^2(M,\Z)  = \text{Span}_{\Z} \{ \upomega^1_{\text{p}}, \upomega^2_{\text{p}}, \upchi_{1},\upchi_{2},\ldots, \upchi_{p}\}~
\end{align}
which will be convenient for our analysis.  We obtain special cases by setting $m_2=0$ or $m_1=0$:  the former corresponds to $X = Y\times S^1$, where $Y$ is a principal circle bundle over $M$, and the latter to the trivial fibration  $X = M \times T^2$.

The homotopy groups of $X$ are determined by the long exact sequence of homotopy groups for a fiber bundle~\cite{Davis:2001at,Hofer:1993tpb}, combined with the Hurewicz theorem and the fact that $M$ is simply connected.  The result is that for $i\ge 3$ $\pi_i(X) = \pi_i(M)$, while $\pi_2(X)$ and $\pi_1(X)$ fit into the exact sequence
\begin{equation}
\begin{tikzcd}
1 \ar[r] &
\pi_2(X) \ar[r] &
H_2(M,\Z) \ar[r,"f"] &
H_1(T^2,\Z) \ar[r] &
\pi_1(X) \ar[r] &
1~.
\end{tikzcd}
\end{equation}
The map $f$ is given by integration of the two classes $\upomega^I$, or, using the isomorphism $H_2(M,\Z) \simeq H^2(M,\Z)$, by intersection of cycles:  for any class $C \in H^2(M,\Z)$
\begin{align}
f(C) = \upomega^1.C  \boldsymbol{t}_1 + \upomega^2.C  \boldsymbol{t}_2~,
\end{align}
where $\boldsymbol{t}_I$ are the generators of $H_1(T^2,\Z) \simeq \Z \times \Z$.   The non-trivial homotopy groups of $X$ are then
\begin{align}
\pi_1(X) &\simeq H_1(T^2,\Z)/\text{im}(f)~,&
\pi_2(X) &\simeq \ker f~.
\end{align}
By using~(\ref{eq:torusclasses}) and~(\ref{eq:specialbasis}), it is easy to see that
\begin{align}
\pi_1(X) & = \Z_{m_1} \times \Z_{m_1m_2}~.
\end{align}

The integral cohomology of $X$ is easy to determine by the Leray--Serre spectral sequence because the base space is simply connected,  and the fiber does not degenerate.\footnote{Equivalently we could also apply the Gysin spectral sequence in two steps, as discussed in~\cite{Goldstein:2002pg}, for example.}  We summarize the results here and present the details in appendix~\ref{app:LeraySerre}.  Denoting for brevity $H^{\bullet}(X,\Z)$ by $H^{\bullet}(X)$, we find that for $m_1, m_2>0$
\begin{align}
H^0(X) & = \Z~, &
H^1(X) & = 0~,&
H^2(X) & = \Z^{20} \times \Z_{m_1} \times \Z_{m_1m_2}~,\nonumber\\
H^3(X) & = \Z^{42} \times \Z_{m_1}~, &
H^4(X) & = \Z^{20} \times \Z_{m_1}~,&
H^5(X) & = \Z_{m_1} \times \Z_{m_1m_2}~, \nonumber\\
H^6(X) & = \Z~.
\end{align}
When $m_2 =0$, then after factoring out a circle in $X = Y \times S^1$, we find for $m_1>0$
\begin{align}
H^0(Y) & = \Z~, &
H^1(Y) & = 0~,&
H^2(Y) & = \Z^{21} \times \Z_{m_1}~,\nonumber\\
H^3(Y) & = \Z^{21}~, &
H^4(Y) & = \Z_{m_1}~,&
H^5(Y) & = \Z~.
\end{align}
In either case it is not hard to check that these results are consistent with the universal coefficients theorem and Poincar{\'e} duality.  

When both circles are non-trivially fibered there are two independent torsion subgroups: those in $H^2(X)$ and $H^3(X)$.  The former is isomorphic to $\pi_1(X)$, while the latter characterizes flat abelian gerbes over $X$.  We have the important result that the universal cover  $\nu : \Xt \to X$ is itself a $T^2$ principal bundle over the same base~\cite{Hofer:1993tpb}.  The same result holds for $Y$ when $m_2=0$: its universal cover $\Yt$ is an $S^1$ principal bundle over $M$.  In either situation the cohomology is torsion-free whenever the space is simply connected.

We end this topological review with a description of line bundles over $X$ with $m_1,m_2>0$.  A line bundle $ L \to X$ is topologically characterized by its first Chern class $c_1(L) \in H^2(X)$.  As we show in the appendix,
\begin{align}
H^2(X) = H^2(M) /\text{Span}_{\Z} \{ \upomega^1, \upomega^2\}~,
\end{align}
which shows that every line bundle over $X$ is topologically equivalent to a pullback bundle $p^\ast(\cL)$ where $\cL \to M$ is a line bundle over $M$, and two such pulled-back bundles $p^\ast(\cL_1)$ and $p^\ast(\cL_2)$ are isomorphic if and only if there exist integers $k_I$ such that $\cL_1 \simeq \cL_2 \otimes \cT_1^{\otimes k_1} \otimes \cT_2^{\otimes k_2}$.   In particular, each $\cT_I$ pulls back to a trivial bundle, and it will be useful for us to demonstrate this explicitly.

Fix a good cover $\mathfrak{U}$ for $M$ consisting of open sets $U_a$ with $U_{ab}$ denoting non-trivial double intersections.  Then the line bundles $\cT_I$ have transition functions $\tau_{ab}^I$, and if we denote by $\theta^I$ the local fiber coordinates, then on each $U_{ab}$ these satisfy 
\begin{align}
e^{i (\theta_{a}^I - \theta_{b}^I)} = p^\ast(\tau_{ab}^I)~.
\end{align}
Expressed in terms of the \v{C}ech coboundary operator $\delta$, we have $\delta\theta^I = -i\log p^\ast (\tau^I)$, so that the torus coordinates provide the trivialization of the transition functions.   If we denote by $\bA^I$ connections the $\cT_I$ bundles, then 
\begin{align}
\delta\, p^\ast(\bA^I)  = i \text{d} \log  p^\ast ( \tau^I) = - \text{d}\delta\theta~,
\end{align}
so that 
\begin{align}
\bTheta^I = \text{d}\theta^I + p^\ast \bA^I
\end{align}
are $1$-forms on $X$.  These nowhere vanishing $1$-forms are dual to the two global commuting vector fields $\frac{\p}{\p\theta^I}$, and the curvature $\bF^I = d\bA^I$ measures their non-closure:
\begin{align}
{\rm d}\bTheta^I = p^\ast(\bF^I)~.
\end{align}
The classes $\upomega^I$ are identified with $[\bF^I/2\pi] \in H^2(M,\R)$ under the natural inclusion $H^2(M,\Z) \hookrightarrow H^2(M,\R)$.

More generally, if $(\lambda,\cA)$ denote the transition functions and compatible connection for a line bundle $L \to X$, with $\delta \cA = i \text{d}\log \lambda$, then for any integer $k_I$ there is gauge-equivalent data
\begin{align}
(\lambda,\cA) \sim (\lambda e^{i k_I \delta \theta^I}, \cA - k_I {\rm d}\theta^I)~.
\end{align}
These large gauge transformations are the geometric origin of the Wilson line shifts of Narain T-duality that we will study below.

\subsection{The geometry of heterotic flux vacua}
We now turn to the geometric structure underlying a heterotic flux vacuum based on the torus bundle $p:X\to M$.  We will first describe the generic fibration where the $\upomega^I$ are linearly independent and hence both circles are non-trivially fibered.

Fix a hyper-K\"ahler structure on $M$ with K\"ahler form $J$ and holomorphic (2,0) form $\Omega$.  The complex structure on $X$ is then determined by writing a holomorphic (1,0) form 
\begin{align}
\Theta = \bTheta^1 + \tau \bTheta^2~,
\end{align}
where we introduced $\tau = \tau_1+i\tau_2$---a complex constant valued in the upper half-plane, and we take the (3,0) form on $X$ to be proportional to $\Theta \wedge \Omega$.  The $\SU(3)$ structure on $X$ is then given in terms of a smooth real  function $\phi$ on $M$---this is the heterotic dilaton---and the forms
\begin{align}
J_X &= e^{2\phi} J + i\,\frac{\alpha'\rho_2}{2 \tau_2}\, \Theta\wedge \Thetab~, &
\Omega_X & = e^{2\phi} \sqrt{\frac{\alpha'\rho_2}{\tau_2}}\, \Omega \wedge \Theta~.
\end{align}
Here we also introduced $\rho_2$---a positive parameter that determines the volume of the $T^2$ fiber, as well as an explicit factor of the string tension $\alpha'$.\footnote{The explicit appearance of $\alpha'$ signals an important subtlety in the construction: whenever the torus fibration is non-trivial, there will be string-scale cycles, so that the supersymmetry conditions and equations of motion must be treated formally as a power-series in $\alpha'$.  A discussion of this expansion and its consequences for solutions can be found in~\cite{Melnikov:2014ywa}.}

In order to satisfy the differential conditions for minimal supersymmetry, the complex structure on $M$ must be constrained so that ${\rm d}\Theta$ has no (0,2) component.   To describe this more concretely, let
\begin{align}
\bF = \bF^1 + \tau \bF^2~,
\end{align}
and now decompose this complex $2$-form by type:
\begin{align}
\bF &= F + F' + F''~, &  F &\in \cA^{(1,1)}(M)~,&  F' & \in \cA^{(2,0)}(M)~,  & F'' & \in\cA^{(0,2)}(M)~.
\end{align}
Here we denote by $\cA^{p,q}(M)$ the space of $(p,q)$ forms on $M$.  Supersymmetry then requires $F'' = 0$, or, equivalently,
\begin{align}
\label{eq:TorusHYM}
J \wedge \bF & = 0~, &
\Omega \wedge \bF & = 0~.
\end{align}
The complex conjugate $2$-form $\overline{\bF} = \bF^1 +\taub \bF^2$ has $(1,1)$ and $(0,2)$ components, which we denote, respectively, by $\Fb$ and $\Fb'$.  When $F'' =0$, we can write these various forms explicitly in terms of the components of the $\bF^I$:
\begin{align}
F & = (\bF^1)^{1,1} + \tau (\bF^2)^{1,1}~,&
F' & = (\tau-\taub) (\bF^2)^{2,0}~, \nonumber\\
\Fb & = (\bF^1)^{1,1} + \taub (\bF^2)^{1,1}~,&
\Fb' & = (\taub-\tau) (\bF^2)^{0,2}~.
\end{align}
We used $F'' = (\bF^1)^{0,2} + \tau (\bF^2)^{0,2} = 0$ to eliminate the $(\bF^1)^{0,2}$ and $(\bF^1)^{2,0}$ components.  Although~(\ref{eq:TorusHYM}) superficially resembles zero-slope HYM conditions, it should be borne in mind that because $\bF$ is a complex $2$-form, the equations are of a very different character from HYM when $F' \neq 0$ and are not in any obvious sense a deformation of the HYM conditions.  

The heterotic $H$-flux is also fixed by supersymmetry:
\begin{align}
H & = i (\pb - \p) J_X = iJ\wedge(\pb-\p)  e^{2\phi} + \frac{\alpha'\rho_2 }{2\tau_2} \left [(\Fb' -\Fb)\wedge \Theta + (F'-F)\wedge\Thetab\right]~\nonumber\\
& = H_{\text{h}} + H_{\text{v},I} \wedge \bTheta^I~,
\end{align}
where the ``horizontal'' and ``vertical'' components of $H$ are
\begin{align}
\label{eq:SUSYH}
H_{\text{h}} & =  iJ\wedge (\pb-\p)  e^{2\phi}~,\nonumber\\
H_{\text{v},I} & =  \alpha' \cG_{IJ} \left( (\bF^I)^{2,0} + (\bF^I)^{0,2} - (\bF^I)^{1,1} \right)~,
\end{align}
and $\cG_{IJ}$ is the metric on the torus
\begin{align}
\cG & = \frac{\rho_2}{\tau_2} \begin{pmatrix} 1 & \tau_1 \\ \tau_1 & |\tau|^2\end{pmatrix}~.
\end{align}

\subsection{The gauge bundle}
To complete the specification of the heterotic background we need to describe the gauge bundle, which we will choose as follows.  Let $\cP\to M$ be a principal $\GE_8\times\GE_8$ bundle with a connection $\cA$ and structure group $G_{\cP} \subseteq \GE_8\times\GE_8$.  Denote the associated adjoint bundle $\text{Ad}(\cP)$.  The pullback bundle $p^\ast(\cP) \to X$ then has the connection $\cA_{\text{h}} = p^\ast(\cA)$ and transition functions that only depend on the base coordinates.  We will consider a more general family of connections on $p^\ast(\cP)$ of the form
\begin{align}
\cA_{X} = \cA_{\text{h}} + a_I \bTheta^I~,
\end{align}
where the $a_I$ are sections of $p^\ast(\text{Ad}(\cP))$.  Such a connection has curvature
\begin{align}
\cF_{X} = \cF_{\text{h}} + a_I \bF^I + (\cD_{\text{h}} a_I) \wedge \bTheta^I + \ff{1}{2} \CO{a_I}{a_J} \bTheta^I \wedge \bTheta^J~,
\end{align}
where $\cD_{\text{h}}$ is the pullback of the covariant derivative on the vector bundle $\text{Ad}(\cP)$ with connection $\cA$.  Supersymmetry requires $\cA_{X}$ to be a zero-slope HYM connection, i.e. 
\begin{align}
J_{X}\wedge J_X \wedge \cF_{X} & = 0~, &
\Omega_X \wedge \cF_{X} & = 0~, &
\Omegab_X \wedge \cF_{X} & =0~.
\end{align}
The geometry---in particular~(\ref{eq:TorusHYM})---implies that these conditions hold if and only if
\begin{align}
\label{eq:susybundleconditions}
\CO{a_I}{a_J} & = 0~, &
\cD_{\text{h}} a_I & =0~,&
J \wedge \cF_{\text{h}} & = 0~,&
\Omega \wedge\cF_{\text{h}} & =0~,&
\Omegab \wedge(\cF_{\text{h}} +  a_I \bF^I) & =0~.
\end{align}
Just as in our discussion of line bundles on $X$, $p^\ast(\cP)$ may admit large gauge transformations that are not pulled back from $M$.  Namely, let $\kappa_I  \in \cA^{0,0}(M,\Le_8\oplus\Le_8)$ be two covariantly constant and commuting sections such that $\exp[2\pi \kappa_I] =\text{id} \in \GE_8\times\GE_8$.\footnote{In order for such sections to exist it must be that $G_{\cP}$ is strictly contained in $\GE_8\times\GE_8$.}  Then, if $g_{ab}$ denote the transition functions of $p^\ast(\cP)$ relative to the cover $\mathfrak{U}$ described above, so that the connection transforms according to
\begin{align}
\cA_{\text{X},a} = g_{ab} \cA_{\text{X},b} g_{ab}^{-1} + g_{ab} d g_{ab}^{-1}~,
\end{align}
then we have a large gauge transformation
\begin{align}
(g, \cA_{\text{X}}) \to ( f g f^{-1}, \cA_{\text{X}} + f {\rm d}f^{-1} )~,
\end{align}
where $f = e^{\kappa_I \theta^I}$~.  In terms of our horizontal--vertical decomposition we then obtain the gauge transformation
\begin{align}
\cA_{\text{h}} + a_I\bTheta^I  \to (2\cA_{\text{h}} - f \cA_{\text{h}} f^{-1} + \kappa_I \bA^I) + (a_I -\kappa_I) \bTheta^I~. 
\end{align}
With that preparation we can finally state our assumptions on $\cA_X$:  we will restrict attention to $\cA_{X}$ that are gauge-equivalent to a configuration where the horizontal component $\cA_{\text{h}}$ is given by the pullback $p^\ast(\cA)$ of a HYM connection from the base.  This leads to a delineation of solutions into two classes.
\begin{enumerate}
\item The complexified torus curvature $\bF$ has no $(2,0)$ component.  In this case the conditions in~(\ref{eq:susybundleconditions}) imply that the connection $\cA$ is HYM, and the geometry preserves $8$ supercharges.  The $a_I$ then encode the Wilson line parameters, and after a suitable gauge fixing their expectation values correspond to scalar moduli in the vector multiplets of the effective four-dimensional theory.
\item $\bF$ has a non-zero $(2,0)$ component.  In this case~(\ref{eq:susybundleconditions}) can be solved by taking $\cA$ to be HYM and setting $a_I =0$, or by taking a large gauge transformation of that configuration.  Having fixed such a connection $\cA_X$, supersymmetry prevents continuous deformations $\cA_X \to \cA_X + a_I \bTheta^I$.
\end{enumerate}
Having discussed the general structure of the gauge bundle, we will now restrict the structure group $G_{\cP}$ to be a subgroup of the Cartan torus $T_E \subset \GE_8\times\GE_8$.  This will allow us to avoid many technical and expository complications, but at least in the case of theories with $8$ supercharges it also captures the essential physics of interest to us, because it should be possible to reach more general configurations by moving in the moduli space.

We do not have a general proof of the last claim, but we believe the following example offers a persuasive picture.   Suppose we compactify the heterotic string on $M$, and we choose an irreducible HYM connection with gauge group $\GE_8\times\GE_8$ and instanton number $12$ in each $\GE_8$ factor.   The moduli space of the instantons includes reducible connections, and we can go to a degeneration locus where the connection is non-zero just in an $\su(2)\oplus\su(2) \subset \Le_8 \oplus \Le_8$ subalgebra, with instanton number $12$ for each $\su(2)$ factor.  An $\su(2)$ connection can in turn become reducible, where the $\su(2)$--valued curvature takes the form
\begin{align}
\cF_{\su(2)} & = \begin{pmatrix} 0 &  \cF_{\Lu(1)} \\ -\cF_{\Lu(1)} & 0 \end{pmatrix}~.
\end{align} 
If $M$ is at a generic point in its moduli space, then the resulting connection must necessarily be singular, and supergravity may not be a good guide to the physics.  But, if on the other hand we tune $M$ to a special point in the moduli space where the abelian curvature $\cF_{\Lu(1)}/2\pi$ is not only HYM but also an integral class, then we can obtain a perfectly smooth supergravity description of the reducible connection.  It would be interesting to make this example rigorous, generalize it to other gauge configurations, and then relate it to earlier studies of K3 with abelian gauge sector such as~\cite{Honecker:2006dt,Honecker:2006qz,Louis:2011hp,Kumar:2009zc}.  

The situation in the case of configurations with $4$ supercharges is more subtle since a superpotential may obstruct the deformations.  Nevertheless, we will for the most part confine our discussion to line bundle configurations even in that case.

We now describe the abelian gauge bundle.  Let $T_E = \R^{16}/(\Gamma_8+\Gamma_8)$ be the Cartan torus of $\GE_8\times\GE_8$, and fix $k$ linearly independent lattice vectors in $\bsigma_s \in \Gamma_8+\Gamma_8$.  Such a choice determines a homomorphism $\Z^k \to T_E$ and therefore an embedding of a torus $T^k \hookrightarrow T_E$:\begin{align}
(\phi^1,\ldots,\phi^k) \to \phi^s \bsigma_s~.
\end{align}
We then set the connection to be 
\begin{align}
\cA =  \cA^s \bsigma_s~,
\end{align}
where $\cA^s$ is a HYM connection on the $\GU(1)$ bundle over $M$ corresponding to the fiber coordinate $\phi^s$.  Denote by $\Uplambda^s$ the corresponding first Chern class, so that the image of $\Uplambda^s$ in $H^2(M,\R)$ coincides with $[\cF^s/2\pi]$.  The lattice vector
\begin{align}
\bUplambda = \Uplambda^s \bsigma_s \in (\Gamma_8 + \Gamma_8) \otimes_{\Z} H^2(M,\Z)~
\end{align}
determines the topology of this abelian $\GE_8\times\GE_8$ principal bundle $\cP\to M$.   In fact the isomorphism class of a $\GE_8\times\GE_8$ bundle $\cP\to M$ is specified by the Pontryagin classes of the two simple factors.\footnote{See, for example, the appendix of~\cite{MR658473} for a proof.}   To express these it is convenient to expand $\cA$ and $\bUplambda$ in a basis of simple roots for $\Gamma_8+\Gamma_8$:
\begin{align}
\bUplambda &= \bUplambda_1+ \bUplambda_2~,&
\bUplambda_1 &= \sum_{i=1}^{8} \Uplambda^i_1  \balpha_i~, &
\bUplambda_2 &= \sum_{i=9}^{16} \Uplambda^i_2 \balpha_i~,
\end{align}
The isomorphism class of the corresponding $\GE_8\times\GE_8 \to M$ bundle is determined by the two Pontryagin classes:
\begin{align}
p_1(\cP_1) & = \sum_{i,j=1}^8 \Uplambda^i_1 \cup \Uplambda^j_1 \balpha_i \cdot \balpha_j~,&
p_1(\cP_2) & = \sum_{i,j=9}^{16} \Uplambda^i_2 \cup\Uplambda^j_2 \balpha_i \cdot \balpha_j~,&
\end{align}
and the Pontryagin class of $\cP$ is given by the sum of these classes.

Similarly, we expand the Wilson line parameters as $a_I = \ba_I^{i} \balpha_i$, so that the full connection is expressed as
\begin{align}
\cA_X = \textstyle\sum_{i=1}^{16}  \left(\cA^i + \ba_I^i \Theta^I\right) \balpha_i~.
\end{align}
Large gauge transformations are generated by lattice vectors $\bkappa_I \in \Gamma_8 +\Gamma_8$, giving an equivalence 
\begin{align}
(\cA, \ba_I) \sim (\cA+ \bkappa_I \bA^I, \ba_I-\bkappa_I)~.
\end{align}

We note that this formulation automatically ensures that the bundle satisfies the global anomaly constraint familiar from heterotic string Calabi--Yau compactification based on a Hermitian vector bundle $\cE$, which is required to satisfy $c_1(\cE) = 0 \mod 2$~\cite{Witten:1985ga,Freed:1986zx}.  This is really nothing other than the requirement that the connection can be lifted to a connection on a principal $\GE_8\times\GE_8$ bundle, and this is ensured by picking the $\bsigma_s$ to be lattice vectors.

\subsection{The Bianchi identity and quantization of moduli}
\label{subsec:quantiz}
The data described above satisfies the supersymmetry conditions for any choice of dilaton profile $\phi$ over the base $M$.
The dilaton profile is determined, up to a constant that we identify with the string coupling, by solving the Bianchi identity, which in our background reduces to a scalar equation on $M$:
\begin{align}
{\rm d}H_{\text{h}} + {\rm d}\left( H_{\text{v},I} \wedge \bTheta^I\right) = \frac{\alpha'}{4} \left[ \tr R^2_+ -\tr \cF_X^2\right]  + \GO(\alpha'^2)~.
\end{align}
Here $R_+$ is the curvature of the $H$-twisted connection $\cS_{+}$ on the tangent bundle with $\tr$ taken in the fundamental representation of $\so(6)$, while the trace for the gauge curvature is defined to be in the fundamental of $\so(16)\oplus\so(16) \subset \Le_8\oplus\Le_8$.  The resulting PDE for $\phi$ is quite non-trivial, since, for example, it depends implicitly on a choice of Ricci-flat metric on $M$.  In some cases existence and uniqueness of solution can be demonstrated---such results go back to~\cite{Fu:2006vj}---but, as discussed in~\cite{Melnikov:2014ywa}, these results should be considered in the $\alpha'$ expansion.   One consequence of that treatment is that the (2,0) component of the curvature $\bF$, i.e. $F'$, must be $\pb$-closed and therefore must be a constant multiple of the holomorphic (2,0) form $\Omega$.  As a result, $(\bF^I)^{1,1}$ and $(\bF^I)^{2,0}$ are separately closed, and ${\rm d} H_{\text{v},I} = 0$. Since ~(\ref{eq:TorusHYM}) implies that the $(\bF^I)^{1,1}$ curvature components are anti-self-dual (ASD) with respect to the Ricci-flat metric on $M$,  the Bianchi identity can be rewritten in terms of the Hodge star $\ast$ on $M$---constructed using the hyper K\"ahler metric---as 
\begin{align}
\label{eq:bianchi_id}
\alpha'\cG_{IJ} (\ast \bF^I) \wedge \bF^J = \frac{\alpha'}{4} \left[ \tr R^2_+ -\tr \cF_X^2\right]  + \GO(\alpha'^2)~.
\end{align}

A crucial aspect of the heterotic compactification is that the $H$-flux has a specific form in terms of the $B$-field and the Chern--Simons forms for the connections $\cA_X$ and $\cS_{+}$:
\begin{align}
H & = {\rm d} B -\frac{\alpha'}{4} \text{CS}_3 (\cA_X) + \frac{\alpha'}{4} \text{CS}_3 (\cS_{+})~.
\end{align}
We take the $B$-field to have the following decomposition into vertical and horizontal terms:
\begin{align}
 B = p^\ast(B_{\text{h}}) + p^\ast(B_{\text{v},I}) \wedge \bTheta^I + \ff{1}{2} \alpha' \bb \ep_{IJ} \bTheta^I \wedge \bTheta^J~.
\end{align}
The parameter $\bb$ must be a constant to be consistent with the form of $H$ required by supersymmetry. We set $\rho_1=\bb$ and introduce the parameter $\rho=\rho_1+i\rho_2$ valued in the upper half-plane, which corresponds to the torus K\"ahler modulus. Plugging the expansion of $B$ into $H$ and comparing to the constraint from supersymmetry, we can obtain explicit expressions for ${\rm d}B_{\text{v},I}$ and $B_{\text{h}}$.

So far the construction has introduced a number of geometric ingredients with continuous deformation parameters.  In particular, we have:
\begin{enumerate}
\item a choice of hyper-K\"ahler metric on $M$, accompanied by a shift of $B_{\text{h}}$ by a form in $H^2(M,\R)$; 
\item the constant shift of the dilaton, which determines the string coupling;
\item the parameters associated with the $T^2$ fiber, i.e. the choice of torus metric $\cG$, the torus B-field $\bb$, and the Wilson line parameters $\ba_I$.
\end{enumerate}
A telltale feature of a flux vacuum is that the integrality conditions on the choice of flux constrain the deformations to lie in a sub-locus of the naive parameter space of supergravity solutions.  This holds in the vacua at hand as well~\cite{Melnikov:2012cv}, which can be seen by showing that the $1$-forms
\begin{align}
B'_{\text{v},I} = B_{\text{v},I} + \frac{\alpha'}{4} \tr\{a_I \cA\} 
\end{align}
transform as connections on two line bundles under a combination of gauge and gerbe transformations, and with
$\ff{1}{2\pi\alpha'} {\rm d}B'_{\text{v},I}$ 
having integral periods~.
Plugging the explicit form of $B$ into $H$ and using~(\ref{eq:SUSYH}), we find that the closed $2$-form
\begin{align}
 \frac{1}{2\pi\alpha'} \left\{\alpha'\cG_{IJ} (-\ast \bF^J) - \alpha'\bb \ep_{IJ} \bF^J- \frac{\alpha'}{4} \tr\{a_I a_J\} \bF^J - \frac{\alpha'}{2} \tr\{ a_I \cF\}\right\}
\end{align}
must also have integral periods, or, equivalently its cohomology class must be the image of an integral class under the inclusion $H^2(M,\Z) \to H^2(M,\R)$.  We will denote that integral class by $\upnu_I$ and abuse notation by identifying $\upnu_I$ with its image in $H^2(M,\R)$.  With similar abuse for the classes $\upomega^I = [\bF^I/2\pi]$, as well as $\bUplambda$, we have
\begin{align}
\label{eq:ModuliQuantization}
\upnu_I & = \cG_{IJ} (-\ast \upomega^J) + \left( \frac{1}{2}\ba_I \cdot \ba_J - \bb \ep_{IJ} \right)\upomega^J +\ba_I \cdot \bUplambda~.
\end{align}
From this we immediately see that the parameters associated to the torus fiber must obey stringent quantization conditions.  We will explore these in more detail below.

Coming back to the Bianchi identity and making a horizontal/vertical expansion while dropping $\GO(\alpha'^2)$ terms, we find the topological requirement
\begin{align}
-\upnu_I \cup \upomega^I = -\frac{1}{2} p_1(T_M) + \frac{1}{2} p_1(\cP)~.
\end{align}
Since $H^4(M,\Z) = \Z$ we lose no topological information by simply integrating both sides to get a single integral condition.
We can follow~\cite{Evslin:2008zm} to make this structure $\GO(\Gamma_{d,d+16})$--covariant by noting that the  topological and quantization conditions are encoded via the vector
\begin{align}
v = \begin{pmatrix} \upomega \\ \upnu \\ \bUplambda \end{pmatrix} \in \Gamma_{d,d+16} \otimes_{\Z} H^2(M,\Z)~.
\end{align}
Denote by $\bullet$ the tensor inner product inherited from the ``$\cdot$'' inner product on $\Gamma$ and the ``$.$'' inner product on $H^2(M,\Z)$.  The topological Bianchi identity then takes the form
\begin{align}
-\frac{1}{2} v \bullet v & = 24~,
\end{align}
while $v$, the torus parameters, and the K3 metric are further constrained by the quantization conditions~(\ref{eq:ModuliQuantization})~.   In particular, for a fixed choice of $v$ that satisfies the topological Bianchi identity there may not exist continuous parameters that satisfy~(\ref{eq:ModuliQuantization})~.

Before we leave the general discussion, we should make a comment on the existence of smooth solutions and string-scale cycles.  The latter are a necessary result of the quantization of the torus moduli.  On the other hand, the volume of the K3 base $M$ can be taken to be arbitrarily large, meaning that the background has a lift to $8$ dimensions.  This is a key point, since it allows us to use $8$-dimensional intuition to discuss properties of the solution.  For example that perspective will largely motivate our T-duality discussion.   We note that for a class of backgrounds that admit a gauged linear sigma model description the T-duality transformations can be lifted to $(0,2)$ dualizations {\it \`a la} Buscher that do not receive corrections from worldsheet instantons~\cite{Israel:2013hna}, so at least for this class of vacua there is good reason to believe that the T-dualities indeed relate isomorphic heterotic worldsheet theories.  We will assume that the same holds for all of the heterotic flux vacua we describe, whether or not there is a linear sigma model realization.

Because we require $M$ to support a number of abelian HYM connections, there are non-trivial constraints on the geometry of $M$, and we might worry that these constraints force $M$ to be singular.  Recall that a volume $1$ hyper-K\"ahler metric on $M$ is singular if and only if there is a zero-size $-2$--curve, i.e. a class $C \in H^2(M,\Z)$ with $C.C = -2$ that is annihilated by $J$ and $\Omega$:  equivalently, $C$ is ASD. Thus, a necessary condition for a smooth geometry is that  $\Uplambda^i.\Uplambda^i \le -4$ for each $i$ , and, similarly, $\upomega^I . \upomega^I \le -4$ for $I =1,2$.

\subsection{Quantization and rationality: 8 supercharges}
For backgrounds that preserve $8$ supercharges the quantization conditions take a particularly simple form because $\ast \upomega^I = -\upomega^I$.  Using that in~(\ref{eq:ModuliQuantization}) and comparing to~(\ref{eq:NarainCoset})~, we find that~(\ref{eq:ModuliQuantization}) reduces to
\begin{align}
\label{eq:ModuliQuantization8}
\bpit_I \cdot v = 0~.
\end{align} 
This has an important consequence for the decompactification limit of the theory to $8$ dimensions: the $8$--dimensional theory is a heterotic compactification on $T^2$ with parameters chosen so that there exist non-zero lattice vectors $\bp \in \Gamma_{2,18}$ with $h_{\sright}(\bp) = 0$:  thus, the SCFT has an enhanced chiral algebra of holomorphic higher spin currents.

For example, if we set $\ba_I =0$ the complexified K\"ahler parameter $\rho$ and  the complex structure $\tau$ 
should obey, following from~(\ref{eq:ModuliQuantization8}), the pair of equations in $H^2(M,\mathbb{Z})$:
\begin{subequations}
\label{eq:przero}
\begin{align}
0= & \tau_2 \upnu_1 - \rho_2 \upomega^1 + (\rho_1 \tau_2-\tau_1 \rho_2) \upomega^2~,\\ 
0= & -\tau_1 \upnu_1 + \upnu_2 - \rho_1 \upomega^1 - (\rho_1 \tau_1+ \rho_2 \tau_2) \upomega^2 ~.
\end{align}
\end{subequations}
Whenever the classes $\upomega^I \in H^2(M,\mathbb{Z})$ are linearly independent, these equations can be solved for $\upnu_I \in  H^2(M,\mathbb{Z})$ if and only if 
$\tau_1$, $\rho_1$, $\tau_2^2$, $\rho_2^2$ and $\rho_2 /\tau_2$ are valued in $\mathbb{Q}$. It implies that $\tau$ should solve a quadratic equation 
$a\tau^2 + b\tau +c=0$ with integer coefficients $a,b,c$ and discriminant $D= b^2-4ac<0$; in other words $\tau$ should belong to the imaginary quadratic 
number field $\mathbb{Q}[\sqrt{D}]$. Due to the condition $\rho_2/\tau_2 \in \mathbb{Q}$, $\rho \in \mathbb{Q}[\sqrt{D}]$ as well. These 
are precisely the conditions under which the bosonic CFT associated to $T^2$ is rational~\cite{Gukov:2002wpj}. The lattices of purely left- and right-moving momenta have maximal rank two and the spectrum 
decomposes into a finite number of primary fields with respect to the extended chiral algebra.\footnote{It was observed some time ago in~\cite{Israel:2013hna} that T-duality 
covariance of the flux geometries implies such rationality, but here we see it directly from the quantization conditions.}

Even when $\ba_I \neq 0$~(\ref{eq:ModuliQuantization8}) implies that the $T^2$ CFT has an infinite set of operators with purely left-moving momentum whenever $\upomega^I \neq 0$.  Thus, while the anti-holomorphic sector of the CFT may not have an enhanced chiral algebra, the holomorphic sector remains highly constrained.   The significance of this condition is not clear to us, nor is it clear what implications it has for the full $c_{\sleft} = 15$, $c_{\sright} = 9$ SCFT describing the compactification to $4$ dimensions:  for example, the vertex operators of the $T^2$ theory that generate the higher spin currents must presumably be dressed by fields from the base geometry in order to yield well-defined fields, and it is not clear how the dressing will affect the dimensions of the full operator.

\section{Five-dimensional flux vacua} \label{s:5dimflux8supercharge}
In this section we specialize the general framework to the case where one circle is trivially fibered.  Taking the decompactification limit we then obtain a $5$--dimensional geometry $Y \to M$---a circle bundle determined by a single ASD class $\upomega \in H^2(M,\Z)$.  The quantization conditions reduce to
\begin{align}
\label{eq:ModuliQuantization5d}
\upnu & = \left( r^2 + \ff{1}{2} \ba\cdot\ba\right) \upomega + \ba \cdot \bUplambda \in H^2(M,\Z)~,
\end{align}
which is equivalent to $\bpit \cdot v = 0$.

Expanding the classes in terms of the special basis introduced in~(\ref{eq:specialbasis}) as
\begin{align}
\upomega &=  k \upomega_{\text{p}}~,&
\Uplambda^i & = M^i_{\text{p}} \upomega_{\text{p}}+ M^{i\alpha} \upchi_\alpha~,&
\upnu & = l \upomega_{\text{p}}+ N^\alpha \upchi_\alpha
\end{align}
leads to the quantization conditions
\begin{align}
k\left(r^2 +\ff{1}{2} \ba\cdot\ba\right) + M^i_{\text{p}} (\balpha_i \cdot \ba) & = l~, &
(\balpha_i\cdot\ba) M^{i\alpha} & = {N}^\alpha~.
\end{align}
Let us denote the corresponding worldsheet CFT by $\cC_{v}(r,\ba)$.

\subsection{Trivial gauge bundle}
There are examples of flux compactifications with $p_1(T_Y) = 0$, where the gauge bundle can be taken to be trivial: 
\begin{align}
v =  \begin{pmatrix} \upomega \\ \upnu\\ 0 \end{pmatrix}~,
\end{align}
and the quantization conditions are
\begin{align}
\upomega & = k \upomega_{\text{p}}~,&
\upnu & = l \upomega_{\text{p}}~,&
k \left(r^2 +\ff{1}{2} \ba\cdot\ba\right)  & = l~,&
kl (-\upomega_{\text{p}}. \upomega_{\text{p}}) = 24~.
\end{align}
Since $ -\upomega_{\text{p}}. \upomega_{\text{p}} \ge 4$, there is a finite number of solutions to the Bianchi identity, and for each of them the combination of moduli $r^2 + \ff{1}{2}\ba\cdot \ba$ is quantized.  By making the volume of $M$ large, we see that the theory has a $9$-dimensional decompactification limit to a heterotic string compactified on a circle.    
We can perform a T-duality in the $9$--dimensional theory, and the vector $v$ transforms in the fundamental representation of $\GO(\Gamma_{1,17})$~\cite{Evslin:2008zm}.  For example, using the generators defined in section~\ref{s:lattice}, we have the factorized duality
\begin{align}
g_{\text{i}} v & = g_{\text{i}}  \begin{pmatrix} k\upomega_{\text{p}} \\ l \upomega_{\text{p}} \\ 0 \end{pmatrix} = \begin{pmatrix} l\upomega_{\text{p}} \\  k\upomega_{\text{p}} \\ 0 \end{pmatrix} ~.
\end{align}
The moduli are transformed as
\begin{align}
\mu_{g_{\text{i}}}(r,\ba) = \left( \frac{k r}{l}~, -\frac{k \ba}{l}\right)~.
\end{align}
When interpreted in terms of the $5$-dimensional geometry this in general leads to a topology change.  For example, the original background has $\pi_1(Y) = \Z_k$, while the dual one has fundamental group $\Z_l$.  Of course the corresponding CFTs are isomorphic:
\begin{align}
\cC_{g_{\text{i}} v}\left(\mu_{g_{\text{i}}}^{-1}(r,\ba) \right) \simeq \cC_{v} \left(r,\ba\right)~.
\end{align}

Whenever $k>1$, the non-simply connected background $Y$ can be constructed as a freely-acting $\Z_{k}$ orbifold of a simply-connected background $Y$ with $k=1$.  We will discuss the structure of the orbifolds in more detail in the following section, but for now let us focus on the case of a simply connected geometry with topology specified by the vector
\begin{align}
v & = \begin{pmatrix} \upomega_{\text{p}} \\ l\upomega_{\text{p}} \\ 0 \end{pmatrix}~
\end{align}
and quantization condition $r^2 +\ff{1}{2}\ba\cdot\ba = l$.

Consider now the action
\begin{align}
g_{\text{i}} g_{\text{s}}[\bkappa]  v & = 
\begin{pmatrix}
-\ff{1}{2} \bkappa\cdot\bkappa & 1 & \bkappa \cdot \\
1 & 0 & 0 \\
-\bkappa& 0  & \iden_{16}
\end{pmatrix} 
\begin{pmatrix}  \upomega_{\text{p}} \\ l\upomega_{\text{p}} \\ 0 \end{pmatrix}~
= \begin{pmatrix} (l-\ff{1}{2}\bkappa\cdot\bkappa) \upomega_{\text{p}} \\ \upomega_{\text{p}} \\ -\bkappa \upomega_{\text{p}} \end{pmatrix} = v'~.
\end{align}
If it is possible to find $\bkappa \in \Gamma_8+\Gamma_8$ so that $\bkappa\cdot\bkappa = 2l$, then we obtain an isomorphism of CFTs
\begin{align}
\cC_{v}(r,\ba) \simeq \cC_{v'} (r',\ba')~,
\end{align}
with $(r',\ba') =   \mu_{g_{\text{i}}}^{-1} (\mu_{g_{\text{s}}}[\bkappa]^{-1}(r,\ba))$~,
\begin{align}
v' & = \begin{pmatrix} 0\\ \upomega_{\text{p}} \\ -\bkappa \upomega_{\text{p}} \end{pmatrix}~,&
r' & =  \frac{r}{2l+\ba\cdot\bkappa}~,&
\ba' & =-\frac{\ba+\bkappa}{2l+\ba\cdot\bkappa}~.
\end{align}
We conclude that Narain T-duality can unwind the circle fibration.

It is perhaps not obvious that we can always find a $\bkappa$ satisfying $\bkappa\cdot\bkappa = 2l$.  This is possible because the Bianchi identity and smoothness of $M$ restrict $l \le 6$, and it is not hard to see that by choosing $\bkappa$ to lie in an $\Gamma_{\mathfrak{a}_3}$ sublattice we can generate all $l \le 13$.\footnote{We can also obtain all $l\le 6$ by taking $\bkappa \in \Gamma_{\mathfrak{a}_1}+\Gamma_{\mathfrak{a}_1}+\Gamma_{\mathfrak{a}_2}$ or $\bkappa \in \Gamma_{\mathfrak{a}_1}+\Gamma_{\mathfrak{a}_1}+\Gamma_{\mathfrak{a}_1}+\Gamma_{\mathfrak{a}_1}$.}

The quantization conditions on the parameters of $\cC_{v'}$ is $\ba' \cdot \bUplambda' = \upomega_{\text{p}}$, and this is indeed satisfied because $\ba'\cdot\bkappa =-1$.  On the other hand, $r'$ is now unconstrained, and there is a decompactification limit to $6$ dimensions.  Inverting the map, we obtain
\begin{align}
r & = \frac{r'}{r'^2+\ff{1}{2}\ba'\cdot\ba'}~,&
\ba & = - \bkappa - \frac{\ba'}{r'^2+\ff{1}{2} \ba'\cdot\ba'}~,
\end{align}
so we see that $r'\to\infty$ corresponds in the original parameters to taking $r\to 0$ and $\ba \to -\bkappa$.  We can also see that $r'\to 0$ sends $r \to 0$, but now $\ba \to -\bkappa - 2\ba'/(\ba'\cdot\ba')$.

\subsection{Non-trivial gauge bundle}
A simple example of a non-trivial bundle is given by
\begin{align}
\bUplambda_1 & = \Uplambda_1 \balpha_1~,&
\bUplambda_2 & = \Uplambda_2  \balpha_9~,
\end{align}
in which case
\begin{align}
\bUplambda_1 \bullet \bUplambda_1 & =  2 \Uplambda_1.\Uplambda_1~,&
\bUplambda_2 \bullet \bUplambda_2 & = 2 \Uplambda_2.\Uplambda_2~,
\end{align}
and the Bianchi identity for a simply--connected $Y$ is
\begin{align}
-l \upomega_{\text{p}}. \upomega_{\text{p}} - \Uplambda_1.\Uplambda_1 -\Uplambda_2.\Uplambda_2 = 24~.
\end{align}
Smoothness of $M$ now puts even stronger constraints on $l$.  For example, if both $\Uplambda_1$ and $\Uplambda_2$ are non-zero, then each contributes at least $+4$ to the left-hand-side of the Bianchi identity, meaning $l \le 4$.

We can now again attempt to T-dualize $\cC_{v}$ to a product geometry.  Starting with
\begin{align}
v & = \begin{pmatrix} \upomega_{\text{p}} \\ l\upomega_{\text{p}} \\ \bUplambda \end{pmatrix}~,
\end{align}
we find
\begin{align}
 v' = g_{\text{i}} g_{\text{s}}[\bkappa] v = \begin{pmatrix} (l-\ff{1}{2}\bkappa\cdot\bkappa) \upomega_{\text{p}}+\bkappa\cdot \bUplambda \\ \upomega_{\text{p}} \\ \bUplambda-\bkappa \upomega_{\text{p}} \end{pmatrix}~.
\end{align}
If we can choose $\bkappa$ such that $\bkappa\cdot \bUplambda = 0$ and $\bkappa\cdot\bkappa = 2l$, then the underlying geometry of $\cC_{v'}$ will be a straight product.  Clearly for this configuration there is no problem in finding such a $\bkappa$ because $\bUplambda$ only has non-zero entries in two of the simple roots of $\Gamma_8+\Gamma_8$.

This simple ``unwinding'' duality generalizes to many other gauge configurations, involving both abelian and non-abelian factors.  Moreover, it is quite plausible that a smooth $5$-dimensional heterotic flux compactification with any other choice of gauge bundle  can be deformed to this configuration by moving in the moduli space of connections with fixed $p_1(\cP_1)$ and $p_1(\cP_2)$, while keeping the heterotic string weakly coupled.  If this is the case then every five-dimensional heterotic flux vacuum is equivalent to a compactification on $M\times S^1$.

\subsection{Orbifolds} \label{s:orbifolds5dim}
So far we restricted attention to the simply connected $5$-dimensional geometries.   Their non-simply connected relatives are obtained as freely-acting orbifolds, and, as is familiar in geometry and topology, in taking the quotient we can introduce new non-trivial structures.  For instance, given a covering $\Yt \to Y$, we know that $Y$ can support flat but topologically non-trivial circle bundles classified by a choice of torsion class in $H^2(M,\Z)$, or, equivalently, a class in $\Hom(\pi_1(Y),\GU(1))$.  In this section we will describe such quotients in the context of $5$--dimensional flux compactifications and their ``unwound'' duals with geometry $M\times S^1$.

For simplicity we will work with a trivial gauge bundle, but the results can be generalized in a straightforward (but somewhat cumbersome) fashion.  In this situation the topology of the simply connected $Y_1$ is specified by the vector
\begin{align}
v_1 & = \begin{pmatrix}  \upomega_{\text{p}} \\ l_1 \upomega_{\text{p}} \\ 0 \end{pmatrix}~,
\end{align}
and the Bianchi identity constrains
\begin{align}
l_1 = \frac{24}{-\upomega_{\text{p}}.\upomega_{\text{p}}}~.
\end{align}
$Y_1$ has a continuous freely-acting $\GU(1)$ isometry that shifts the fiber coordinate $\theta\to \theta + \alpha$.  Consider a $\Z_k$ subgroup of this isometry group generated by $\alpha = 2\pi /k$.  To describe the quotient geometry $Y =Y_1/\Z_k$ we can  work with an invariant fiber coordinate $\theta' = k \theta$, with identification $\theta' \sim\theta' + 2\pi$.  On overlaps for the circle bundle these satisfy $e^{i\theta'_a} = (\tau_{ab})^k e^{i\theta'_b}$, showing that the fibration $p: Y \to M$ is determined by the first Chern class $ k\upomega_{\text{p}}$, i.e. one that is $k$ times larger than that of our original simply connected space $\pt : Y_1 \to M$.  The metric and connection on $Y$ can be expressed in terms of the torus parameters $r_1$ and $\ba_1$ associated to the $Y_0$ compactification as
\begin{align}
r & = \frac{r_1}{k}~,&
\ba & = \frac{\ba_1}{k}~.
\end{align}
While the geometry would allow a quotient for any $k$, the flux quantization conditions require $k$ to divide $\upnu = l_1 \upomega_{\text{p}}$.  Setting $l = l_1/k$, we then obtain the theory $\cC_{v} = \cC_{v_1}/\Z_k$, with 
\begin{align}
v & = \begin{pmatrix}  k\upomega_{\text{p}} \\ l \upomega_{\text{p}} \\ 0 \end{pmatrix}~.
\end{align}
It is instructive to consider this orbifold in the $9$-dimensional decompactification limit.  There we see that we are performing a shift orbifold by group $G_{\text{shift}} \simeq \Z_k$, with generator $g_{\text{shift}}$, which acts on the vertex operators $\cV_{\bp}$ by
\begin{align}
g_{\text{shift}} \circ \cV_{\bp} = e^{2\pi i \text{n}/k} \cV_{\bp}~.
\end{align}
The orbifold theory has a quantum symmetry $G_{\text{gerbe}}$ with generator $g_{\text{gerbe}}$ that acts as:\footnote{Our name for this symmetry follows~\cite{Cheng:2022nso}, since in a higher-dimensional setting such an action can be interpreted as turning on a flat non-trivial gerbe for the B-field along the lines discussed in~\cite{Sharpe:2003cs}.}
\begin{align}
g_{\text{gerbe}} \circ \cV_{\bp} = e^{2\pi i \text{w}/k} \cV_{\bp}~,
\end{align}
leading to the CFT equivalence $\cC_{v}/G_{\text{gerbe}} \simeq \cC_{v_1}$.

But, we argued above that $\cC_{v_1}$ is T-dual to $\cC_{v'}$ with $v' = g_{\text{i}} g_{\text{s}}[\bkappa] v_1$, with $\bkappa$ chosen so that $\bkappa\cdot\bkappa = 2l_1$.  Thus, on the ``unwound'' side we can find a T-dual symmetry group $G'_{\text{shift}}$, with generator
\begin{align}
g'_{\text{shift}} & = \cT g_{\text{shift}} \cT^{-1}~,
\end{align}
where $\cT = g_{\text{i}} g_{\text{s}}[\bkappa]$.  Using the phase factors described in section~\ref{s:lattice}, we find
\begin{align}
g'_{\text{shift}} \circ \cV_{\bp} = U(g_{\text{shift}}, \varphi_{\cT^{-1}}(\bp)) \cV_{\bp} = e^{2\pi i \text{w}/k } e^{-2\pi i \bkappa\cdot\bL/k} \cV_{\bp}~.
\end{align}
This is also a shift orbifold, but now without a simple geometric interpretation since it acts on the winding modes of the string: the action is $g_{\text{gerbe}}$ accompanied by an additional shift on the gauge degrees of freedom. The bottom line is that once we know how to relate a simply connected heterotic flux geometry and corresponding CFT $\cC_{v_1}$ to a compactification on $M\times S^1$, we can also obtain a dual description of the more general $\cC_{v}$ CFT by taking an orbifold $\cC_{v'}/G'_{\text{shift}}$, but the latter quotient does not have a geometric interpretation.

As another example of an orbifold action that preserves spacetime supersymmetry, we suppose that $l_1$ is divisible by $2$, so that $\cC_{v_1}$ has an order $2$ shift symmetry.  The quotient geometry will then have $\pi_1(Y) = \Z_2$, which is just right to turn on a $\Z_2$ holonomy for the principal $(\GE_8\times\GE_8)\rtimes\Z_2$ bundle and thereby reduce the rank of the spacetime gauge group to $8$.  We can implement this by taking a CHL orbifold of $\cC_{v_1}$ by combining the shift symmetry with the $g_{\text{flip}}$ action on the bundle given in~(\ref{eq:CHLflip}).  In order for this to be a symmetry of $\cC_{v_1}$, we must choose the Wilson lines symmetrically, i.e.
\begin{align}
\ba = \sum_{i=1}^8 a^i \left( \balpha_{i} +\balpha_{i+8} \right)~.
\end{align}
We take the CHL action to be generated by 
\begin{align}
g_{\text{chl}} = g_{\text{shift}} g_{\text{flip}}~,
\end{align}
and the resulting quotient $\cC_{v_1}/G_{\text{chl}}$ will lead to a $5$--dimensional compactification with gauge group of rank $8$.

There is a T-dual ``unwound'' theory $\cC_{v'_1}$ obtained by using a vector $\bkappa$ that is invariant under the flip $\varphi_{g_{\text{flip}}} (\bkappa) = \bkappa$ and satisfying $\bkappa\cdot\bkappa =2 l_1$.  We can find such a vector because $l_1$ is even, so that $\bkappa$ can be expressed in the same form as the symmetric Wilson line above.

In the T-dual ``unwound'' geometry $\cC_{v'_1}$ we can again find the corresponding dual group $G'_{\text{chl}}$, with generator $g'_{\text{chl}} =\cT g_{\text{chl}} \cT^{-1}$.  We can show that $\cT$ commutes with $g_{\text{flip}}$ whenever $\bkappa$ is invariant, so that 
\begin{align}
g'_{\text{chl}} \circ \cV_{\bp} = g_{\text{flip}} g'_{\text{shift}} \circ \cV_{\bp} = e^{2\pi i \text{w}/k } e^{-2\pi i \bkappa\cdot\bL/k} \cV_{\varphi_{g_{\text{flip}}}(\bp)}~.
\end{align}
So, we find a dual description of a CHL orbifold of a flux vacuum as a CHL--like orbifold of a compactification on $M\times S^1$, but once again the orbifold does not have a simple geometric interpretation because the shift acts non-trivially on the winding sectors.

\section{N=2 flux vacua in \texorpdfstring{$4$}{4} dimensions} \label{s:4dimflux8supercharge}

We now turn our attention to $4$-dimensional heterotic compactifications preserving $8$ supercharges. Consider the $8$-dimensional theory obtained by compactifying the $\GE_8\times\GE_8$ heterotic theory on a torus $T^2$. The parameters associated to this theory include the $T^2$ complex structure and (complexified) K\"ahler moduli $\tau$ and $\rho$, valued in the upper half-plane, as well as two continuous Wilson line parameters $\ba_I$. The $4$-dimensional theory is constructed by compactifying again on a K3 surface $M$ while solving the $8$-dimensional heterotic Bianchi identity with a choice of ASD classes on $M$: two classes $\upomega^I\in H^2(M,\Z)$ identified with the $(1,1)$ primitive curvatures of the $T^2$ bundle, and a lattice vector $\bUplambda\in(\Gamma_8+\Gamma_8)\otimes H^2(M,\Z)$ which encodes the topology of the gauge bundle. 

\subsection{Simply connected geometries}
\label{subsec:n=2duals}
The quantization conditions~(\ref{eq:ModuliQuantization8}) are most easily understood for backgrounds with Wilson line parameters satisfying $\ba_I\cdot\bUplambda=0$. Consider a simply-connected space $X_v$ built using two primitive elements $\upomega^I_{\text{p}}$ in the lattice $H^2(M,\Z)$.  The classes $\upnu_I$  are then required to be integral combinations of the $\upomega^I_{\text{p}}$, and the vector $v$ takes the following form:
\begin{align}
\label{eq:4dconfiguration}
v = \begin{pmatrix} \upomega_{\text{p}}^1 \\ \upomega_{\text{p}}^2 \\ l_{11}\upomega_{\text{p}}^1+l_{12}\upomega_{\text{p}}^2 \\ l_{21}\upomega_{\text{p}}^1+l_{22}\upomega_{\text{p}}^2 \\ \bUplambda \end{pmatrix}~,
\end{align}
with $l_{IJ}\in\Z$. The quantization conditions can be expressed in terms of those four integers as
\begin{align}
\rho&=l^\ast_{11}\tau-l^\ast_{12}~, & \rho\tau&=l^\ast_{21}\tau-l^\ast_{22}~,
\end{align}
where we introduced $l^\ast_{IJ}=l_{IJ}-\tfrac{1}{2}\ba_I\cdot\ba_J$. For these equations to admit solutions in the upper half-plane, the $l^\ast_{IJ}$ must satisfy
\begin{align}
l^\ast_{11}&>0~,& l^\ast_{22}&>0~,& (l^\ast_{12}+l^\ast_{21})^2&< 4l^\ast_{11}l^\ast_{22}~.
\end{align}
Similar (although less constraining) inequalities hold for the integers $l_{IJ}$. The topological Bianchi identity reads
\begin{align}
-l_{IJ}\upomega^I_\text{p}.\upomega^J_\text{p}-\tfrac{1}{2}\bUplambda\bullet\bUplambda=24~.
\end{align}
Notice that the combination $l_{21}-l_{12}$ disappears from the Bianchi identity, as it encodes the $\bTheta^1\wedge\bTheta^2$ component of the $B$-field $\bb=\tfrac{1}{2}(l^\ast_{21}-l^\ast_{12})$.

We observe that the Bianchi identity admits a limited number of solutions, extending the findings already obtained for five-dimensional vacua. Consider for example two classes $\upomega^I_{\text{p}}$ of equal self-intersection $-\upomega^I_{\text{p}}.\upomega^I_{\text{p}}=2k$, with $k\geq2$ a requisite bound for anti-self-duality and smoothness. The requirement of a smooth K3 base $M$ imposes that the two classes can intersect at most at $2(k-1)$ points, i.e. $|\upomega^1_{\text{p}}.\upomega^2_{\text{p}}|\leq 2(k-1)$.\footnote{This condition emerges from the necessity for the ASD classes $\upomega^1_{\text{p}}+\upomega^2_{\text{p}}$ and $\upomega^1_{\text{p}}-\upomega^2_{\text{p}}$ not to be $-2$ curves, leading to $|\upomega^1_{\text{p}}.\upomega^2_{\text{p}}|\leq\tfrac{1}{2}(-\upomega^1_{\text{p}}.\upomega^1_{\text{p}}-\upomega^2_{\text{p}}.\upomega^2_{\text{p}})-2$.} Taking into account the bound on $l_{12}+l_{21}$ coming from the quantization conditions, we arrive at the inequality
\begin{align}
-l_{IJ}\upomega^I_{\text{p}}.\upomega^J_{\text{p}}&> 2k(l_{11}+l_{22})-4(k-1)\sqrt{l_{11}l_{22}}\nonumber\\ &> 2(l_{11}+l_{22})+ 2(k-1)(\sqrt{l_{11}}-\sqrt{l_{22}})^2\nonumber\\ &> 2(l_{11}+l_{22})~,
\end{align}
resulting in a finite set of permissible values for $l_{11}$ and $l_{22}$. In the case of a trivial gauge bundle, we obtain $l_{11},l_{22}\leq10$, while turning on a non-trivial $\bUplambda$ further restricts the range of admissible values. Similar bounds can be derived in cases where $\upomega^1_{\text{p}}$ and $\upomega^2_{\text{p}}$ have different self-intersection, resulting in a limited set of distinct topologies.

The above configurations can be T-dualized to the trivial product $M\times T^2$ by the action of a $\GO(\Gamma_{2,18})$ transformation of the form
\begin{align}
\label{eq:unwindingduality}
\cT=g_{\text{i}}g_{\text{s},1}[\bkappa_1]g_{\text{s},2}[\bkappa_2]g_{\text{b}}[m]~.
\end{align}
Here $g_{\text{i}}=g_{\text{i},1}g_{\text{i},2}$ represents a product of factorized dualities in the two circle directions, $\bkappa_I$ are two lattice vectors taken such that $\bkappa_I\cdot\bUplambda=0$, and $m$ is an integer parametrizing the $B$-field shift. This T-duality transformation generalizes to $\GO(\Gamma_{2,18})$ the ``unwinding'' transformation introduced in section \ref{s:5dimflux8supercharge}. The T-dual configuration is described by the vector
\begin{align}
v'=\cT v=\begin{pmatrix} (l_{11}-\tfrac{1}{2}\bkappa_1\cdot\bkappa_1)\upomega_{\text{p}}^1+(l_{12}+m-\bkappa_1\cdot\bkappa_2)\upomega_{\text{p}}^2 \\ (l_{21}-m)\upomega_{\text{p}}^1+(l_{22}-\tfrac{1}{2}\bkappa_2\cdot\bkappa_2)\upomega_{\text{p}}^2\\ \upomega_{\text{p}}^1 \\ \upomega_{\text{p}}^2 \\ \bUplambda-\bkappa_1\upomega_{\text{p}}^1-\bkappa_2\upomega_{\text{p}}^2\end{pmatrix}~.
\end{align}
We can choose $\bkappa_I$ and $m$ such that
\begin{align}
\bkappa_1\cdot\bkappa_1&=2l_{11}~,&\bkappa_2\cdot\bkappa_2&=2l_{22}~,&\bkappa_1\cdot\bkappa_2&=l_{12}+l_{21}~,&m&=l_{21}~,&
\end{align}
yielding an equivalent description of the compactification as a direct product $M\times T^2$, with torus and Wilson line parameters given by $(\tau',\rho',\ba'_1,\ba'_2) = \mu_{\cT}^{-1}(\tau,\rho,\ba_1,\ba_2)$. 

\subsubsection*{Quantization of Wilson lines}
The unwinding duality gives an isomorphism between the theories defined by $v$ and $v'=\cT v$:
\begin{align}
\cC_{v'}(\tau',\rho',\ba'_1,\ba'_2)\simeq \cC_{v}(\tau,\rho,\ba_1,\ba_2)~,
\end{align}
and the moduli of the two CFTs are related by $(\tau',\rho',\ba'_1,\ba'_2) = \mu_{\cT}^{-1}(\tau,\rho,\ba_1,\ba_2)$. Consider the configuration defined by $v$ in~(\ref{eq:4dconfiguration}). Let us for now set the parameters $\ba_1$ and $\ba_2$ to zero. In the absence of Wilson lines, the $T^2$ parameters of $\cC_v$ are fixed by flux quantization to the values
\begin{align}
\tau&=\frac{l_{(12)}+i\sqrt{l_{11}l_{22}-l_{(12)}^2}}{l_{11}}~,&
\rho&=l_{(12)}-l_{12}+i\sqrt{l_{11}l_{22}-l_{(12)}^2}~,
\end{align}
where $l_{IJ}$ are the integers that appear in the expansion of $\upnu_I$ in a basis of primitive classes, and we have introduced the symmetrized quantity $l_{(12)}=\frac{1}{2}(l_{12}+l_{21})$. The torus parameters of the T-dual theory $\cC_{v'}$ can be obtained using the map $\mu_\cT$, and we find
\begin{align}
\tau'&=\frac{-l_{(12)}+i\sqrt{l_{11}l_{22}-l_{(12)}^2}}{l_{22}}~,&
\rho'&=\frac{i}{4\sqrt{l_{11}l_{22}-l_{(12)}^2}}~.
\end{align}
By construction, this T-dual theory is flux-free and describes a $M\times T^2$ compactification. As a trade-off, the space $X_{v'}$ is endowed with a new abelian instanton configuration 
\begin{align}
\bUplambda'=\bUplambda-\bkappa_1\upomega^1_\text{p}-\bkappa_2\upomega^2_\text{p}~,
\end{align}
with two line bundles over the curves that used to characterize the torus fibration.
In addition, the duality has produced non-trivial Wilson lines in both circle directions:
\begin{align}
\ba'_1&=\frac{-l_{22}\bkappa_1+l_{(12)}\bkappa_2}{2(l_{11}l_{22}-l_{(12)}^2)}~,
&\ba'_2&=\frac{l_{(12)}\bkappa_1-l_{11}\bkappa_2}{2(l_{11}l_{22}-l_{(12)}^2)}~.
\end{align}
Notice that the parameters $\ba'_1$ and $\ba'_2$ satisfy
\begin{align}
\ba'_1\cdot\bUplambda'&=\upomega^1_\text{p}~,&\ba'_2\cdot\bUplambda'&=\upomega^2_\text{p}~.
\end{align}
These are nothing but the quantization conditions~(\ref{eq:ModuliQuantization}) for the T-dual configuration: turning on an abelian curvature for the two $\mathfrak{u}(1)$ factors defined by $\bkappa_1$ and $\bkappa_2$ forbids any continuous Wilson line degree of freedom in the same direction. This condition can be understood in the four-dimensional effective theory as a  Higgsing mechanism: turning on an abelian instanton removes a full vector multiplet from the theory, and the complex scalar of this multiplet corresponds to the Wilson line modulus in the associated $\mathfrak{u}(1)$ factor.  From the point of view of the six-dimensional compactification, we know that turning on a Wilson line leads to a contribution to the vertical component of $H$ from the Chern--Simons term, but since spacetime supersymmetry requires this component to vanish, it must be canceled by a suitable contribution from the vertical component of the $B$-field.  The results of~\cite{Melnikov:2012cv} show that this can only be done for suitably quantized Wilson line parameters.   It should be possible (and very possibly enlightening) to understand this quantization directly from the topological quantization conditions of the heterotic gerbe, as well as from the worldsheet theory.

In the $X_{v'}\simeq M\times T^2$ geometry the classes $\upnu'_I$ encode the integrality conditions on the Wilson lines, and they could be shifted to zero by an extra T-duality. Using two additional lattice  vectors $\bkappa'_1$, $\bkappa'_2$ that satisfy $\bkappa'_I\cdot\bkappa_J=\delta_{IJ}$ and $\bkappa'_I\cdot\bUplambda=0$, we can obtain the configuration 
\begin{align}
\label{eq:wlines_remov}
v''=g_{\text{s},1}[\bkappa'_1]g_{\text{s},2}[\bkappa'_2] v'=\begin{pmatrix}
0\\0\\0\\0\\ \bUplambda-\bkappa_1\upomega^1_\text{p}-\bkappa_2\upomega^2_\text{p}
\end{pmatrix}~.
\end{align}
This additional T-duality does not affect the abelian instanton configuration. However, the Wilson line parameters get shifted to the values
\begin{align}
\label{eq:wilson_finalunwind}
\ba''_1&=\frac{-l_{22}\bkappa_1+l_{(12)}\bkappa_2}{2(l_{11}l_{22}-l_{(12)}^2)}+\bkappa'_1~,
&\ba''_2&=\frac{l_{(12)}\bkappa_1-l_{11}\bkappa_2}{2(l_{11}l_{22}-l_{(12)}^2)}+\bkappa'_2~,
\end{align}
and become orthogonal to $\bUplambda''$.

\subsection{Non-trivial fundamental group}
Although we have not considered non-simply connected geometries in the above discussion, every such configuration can be obtained as a quotient of a simply connected configuration by some freely-acting isometries. Let $X_0$ be a simply connected geometry defined by two primitive classes $\upomega^I_0$. In general, the classes $\upnu_{0,I}$ associated to the vertical $H$-flux components are not primitive elements of $H^2(M,\Z)$. For every pair of integers $(k_1,k_2)$ such that $\upnu_{0,I}/k_I$ belongs to $H^2(M,\Z)$, there exists a $\Z_{k_1}\times\Z_{k_2}$ quotient of $X_0$ compatible with the quantization conditions. Indeed, the space $X_0$ admits a freely acting group $G_{\text{shift}}\simeq\Z_{k_1}\times\Z_{k_2}$ generated by the two isometries $g_{\text{shift},1}$ and $g_{\text{shift},2}$, where $g_{\text{shift},I}$ acts by a translation of $2\pi/k_I$ in the $I$-th circle direction. The quotient space $X=X_0/G_{\text{shift}}$ is characterized by integral classes $\upomega^I=k_I\upomega^I_{\text{p}}$ and $\upnu_I=\upnu_{0,I}/k_I$. This new configuration has a fundamental group $\pi_1(X)\simeq\Z_{k_1}\times\Z_{k_2}$. Its torus and Wilson line parameters are related to the moduli of the simply connected theory as
\begin{align}
\tau&=\frac{k_1\tau_0}{k_2}~,&\rho&=\frac{\rho_0}{k_1k_2}~,&\ba_1&=\frac{\ba_{0,1}}{k_1}~,&\ba_2&=\frac{\ba_{0,2}}{k_2}~.
\end{align}
Just as in our discussion of five-dimensional compactifications, the $G_{\text{shift}}$ orbifold action gives an isomorphic $G'_{\text{shift}}=\cT G_{\text{shift}}\cT^{-1}$ action in the unwound T-dual description. The two $G_{\text{shift}}$ generators act on vertex operators with a phase $U(g_{\text{shift},I},\bp)=e^{2i\pi\text{n}_I/k_I}$, and on the T-dual side this phase becomes
\begin{align}
g'_{\text{shift},I}\circ\cV_\bp=e^{2\pi i\text{w}^I/k_I}e^{-2\pi i\bkappa_I\cdot\bL/k_I}\cV_\bp~. 
\end{align}
Although the quotient theory in the original description can be understood as a compactification on a non-simply connected space $X$, this is not the case anymore in the unwound description: the $G'_{\text{shift}}$ T-dual action does not have a simple geometric interpretation in terms of a quotient of $X'_0\simeq M\times T^2$.

The possibility of CHL-like orbifolds, introduced in section~\ref{s:5dimflux8supercharge}, extends to four-dimensional compactification. The above $G_{\text{shift}}$ action can be combined with an action in the gauge sector. From the point of view of the quotient geometry, the gauge action is understood as turning on a $\Z_{k_1}\times\Z_{k_2}$ holonomy for the gauge bundle, supported on the non-trivial fundamental group. This leads to CHL-like theories with a possibly reduced gauge group rank. A new feature of the six-dimensional geometries, compared to their five-dimensional counterparts, is the possibility of turning on a flat gerbe in the quotient geometry. As described in section~\ref{s:fluxvacua}, a generic flux configuration has a fundamental group $\pi_1(X)\simeq\Z_{m_1}\times\Z_{m_1m_2}$ with $m_1>0$, where the two integers are obtained by fixing a basis in Smith normal form for the torus Chern classes. The space $X$ can support non-trivial flat gerbes, since their topology is classified by the torsion subgroup of $H^3(X,\Z)$, which evaluates to
\begin{align}
\left\lbrace H^3(X,\Z)\right\rbrace_{\text{tors}}=\Z_{m_1}~.
\end{align}
Passing to the simply-connected cover $X_0$ of $X$, any such flat gerbe on $X$ has an equivalent description in terms of a $\Z_{m_1}\times\Z_{m_1m_2}$-equivariant  gerbe on $X_0$~\cite{Cheng:2022nso}. Notice that the group cohomology group $H^2(\Z_{m_1}\times\Z_{m_1m_2},\GU(1))$, characterizing orbifold $\Z_{m_1}\times\Z_{m_1m_2}$ actions on $B$-fields~\cite{Sharpe:2000ki}, is given by
\begin{align}
H^2\left(\pi_1(X),\GU(1)\right)=\Z_{m_1}~.
\end{align}
This motivates the interpretation of flat gerbes on $X=X_0/\pi_1(X)$ as encoding the possibility of discrete torsion in the orbifold of the simply-connected theory~\cite{Vafa:1986wx,MR2128387}.

\subsection{Orbifolds and supersymmetry reduction}
\label{ss:n1quotient}

The orbifold construction described above leads to large families of topologically distinct flux compactifications, with different flat structures and possibly reduced gauge group rank. By construction, the orbifolds always act freely on the space $X_0$, leading to smooth quotient configurations. Any such quotient geometry $X$ is a principal $T^2$ bundle fibered over a K3 surface, albeit with different topological data than the one of $X_0$---this is consistent with the result of~\cite{Hofer:1993tpb}. Consequently, the orbifold compactifications preserve $8$ supercharges in spacetime.
As we will now discuss, the six-dimensional geometries also admit orbifolds that preserve only $4$ supercharges.

Consider the by now familiar six-dimensional space $X_v\to M$, with the corresponding CFT denoted by $\cC_{v}(\tau,\rho,\ba_1,\ba_2)$. As described in~\cite{Becker:2008rc}, this six-dimensional space can be used to construct orbifold geometries that partially break supersymmetry. Indeed, some spaces $X_v$ admit a cyclic group $G$ of isometries such that $X_v/G$ preserves exactly four supercharges. This class of geometries is obtained by tuning the surface $M$ to a specific locus in the K3 moduli space where it admits a non-symplectic group action, and then extending this non-symplectic action to the total space. The quotient N=1 geometries $X_v/G$ are generically singular and, $X$ being non-K\"ahler, little is known about the existence of a smooth supergravity solution upon resolution. A partial answer has been recently discussed in~\cite{Giusti:2023mqz}, where the existence of conformally balanced metrics is shown for crepant resolutions of non-K\"ahler orbifolds with isolated singularities. The belief that $X/G$ could lead to a smooth $\SU(3)$ structure heterotic background after resolution of its singularities is also supported by the existence of a torsional linear sigma model framework for the N=1 quotients~\cite{Israel:2023itj}.

\subsubsection*{Orbifold construction}
Let us describe the orbifold construction in more detail. In general, the quotient of $X$ by the discrete isometry group $G$ will break the hyper-K\"ahler symmetry of the K3 base $M$, while still preserving the $\SU(3)$ structure on $X$---this is the assumption of minimal supersymmetry. We can construct such N=1 orbifolds by lifting the $G$-action on $M$ to an action on the torus bundle $X$, subject to some requirements on the $T^2$ fibration. On the base, $G$ must act as a non-symplectic automorphism of the K3 surface $M$, which is a K\"ahler isometry of $M$ with a non-trivial action on $H^{2,0}(M)$. The non-symplectic group $G_M\subset\text{Aut}(M)$ of a K3 surface is finite and cyclic~\cite{Nikulin:1980,Sterk:1985}. For our purposes, it will be sufficient to consider K3 surfaces $M$ with a non-symplectic group of order two or three, but for now we can allow for the general case $G_M\simeq\Z_{k}$. The cyclic group $G_M$ is generated by an isometry $\sigma~:~M\to M$ such that
\begin{align}
\sigma^k&=\text{id}_{M}~,&\sigma^\ast J&=J~,&\sigma^\ast\Omega&=\zeta_{k}\Omega~,
\end{align}
where $\zeta_{k}=e^{2\pi i/k}$ denotes a $k$-th primitive root of unity. Given such a non-symplectic automorphism of order $k$, the lift of the $G_M=\left\langle\sigma\right\rangle$ isometries of $M$ to $G=\left\langle\hat{\sigma}\right\rangle$ isometries of the total space $X$ is provided by specifying a $\Z_k$ action on the $T^2$ fiber. In order to preserve the holomorphic (3,0) form on $X$, the lift $\hat{\sigma}$ should act by the phase $\hat{\sigma}^\ast\Theta=\bar{\zeta}_{k}\Theta$ on the holomorphic (1,0) form. This constrains the $\Z_k$ action on the $T^2$ fiber to be a rotation, which restricts the order of the orbifold group to $k\in\{2,3,4,6\}$. In addition, the $T^2$ complex structure parameter $\tau$ is fixed, up to a $\SL(2,\Z)$ redefinition, to the value $\tau=\zeta_3$ for $k\in\{3,6\}$ or $\tau=\zeta_4$ for $k=4$. The resulting torus cyclic isometry can be represented by the $\SL(2,\Z)$ action $\hat{\sigma}~:~\theta^I\to\cR^I{}_J\theta^J$, where $\mathcal{R}$ satisfies
\begin{align}
\cR^k&=1_2~,&\cR^t\cG\cR=\cG~.
\end{align}
Of course, defining this $G$-action locally is not sufficient: the isometry should be compatible with the fibration structure of $X$.

Consider the cover $\mathfrak{U}=\{U_a\}_{a\in I}$ of $M$ introduced in section~\ref{s:fluxvacua}, taken such that $\Z_k$ acts on the indexing  set $I$ as $\sigma(U_a)=U_{\sigma(a)}$.\footnote{If $\mathfrak{U}$ does not satisfy this property, one can consider the refined cover $\cap_{g\in G}g(\mathfrak{U})$. We refer to~\cite{YANG2014230} for more details on invariant open covers of $G$-spaces.} Recall that the $T^2$ principal bundle $X$ is defined by local trivializations $\psi_a~:~U_a\times T^2\to\pi^{-1}(U_a)$, with transition functions $\psi_{ab}=\psi_a^{-1}\psi_b$ given by
\begin{align}
\begin{array}{cccccc} \psi_{ab} & : & U_b\times T^2 & \to & U_a\times T^2 &\\ && (p,e^{i\theta_b^I}) & \mapsto & (p,e^{i\theta_a^I}) & =(p,\tau^I_{ab}(p)e^{i\theta^I_b}) \end{array}~.
\end{align}
The bundle isomorphism $\hat{\sigma}$ is specified by the local action $\hat{\sigma}_a=\psi_{\sigma(a)}^{-1}\hat{\sigma}\psi_a$ of the form 
\begin{align}
\begin{array}{cccccc} \hat{\sigma}_a & : & U_{a}\times T^2 & \to & U_{\sigma(a)}\times T^2 &\\ && (p,e^{i\theta_a^I}) & \mapsto & (p',e^{i\theta^{\prime I}_a}) & =(\sigma(p),e^{i\cR^I{}_J\theta_a^J}) \end{array}~.
\end{align}
On all non-empty double overlaps $U_{ab}$, consistency of this local action with the patching requires $\hat{\sigma}_a\psi_{ab}=\psi_{\sigma(a)\sigma(b)}\hat{\sigma}_b$, which translates to the condition 
\begin{align}
\sigma^\ast\tau^I_{\sigma(a)\sigma(b)}=\prod_J(\tau^J_{ab})^{\cR^I{}_J}
\end{align}
on transition functions. Whenever this conditions is satisfied, the connection $1$-forms $\bA^I$ on the line bundles $\cT_I$ can be consistently taken such that $\sigma^\ast\bA^I_{\sigma(a)}=\mathcal{R}^I{}_J\bA^J_a$. The corresponding Chern classes satisfy
\begin{align}
\sigma^\ast \upomega^I=\cR^I{}_J\upomega^J~.
\end{align}
This implies that the nowhere vanishing $1$-forms $\bTheta^I$ transform as $\hat{\sigma}^\ast \bTheta^I=\cR^I{}_J\bTheta^J$ under the lift, and the resulting symmetry of $X$ preserves its $\SU(3)$ structure. 

The $G$-action on $M$ should also lift to an action on the gauge bundle. Generically, the orbifold acts non-trivially in the gauge sector.\footnote{A non-trivial action on the $\GE_8\times\GE_8$ bundle is required in the linear sigma model construction of~\cite{Israel:2023itj}. Indeed, the left-moving fermions associated to the gauge sector should couple to the $(0,2)$ vector multiplets in order to cancel some of the gauge anomalies. The orbifold has a non-trivial action on vector multiplets by construction, so there has to be a corresponding action on charged gauge fermions. This should also hold in the non-linear sigma model.} We choose the lift to act as a global $\GO(\Gamma_8+\Gamma_8)$ order $k$ action $\phi^s \bsigma_s\to R(\phi^s\bsigma_s)$ on the fiber coordinates. Following the same reasoning as that for the $T^2$ lift, we require the Chern class $\bUplambda$ of abelian instantons to satisfy
\begin{align}
\sigma^\ast\bUplambda=R(\bUplambda)~.
\end{align}
In the presence of Wilson lines, the above condition should extend to the component $\ba_I \upomega^I$ of the gauge curvature. This requires the pair of Wilson line parameters $(\ba_1,\ba_2)$ to belong to the sublattice of $(\Gamma_8+\Gamma_8)^2$ defined by $R(\ba_I)=\cR^J{}_I\ba_J$. From the quantization conditions~(\ref{eq:ModuliQuantization}) we then easily see that the $2$-forms $\upnu_I$ obey $\sigma^\ast\upnu_I=(\cR^{-1})^J{}_I\upnu_J$.

We can summarize the above conditions as follows. The lift of the $G$-action on the $T^2$ and gauge bundles is represented by the $\Z_k\subset\GO(\Gamma_{2,18})$ subgroup generated by
\begin{align}
\label{eq:Zkaction}
g_k = g_{\text{t}}[\cR]g_{\text{g}}[R] = \begin{pmatrix} \cR & 0 & 0 \\ 0 & \cR^{-t} & 0 \\ 0 & 0 & R \end{pmatrix}~.
\end{align}
Let $X_v\to M$ be a six-dimensional heterotic flux geometry, fibered over a K3 base equipped with a non-symplectic automorphism $\sigma$. This non-symplectic symmetry of $M$ lifts to a supersymmetric action on $X_v$ if the vector $v\in\Gamma_{1,17} \otimes H^2(M,\Z)$ satisfies
\begin{align}
\label{eq:Zklift}
\sigma^\ast v=g_k v~,
\end{align}
and the $T^2$ and Wilson line parameters obey $(\tau,\rho,\ba_1,\ba_2)=\mu_{g_{k}}(\tau,\rho,\ba_1,\ba_2)$.

The N=1 quotient geometries $X_v/G$ obtained by this orbifold construction are generically singular. Indeed, the $\SL(2,\Z)$ rotation of the torus fiber leaves a number of points of $T^2$ invariant. The quotient geometry will thus inherit the same number of singularities based at each point of $M$ that is fixed by $\sigma$. Requiring that $\sigma$ acts freely is too restrictive: any non-symplectic automorphism of order $k\geq3$ has a non-empty fixed locus, whose topological structure is constrained by both the topological Lefschetz fixed point formula and the holomorphic Lefschetz formula (see, for example~\cite{Artebani:2011}). This singles out $k=2$, as freely-acting K3 non-symplectic automorphisms of order two exist---they are nothing but Enriques involutions.  For $k\ge3$ the quotients $X_v/G$ unavoidably contain a singular locus, although the situation with $k=3$ is much simpler since there exist $\Z_3$ non-symplectic actions on K3 with only three isolated fixed points.

As a last general remark, note that any non-symplectic automorphism $\sigma$ of $M$ naturally provides an isometry of the second integral cohomology $H^2(M,\Z)\simeq\Gamma_{3,19}$ given by the pullback $\sigma^\ast$. The curves left invariant by $\sigma^\ast$ form the sublattice $\Gamma_{\text{inv}}\subset H^2(M,\Z)$. Since the (2,0) holomorphic form transforms non-trivially, $\Gamma_{\text{inv}}$ has to be contained in the Picard lattice of $M$. We denote by $(\Gamma_{\text{inv}})^\perp$ its orthogonal complement in $\Gamma_{3,19}$. Remarkably, the topology of the fixed locus of $M$ uniquely determines both $\Gamma_{\text{inv}}$ and $(\Gamma_{\text{inv}})^\perp$ as well as the action of $\sigma^\ast$ on the two sublattices (the details of these actions can be found in \cite{Nikulin:1980,Artebani:2011}). The vector $v$ specifying the bundle $X_v$ can thus be constructed from the explicit decomposition of $\Gamma_{\text{inv}}+(\Gamma_{\text{inv}})^\perp\subset H^2(M,\Z)$.

\subsubsection*{T-dual configurations and their quotients}

Consider the condition $\sigma^\ast v=g_k v$ required to lift the non-symplectic action on $M$ to a $G\simeq\Z_k$ orbifold of $X_v\to M$. For every $\GO(\Gamma_{2,18})$ element $\cT$, the vector $v'=\cT v$ describes an isomorphic compactification $X_{v'}\to M$. This T-dual configuration admits a $G'\simeq\Z_k$ symmetry acting on the base $M$ by the non-symplectic automorphism $\sigma$, and on the $T^2$ and gauge bundle by the $\GO(\Gamma_{2,18})$ action
\begin{align}
g'_k = \cT g_k \cT^{-1}~.
\end{align}
This action is consistent with the fibration structure of the T-dual configuration $X_{v'}$, since the vector $v'$ obeys $\sigma^\ast v'=g'_k v'$. 

Generically, $g'_k$ does not belong to $\GO(\Gamma_{2,2})\times\GO(\Gamma_8+\Gamma_8)$, and the T-dual action is non-geometric. Specifically, for the ``unwinding'' T-duality $\cT=\cT[\bkappa_1,\bkappa_2,m]$ introduced in~(\ref{eq:unwindingduality}), we can read off the dual $\GO(\Gamma_{2,18})$ action
\begin{align}
g'_k=\cT[\bkappa_1,\bkappa_2,m]\cT[\bkappa'_1,\bkappa'_2,m]^{-1}g_{\text{t}}[\cR']g_{\text{g}}[R]~,
\end{align}
where $\cR'=\cR^{-t}$ and $\bkappa'_I=(\cR^{-1})^J{}_IR(\bkappa_J)$. For a generic choice of the lattice vectors parametrizing $\cT$, the T-dual $G'$ action mixes the $T^2$ and gauge degrees of freedom. However, if the $\bkappa_I$ are such that $R(\bkappa_I)=\cR^J{}_I\bkappa_J$, we are led to the much simpler T-dual action
\begin{align}
g'_k=g_{\text{t}}[\cR']g_{\text{g}}[R]~.
\end{align}
This form of $g'_k$ now has the exact same geometric interpretation as $g_k$: a $\SL(2,\Z)$ isometry of the torus by $\cR'=\cR^{-t}$ accompanied by a global rotation $R$ in the gauge sector. However, recall that the $G$ action on vertex operators also involves a phase $U(g_k,\bp)$. In the T-dual description, the symmetry group $G'$ acts with a different phase $U(g'_k,\bp)$.\footnote{Details on the computation of this $U(g'_k,\bp)$ are spelled out in appendix~\ref{app:Tdualcocycles}.} It might be that the resulting $G'$ action does not have a simple geometric interpretation but involves, for example, an action on winding modes, similarly to what we found for CHL quotients in section~\ref{s:5dimflux8supercharge}.

\subsection{Examples}
We now examine some examples of quotient geometries that preserve $4$ supercharges.
\subsubsection*{Smooth $\Z_{2}$ quotients}
Let us first focus on $\Z_2$ orbifolds. The construction starts from a K3 base $M$ tuned to the Enriques locus. The Enriques involution $\sigma_{2}$ acts freely on $M$, and the holomorphic $(2,0)$ form transforms as $\sigma_2^\ast\Omega=-\Omega$. We can lift the Enriques involution to a $G\simeq\Z_2$ action on the configuration $X_v\to M$ using the $\GO(\Gamma_{2,18})$ element\footnote{In the above orbifold set up, the reflection $g_{\text{ref}}$ corresponds to the $\SL(2,\Z)$ action $g_{\text{ref}}=g_{\text{t}}[-1_2]$.}
\begin{align}
g_{2}=g_{\text{ref}}\,g_{\text{flip}}~.
\end{align}
In other words, the action on $X$ combines the Enriques involution on the base with a reflection on both $T^2$ coordinates and an exchange $g_{\text{flip}}=g_{\text{g}}[R_{\text{flip}}]$ of the two $\GE_8$ factors of the gauge bundle. As explained above, this $\Z_2$ lift is compatible with the fibration structure of $X_v$ if the vector $v$ satisfies the condition~(\ref{eq:Zklift}). Here this condition amounts to $\sigma_2^\ast v=g_{\text{ref}}\,g_{\text{flip}}v$. We can solve this requirement in the following way. Recall the Enriques action on the second integral cohomology of $M$:
\begin{align}
\Gamma_{\text{inv}}&=(2\Gamma_{1,1})+(2\Gamma_8)~,&(\Gamma_{\text{inv}})^\perp&=\Gamma_{1,1}+(2\Gamma_{1,1})+(2\Gamma_8)~,
\end{align}
where the two sublattices correspond to even and odd classes under the pullback by $\sigma_2$.
The condition on $v$ constrains $\upomega^I$ and $\upnu_I$ to belong to $(\Gamma_{\text{inv}})^\perp$, while $\bUplambda$ should split as $\bUplambda=\bUplambda_{(+)}+\bUplambda_{(-)}$ where the two factors $\bUplambda_{(+)}$ and $\bUplambda_{(-)}$ are respectively elements of $\Gamma_{\text{inv}}$ and $(\Gamma_{\text{inv}})^\perp$ and satisfy $R_{\text{flip}}(\bUplambda_{(\pm)})=\pm\bUplambda_{(\pm)}$. The action of $\mu_{g_2}$ on the moduli
\begin{align}
\mu_{g_2}(\tau,\rho,\ba_1,\ba_2)=(\tau,\rho,-R_{\text{flip}}(\ba_1),-R_{\text{flip}}(\ba_2))~,
\end{align}
constrains the Wilson line parameters $\ba_I\in\Gamma_8+\Gamma_8$ to be antisymmetric under the exchange of their two $\Gamma_8$ components.

We can of course consider $\Z_2$-symmetric configurations $X_v$ that are T-dual to direct product geometries $X_{v'}\simeq M\times T^2$, using the $\GO(\Gamma_{2,18})$ element $\cT$ defined in~(\ref{eq:unwindingduality}). In order for the T-dual configuration to admit a geometric $\Z_2$ quotient $g'_2$, the transformation $\cT$ must be parametrized by lattice vectors $\bkappa_I$ such that $R_{\text{flip}}(\bkappa_I)=-\bkappa_I$. This condition is easily satisfied for most configurations, and the T-duality leads to a trivial product configuration with 
\begin{align}
\bUplambda'=\bUplambda-\bkappa_1\upomega^1-\bkappa_2\upomega^2~.
\end{align}
We see that the T-duality introduces abelian instantons in configurations that are antisymmetric with respect
to the exchange of the two $\GE_8$ factors. However, this antisymmetry is compensated by the non-trivial transformation of the classes $\upomega^I$ under the Enriques involution, resulting in $\sigma_2^\ast\bUplambda'=R_{\text{flip}}(\bUplambda')$. The T-dual theory corresponds, after orbifolding, to a compactification on the Enriques Calabi--Yau manifold $(M\times T^2)/\Z_2$, endowed with an abelian gauge bundle with non-trivial holonomy turned on for the $\Z_2$ factor of $(\GE_8\times\GE_8)\rtimes\Z_2$.

When viewed from the $8$-dimensional perspective, this example may at first appear to be slightly puzzling, since it appears that we are able to put the $8$-dimensional theory on a non-spin manifold---the Enriques surface---while preserving supersymmetry.   This is reminiscent of the six-dimensional IIB orientifold compactification recently discussed in~\cite{Cheng:2023owv}, and the resolution to the seeming puzzle is similar in spirit: the $8$--dimensional gravitinos are charged under a lift of the $\Z_2$ reflection symmetry inherited from the torus, and by turning on a non-trivial holonomy for this symmetry supported by the fundamental group of the Enriques surface, we obtain well-defined gravitinos and preserve four supercharges in four dimensions.

\subsubsection*{Singular $\Z_{3}$ orbifolds}
We now turn our attention to $\Z_3$ N=1 orbifolds. 
The construction requires a K3 surface $M$ admitting a non-symplectic automorphism of order three, with only three isolated fixed points. An example of such K3 manifold is obtained by taking $M$ to be a double cover of a del Pezzo surface $\text{dP}_{6}$ of degree six, branched over a curve of genus seven. Such a construction yields a K3 surface, as can be checked from the Euler characteristic $\chi(M)=2\chi(\text{dP}_{6})-(2-2\times 7)=24$. The surface $\text{dP}_{6}$ can be represented as a  complete intersection of tridegree $(1,1,1)$ in a projective ambient space $\P^1\times\P^1\times\P^1$. Denoting the projective coordinates of the $a$-th $\P^1$ by $y^a=[y_{0}^{a}:y_{1}^{a}]$ and by $x$ the homogeneous coordinate describing the branched cover, the K3 surface $M$ admits the algebraic description
\begin{align}
\label{eq:Z3symmetricK3}
p_{1}(y^1,y^2,y^3)&=0~,&
x^{2}&=p_{2}(y^1,y^2,y^3)~,
\end{align}
where $p_{1}$ and $p_{2}$ are polynomials of respective degree $(1,1,1)$ and $(2,2,2)$ in the $\P^1$ homogeneous coordinates. The polynomials can be tuned to the symmetric locus $p_{1}(y^2,y^3,y^1)=p_{1}(y^1,y^2,y^3)$ and $p_{2}(y^2,y^3,y^1)=\zeta_{3}p_{2}(y^1,y^2,y^3)$ while keeping the hypersurface $M$ smooth. On this locus, the K\"ahler isometry of the projective space
\begin{equation}
\label{eq:nonsymplecticZ3}
\sigma_3\cdot[x:y_0^1:y_1^1:y_0^2:y_1^2:y_0^3:y_1^3]=[\zeta_3^2 x:y_0^2:y_1^2:y_0^3:y_1^3:y_0^1:y_1^1]
\end{equation}
descends to $M$, providing a non-symplectic automorphism of order three, with $\sigma_3^\ast\Omega=\zeta_3\Omega$. The three isolated fixed points of $\sigma_3$ are located at $y^{1}=y^{2}=y^{3}$ and $x=0$. This example is the one introduced in~\cite{Israel:2023itj}, and is amenable to a torsional linear sigma model description of the orbifold.

We lift the action of $\sigma_3$ to the $T^2$ bundle by the $\SL(2,\Z)$ action
\begin{align}
\cR = \begin{pmatrix} -1 & 1 \\ -1 & 0 \end{pmatrix}~.
\end{align}
Note that this $T^2$ rotation leaves the three points $(\theta^{1},\theta^{2})=(0,0),(\tfrac{2\pi}{3},\tfrac{4\pi}{3}),(\tfrac{4\pi}{3},\tfrac{2\pi}{3})$ invariant. The quotient space $X/G$ will consequently inherit nine isolated $\C^3/\Z_{3}$ singularities. 

The $\Z_3$ action on the gauge bundle should also be specified. In contrast to the $\Z_2$ outer automorphism of $\GE_8\times\GE_8$, there is no canonical choice of $\Z_3$ action on $\GE_8\times\GE_8$, so we need to fix an embedding $\Z_3\subset\GO(\Gamma_8+\Gamma_8)$. A convenient choice in the following will be to pick two roots $\bbeta_1$ and $\bbeta_2$ of $\Gamma_8+\Gamma_8$ at an angle $\bbeta_1\cdot\bbeta_2=-1$. The product of Weyl reflections
\begin{align}
R=R_{\bbeta_2}R_{\bbeta_1}
\end{align}
provides an order three lattice isometry, with $R^2=R_{\bbeta_1}R_{\bbeta_2}$. An advantage of such construction is that the phase $U(g_{\text{g}}[R],\bp)$ associated to this $\Z_3$ action can be explicitly computed. We give a choice of representative in appendix~\ref{app:Tdualcocycles}.

\subsubsection*{An example with trivial gauge bundle}
Consider a simply-connected configuration $X_v\to M$ with trivial gauge bundle, fibered over the K3 surface $M$ introduced above. The bundle is described by a vector $v$ of the form~(\ref{eq:4dconfiguration}) with $\bUplambda=0$. The lift of the $\Z_3$ action to $X_v$ requires $\sigma_3^\ast v=g_3v$, where $g_3$ is the $\GO(\Gamma_{2,18})$ element defined in~(\ref{eq:Zkaction}). This condition constrains the torus classes $\upomega^I_{\text{p}}$ to obey $\sigma_3^\ast \upomega^1_{\text{p}}=-\upomega^1_{\text{p}}+\upomega^2_{\text{p}}$ and $\sigma_3^\ast \upomega^2_{\text{p}}=-\upomega^1_{\text{p}}$. In terms of these classes, the vector $v$ must be of the form
\begin{align}
v = \begin{pmatrix} \upomega_{\text{p}}^1 \\ \upomega_{\text{p}}^2 \\ l_1 \upomega_{\text{p}}^1+l_2 \upomega_{\text{p}}^2 \\ (-l_1-l_2)\upomega_{\text{p}}^1+l_1 \upomega_{\text{p}}^2 \\ 0 \end{pmatrix}~.
\end{align}
The Bianchi identity reads
\begin{align}
-l_1\left(\upomega_{\text{p}}^1.\upomega_{\text{p}}^1+\upomega_{\text{p}}^2.\upomega_{\text{p}}^2-\upomega_{\text{p}}^1.\upomega_{\text{p}}^2\right)=24~,
\end{align}
and for a smooth K3 surface $M$ this restricts the integer $l_1$ to values $1\leq l_1\leq 4$.

The second integral cohomology of $M$ splits under the action of $\sigma_3^\ast$ as~\cite{Artebani:2008}
\begin{align}
\Gamma_{\text{inv}}&=(3\Gamma_{1,1})+(-3\Gamma_{\mathfrak{e}_6}^\ast)~,&(\Gamma_{\text{inv}})^\perp&=\Gamma_{1,1}+(3\Gamma_{1,1})+(-\Gamma_{\mathfrak{a}_2})^5~.
\end{align} 
Every $-\Gamma_{\mathfrak{a}_2}$ factor in the cohomology sublattice $(\Gamma_{\text{inv}})^\perp$ is generated by two curves $x$, $y$ with intersection $x.x=y.y=-2$, $x.y=1$, and on which the non-symplectic automorphism acts as $\sigma_3^\ast x=-x-y$ and $\sigma_3^\ast y=x$. We can thus pick a $(-\Gamma_{\mathfrak{a}_2})^2$ sublattice of $(\Gamma_{\text{inv}})^\perp$ with generators $x_1$, $y_1$ and $x_2$, $y_2$, and set
\begin{align}
\upomega_{\text{p}}^1&=x_1-x_2~,&\upomega_{\text{p}}^2&=-y_1+y_2~.
\end{align}
The Bianchi identity is straightforwardly solved by taking $l_1=4$. Using a $B$-field shift parametrized by $m=-l_2-4$, we can bring $v$ to the simple form
\begin{align}
v=g_{\text{b}}[m] \begin{pmatrix} \upomega_{\text{p}}^1 \\ \upomega_{\text{p}}^2 \\ 4\upomega_{\text{p}}^1+l_2\upomega_{\text{p}}^2 \\ -l_2\upomega_{\text{p}}^1+4\upomega_{\text{p}}^2 \\ 0 \end{pmatrix}
=
\begin{pmatrix} \upomega_{\text{p}}^1 \\ \upomega_{\text{p}}^2 \\ 4\upomega_{\text{p}}^1-4\upomega_{\text{p}}^2 \\ 4\upomega_{\text{p}}^2 \\ 0 \end{pmatrix}~.
\end{align}
The above configuration can then be ``unwound'' to $M\times T^2$ using the transformation $\cT=g_{\text{i}}g_{\text{s},1}[\bkappa_1]g_{\text{s},2}[\bkappa_2]$, with two lattice vectors such that $\bkappa_1\cdot\bkappa_1=\bkappa_2\cdot\bkappa_2=8$ and $\bkappa_1\cdot\bkappa_2=-4$. We easily see that these conditions are satisfied by the two $\Gamma_8+\Gamma_8$ vectors
\begin{align}
\bkappa_1&=2\bbeta_1~,&\bkappa_2&=2\bbeta_2~,
\end{align}
and this is also consistent with the requirement $R(\bkappa_I)=\cR^J{}_I\bkappa_I$ for our choice of order three rotation $R=R_{\bbeta_2}R_{\bbeta_1}$. 

The topology of the T-dual configuration is described by the vector
\begin{align}
v'=\cT v=\begin{pmatrix} 0 \\ 0 \\ \upomega_{\text{p}}^1 \\ \upomega_{\text{p}}^2 \\ -\bkappa_1 \upomega_{\text{p}}^1-\bkappa_2\upomega_{\text{p}}^2 \end{pmatrix}~,
\end{align}
which represents a product geometry $M\times T^2$ equipped with an abelian bundle $\bUplambda'=-\bkappa_I \upomega^I_{\text{p}}$. We can observe that this configuration obeys
\begin{align}
\sigma_3^\ast v'=g'_3 v'~,
\end{align}
with the order three $\GO(\Gamma_{2,18})$ element $g'_3=\cT g_3\cT^{-1}=g_{\text{t}}[\cR^{-t}]g_{\text{g}}[R]$.

The $T^2$ and Wilson line parameters of the two configurations can also be tracked through the duality. Starting from a configuration $v$ without Wilson lines,\footnote{In principle, the condition $(\tau,\rho,\ba_1,\ba_2)=\mu_{g_{3}}(\tau,\rho,\ba_1,\ba_2)$ allows for non-trivial Wilson lines parameters satisfying $R(\ba_1)=-\ba_1-\ba_2$ and $R(\ba_2)=\ba_1$. The following discussion stays valid when such parameters are turned on.} the initial theory is described by the parameters
\begin{align}
\tau&=\zeta_3~,&\rho&=4(\zeta_3+1)~,&\ba_I=0~,
\end{align}
which solve the quantization conditions and are invariant under the action of $\mu_{g_3}$. The  T-dual configuration $v'$ has parameters
\begin{align}
\tau'&=\zeta_3+1~,&\rho'&=\tfrac{1}{24}(2\zeta_3+1)~,&\ba'_1&=-\tfrac{1}{12}(2\bkappa_1+\bkappa_2)~,&\ba'_2&=-\tfrac{1}{12}(\bkappa_1+2\bkappa_2)~.
\end{align}
Notice that the quantization conditions on $v'$ amount to
\begin{align}
\ba'_I\cdot\bUplambda'\in H^2(M,\Z)~,
\end{align}
and those conditions are obeyed for the specific choices of $\ba'_I$ and $\bUplambda'$ of the T-dual geometry. Moreover, we easily see that the action
\begin{align}
\mu_{g_3'}^{-1}(\tau',\rho',\ba'_1,\ba'_2)=\left(\frac{1}{1-\tau'},\rho',-R(\ba'_1)+R(\ba'_2),-R(\ba'_1)\right)
\end{align}
of the T-dual $\Z_3$ symmetry on the moduli leaves invariant the parameters associated to $v'$. Finally, let us note that to fully specify the orbifold of the T-dual worldsheet CFT, it is necessary to compute the phase factor $U(g'_3,\bp)$ that enter into the action of $g'_3$ on vertex operators. For the unwound example described above, the cocycle representatives can be chosen such that $U(g'_3,\bp)=U(g_3,\bp)$. The $\Z_3$ orbifold can therefore be understood in the T-dual description as the same geometric quotient as in the original theory. This feature is not generic: for different configurations, one should expect that the T-dual orbifold does not always have a geometric interpretation.

Having obtained a T-dual K\"ahler description of the $\Z_3$ action, a question naturally arises: how can we understand the $\Z_3$ quotient of the T-dual theory? On the unwound side, the orbifold corresponds to a $\Z_3$ symmetry of the direct product geometry $X'\simeq M\times T^2$ accompanied by an order three action on the gauge bundle. The action on the six-dimensional K\"ahler geometry is by a non-symplectic action on the K3 factor $M$ along with an order three rotation of the torus. This isometry of $X'$ leads to a quotient with nine $\C^3/\Z_3$ singular points. The singular $X'/\Z_3$ geometry admits a smooth Ricci-flat resolution: the orbifold corresponds to the Borcea--Voisin construction of Calabi--Yau threefolds~\cite{Borcea:1996mxz,Voisin:1993mir}. However, the understanding of the resolved geometry necessitates a description of the gauge bundle. Indeed, on the unwound T-dual side, the gauge bundle is non-trivial by construction: the T-duality introduces two abelian instantons fibered over the classes $\upomega^1$ and $\upomega^2$. If those curves were to intersect the singular locus of the order three K3 automorphism, then the resolution of the singularities would also impact the gauge instantons. However, it is possible to construct geometries where this is not the case. Consider the K3 surface $M$ defined by~(\ref{eq:Z3symmetricK3}). This surface inherits three classes $C_1$, $C_2$, $C_3$ from its ambient projective space $\P^1\times\P^1\times\P^1$, cut out by hyperplanes of the form $\P^1\times\P^1$. The curves $C_i$ have zero self-intersection and intersect each other at two points. Under the non-symplectic automorphism~(\ref{eq:nonsymplecticZ3}), they transform by the permutation $\sigma^\ast_3 C_i=C_{i+1}$. The torsional geometry $X\to M$ can be constructed from two classes
\begin{align}
\upomega^1&=C_1-C_2~,&\upomega^2&=C_1-C_3~,
\end{align}
with the ASD condition on $\upomega^I$ requiring the three $\P^1$ factors to be of the same size. A generic curve representing the class $C_i$ does not intersect the fixed locus of $\sigma_3$. Therefore, we can expect that the K\"ahler resolution of the $(M\times T^2)/\Z_3$ orbifold geometry will be unobstructed. It would be interesting to better understand the resolved geometry and its worldsheet realization. If the resolution modes admit a description in terms of twisted marginal operators of the unwound $(0,2)$ CFT, then studying the corresponding modes on the T-dual non-K\"ahler side could give some insight on the existence of a non-K\"ahler $\SU(3)$-structure resolution of the $X/\Z_3$ quotient geometry.

\subsection{Duality and the supersymmetric index} \label{s:torglsm}

The supersymmetric index of N=2 four-dimensional heterotic vacua was introduced in~\cite{Cecotti:1992qh} for $\text{K3}\times T^2$ compactifications and generalized in~\cite{Israel:2015aea,Israel:2016xfu} to heterotic flux vacua. Our aim is to reformulate the index in a duality-covariant way and examine in particular the effect of the unwinding dualities. We refer 
to the aforementioned publications for conventions and details. 

Let us consider for definiteness a compactification to four dimensions endowed with the pullback of a rank $r$ HYM bundle $\cP$ on K3 with structure group in the second $\GE_8$ factor with 
vanishing first Chern class. We will consider later on possible line bundles and Wilson lines in the first $\GE_8$ factor only. Since $c_1(\cP)=0$, 
the $(0,2)$ SCFT associated with the non-linear sigma model contains a non-anomalous left-moving $U(1)_\sleft$ symmetry, associated to a holomorphic current $J(z)$, 
that will eventually be used for the left-moving GSO projection. We also denote by $\bar{J}(\bar z)$ the R-current of the right-moving N=2 superconformal algebra.  

We first define the dressed elliptic genus of the $(0,2)$ two-dimensional SCFT with target space $T^2 \hookrightarrow X \stackrel{p}{\to} M$ and central charges $(c,\bar{c}) = (14+r,9)$ as the following trace in the Ramond--Ramond sector~:
\begin{align}
Z_{\text{deg}}(\boldsymbol{q},\bar{\boldsymbol{q}} ; \boldsymbol{y}) = \frac{1}{\bar{\eta}^2 (\bar{\boldsymbol{q}})} 
\text{Tr}_{\textsc{r}} \left\{ \boldsymbol{y}^{J_0} \bar{J}_0 (-1)^F \boldsymbol{q}^{L_0-c/24} \bar{\boldsymbol{q}}^{\bar{L}_0 - \bar{c}/24} \right\} ~,
\end{align}
where $J_0$ (resp. $\bar{J}_0$) is the zero-mode of the $U(1)_\sleft$ current (resp. the $U(1)_\sright$ current) and $(-1)^F = \exp 
i\pi (J_0 -\bar{J}_0)$. One can then obtain the new supersymmetric index proper as:
\begin{align}
\label{eq:new_susy_index}
Z_{\text{new}} (\boldsymbol{q}, \bar{\boldsymbol{q}}) =  \frac{\bar{\eta}^2 (\bar{\boldsymbol{q}})}{2 \eta (\boldsymbol{q})} 
\sum_{\gamma, \delta = 0}^1 \boldsymbol{q}^{\gamma^2} \left\{ 
\left(\frac{\vartheta_1 (\boldsymbol{q}|\boldsymbol{y})}{\eta (\boldsymbol{q})}\right)^{8-r}Z_{\text{deg}} (\boldsymbol{q},\bar{\boldsymbol{q}} ; \boldsymbol{y}) \right\} 
\Bigg|_{\boldsymbol{y} = \boldsymbol{q}^{\gamma/2}e^{i\pi\delta}} ~.
\end{align}

The dressed elliptic genus of flux N=2 compactifications to four dimensions was obtained in~\cite{Israel:2015aea,Israel:2016xfu} using a UV completion of the corresponding $(0,2)$ non-linear sigma model as a $(0,2)$ gauged linear sigma model~\cite{Adams:2006kb} and supersymmetric localization, building on previous works about the ordinary elliptic genera~\cite{Benini:2013nda,Benini:2013xpa}. From this result a geometric formulation of the dressed elliptic genus was derived for any compactification such that the K3 base $M$ is 
realized as a subvariety of a weighted projective space.

In the absence of Wilson lines, the $\Gamma_{2,10}$ lattice splits into a $(2,2)$ lattice and the $\GE_8$ lattice. The former is the lattice 
associated with the $T^2$ fiber, spanned by 
\begin{align}
\bp=\text{w}^I\be_I+\text{n}_I\be^{\ast I}=\bp_\sleft+\bp_\sright
\end{align}
in $\mathbb{R}^{2,2}$,  where the $\text{n}_I$ (resp. the $\text{w}^I$) are the momenta (resp. the winding numbers). 
The decomposition in left- and right-moving components is defined through the basis vectors
\begin{align}
\be_\sleft^1&=\frac{1}{\sqrt{2\tau_2\rho_2}}\left(\tau_2\be_1+\rho_2\be^{\ast1}+(\tau_1\rho_2+\tau_2\rho_1)\be^{\ast2}
\right)~,\nonumber\\
\be_\sleft^2&=\frac{1}{\sqrt{2\tau_2\rho_2}}\left(-\tau_1\be_1+\be_2-\rho_1\be^{\ast1}+(\tau_2\rho_2-\tau_1\rho_1)\be^{\ast2}
\right)~,\nonumber\\
\be_\sright^1&=\frac{1}{\sqrt{2\tau_2\rho_2}}\left(-\tau_2\be_1+\rho_2\be^{\ast1}+(\tau_1\rho_2-\tau_2\rho_1)\be^{\ast2}
\right)~,\nonumber\\
\be_\sright^2&=\frac{1}{\sqrt{2\tau_2\rho_2}}\left(\tau_1\be_1-\be_2+\rho_1\be^{\ast1}+(\tau_2\rho_2+\tau_1\rho_1)\be^{\ast2}
\right)~.
\end{align}
This basis satisfies
\begin{align}
\be_\sleft^a\cdot\be_\sleft^b&=\delta^{ab}~,&
\be_\sright^a\cdot\be_\sright^b&=-\delta^{ab}~,&
\be_\sleft^a\cdot\be_\sright^b&=0~,
\end{align}
and the left and right projections are defined by $\bp_\sleft=(\bp\cdot\be_{\sleft a})\be_\sleft^a$ and $\bp_\sright=-(\bp\cdot\be_{\sright a})\be_\sright^a$.
As we have 
seen in subsection~\ref{subsec:quantiz} the moduli of the $T^2$ are quantized in such a way that the $T^2$ CFT is rational, 
or in other words the sublattices $\Gamma_\sleft = \Gamma_{2,2} \cap \mathbb{R}^{2,0}$ and  $\Gamma_\sright = \Gamma_{2,2} \cap \mathbb{R}^{0,2}$ are 
both rank two lattices. Elements of the lattice $\Gamma_{2,2}$ with $\bp_\sright=0$ are such that the momenta $\text{n}_1$ and $\text{n}_2$ 
can be expressed in terms of the winding numbers $\text{w}^1$ and $\text{w}^2$:
\begin{subequations}
\label{eq:momenta_rat}
\begin{align}
\text{n}_1 & = \frac{\rho_2}{\tau_2} (\text{w}^1 + \tau_1 \text{w}^2) - \rho_1 \text{w}^2 \ \in \ \mathbb{Z} \, , \\
\text{n}_2 & =  \frac{\rho_2}{\tau_2} (\tau_1 \text{w}^1 + |\tau|^2 \text{w}^2) + \rho_1 \text{w}^1 \ \in \ \mathbb{Z} \, .
\end{align}
\end{subequations}
Note that any lattice vector in $\Gamma_\sleft$ can be expressed purely in terms of the winding numbers, using~(\ref{eq:momenta_rat}):
\begin{align}
\label{eq:pl_winding}
\bp_\sleft=\sqrt{\frac{2\rho_2}{\tau_2}}\left(\text{w}^1\be_\sleft^1+\text{w}^2(\tau_1\be_\sleft^1+\tau_2\be_\sleft^2)\right)~.
\end{align}
In the geometrical formulation for the dressed elliptic genus derived in~\cite{Israel:2015aea} from the torsional GLSM approach, 
the coupling between the $\Gamma_{2,2}$ lattice of the two-torus and the anti-self-dual (1,1) forms $\upomega^1$ and 
$\upomega^2$ associated with the curvatures of the torus bundle appeared naturally in the formula in terms of a formal extension of the winding lattice 
as a module over $H^2(M,\Z)$:
\begin{align}
\bp_\upomega=\sqrt{\frac{2\rho_2}{\tau_2}}\left(\upomega^1\be_\sleft^1+\upomega^2(\tau_1\be_\sleft^1+\tau_2\be_\sleft^2)\right)~.
\end{align}
As noticed above, this element of the winding lattice can be also viewed as an element of the lattice $\Gamma_\sleft$ of purely left-moving momenta. Adding the 
$\Gamma_{8}$ associated with the first $\GE_8$ factor, one can then use the same lattice formalism  
as in subsection~\ref{subsec:latticesetup} and associate to the principal torus bundle 
the element $v= \upomega^I \be_I + \upnu_I \be^{\ast I}$ of $\Gamma_{2,10} \otimes H^2 (M,\Z)$,
where the classes $\upnu_1$ and $\upnu_2$ are given by~(\ref{eq:przero}). We recall that, for the time being, there are neither Wilson lines nor line bundles in the first $\GE_8$ factor.  

In terms of $v$, one can express the dressed elliptic genus in a manifestly T-duality covariant presentation.   As for the ordinary elliptic genus~\cite{Witten:1986bf} 
we define first the formal power series with bundle coefficients:
\begin{align}
\mathbb{E}_{\boldsymbol{q},\boldsymbol{y}} = \bigotimes_{n=0}^\infty \bigwedge\nolimits_{-\boldsymbol{y} \boldsymbol{q}^n} \cP^* \otimes \bigwedge\nolimits_{-\boldsymbol{y}^{-1} \boldsymbol{q}^n} \cP
\otimes \bigotimes_{n=1}^\infty S_{\boldsymbol{q}^n} T_{M}^* \otimes \bigotimes_{n=1}^\infty S_{\boldsymbol{q}^n} T_{M} ~,
\end{align}
where
\begin{align}
\bigwedge\nolimits_{\boldsymbol{t}} \cP&= \sum_{k=0}^\infty \boldsymbol{t}^k\bigwedge\nolimits^k \cP~,& 
S_{\boldsymbol{t}} T_M &= \sum_{k=0}^\infty \boldsymbol{t}^k S^k T_M ~,
\end{align}
$\bigwedge\nolimits^k$ and $S^k$ being respectively the $k$-th exterior product and the $k$-th symmetric product. One has then
\begin{align}
Z_{\text{deg}} (\boldsymbol{q},\bar{\boldsymbol{q}} ; \boldsymbol{y}) =  \boldsymbol{q}^{\frac{r-2}{12}} \boldsymbol{y}^{r/2} \int_M \, \text{ch}\, \big( \mathbb{E}_{\boldsymbol{q},\boldsymbol{y}} \big)\, 
\text{td} \big( T_M\big) 
\sum_{\bp \in \Gamma_{2,10}} \frac{\boldsymbol{q}^{\frac{1}{2}\bp_\sleft^2}}{\eta^2(\boldsymbol{q})} \frac{\bar{\boldsymbol{q}}^{\frac{1}{2} \bp_\sright^2}}{\bar{\eta}^2 (\bar{\boldsymbol{q}})}\,  e^{-\bp \cdot  v}~, 
\end{align}
where as in subsection~\ref{subsec:latticesetup}, $\cdot$ denotes the inner product on $\Gamma_{2,10}$ preserving the $\R^{2,10}$ metric $\eta$.  

Using the splitting principle, let $c(T_M) = (1+\uppsi_1)(1+\uppsi_2)$ and $c(\cP)= \prod_{a=1}^r (1+ \upxi_a)$. One obtains then the final expression for the dressed elliptic genus, given as a function of the 
vector $v \in\Gamma_{2,10} \otimes H^2 (M,\Z)$:
\begin{equation}
\label{eq:deg_fy}
Z_{\text{deg}} (\boldsymbol{q},\bar{\boldsymbol{q}} ; \boldsymbol{y}|v) = \int_M 
\prod_{a=1}^r \frac{i \vartheta_1 \big(\boldsymbol{q}\big|\boldsymbol{y}^{-1}e^{\, \upxi_a} \big)}{\eta (\boldsymbol{q})} 
\prod_{i=1}^{2} \frac{\eta (\boldsymbol{q}) \uppsi_i}{ i \vartheta_1 \big(\boldsymbol{q}\big| e^{\uppsi_i} \big)}\, \sum_{\bp \in \Gamma_{2,10}} \, 
 \frac{\boldsymbol{q}^{\frac{1}{2}\bp_\sleft^2}}{\eta^2(\boldsymbol{q})} \frac{\bar{\boldsymbol{q}}^{\frac{1}{2} \bp_\sright^2}}{\bar{\eta}^2 (\bar{\boldsymbol{q}})}\,  e^{-\bp \cdot  v}~,
\end{equation}
where it is understood that the integral over the base $M$ selects the top form.  This index, although non-holomorphic, transforms under the action of the (worldsheet) modular group as  
a Jacobi form of weight -2 and rank $r/2$, whenever the Bianchi identity~(\ref{eq:bianchi_id}) is satisfied in cohomology~\cite{Israel:2015aea}.  The dressed 
elliptic genus~(\ref{eq:deg_fy}) is also invariant under the action of a space-time perturbative duality $g \in O(\Gamma_{2,10})$:
\begin{align}
Z_{\text{deg}} (\boldsymbol{q},\bar{\boldsymbol{q}} ; \boldsymbol{y}|gv)=Z_{\text{deg}} (\boldsymbol{q},\bar{\boldsymbol{q}} ; \boldsymbol{y}|v)~,
\end{align}
where it is understood that the index $Z_{\text{deg}} (\boldsymbol{q},\bar{\boldsymbol{q}} ; \boldsymbol{y}|gv)$ is computed using the left/right decomposition $\bp=\bp_\sleft+\bp_\sright$ of $\Gamma_{2,18}$ lattice vectors defined by the T-dual moduli.

\subsubsection*{Unwinding the index}

The dressed elliptic genus~(\ref{eq:deg_fy}) has been given in an $\GO(\Gamma_{2,10})$ covariant way, allowing to understand its transformation under topology-changing perturbative dualities. Explicitly, one would want to map the new supersymmetric index of a torsional compactification to the new supersymmetric index of a $\text{K3} \times T^2$ compactification with an extra abelian gauge bundle. 

The main example discussed in subsection~\ref{subsec:n=2duals} involved a torus fibration built out of two primitive elements of $H^2 (M,\mathbb{Z})$, corresponding to the vector
\begin{align}
\label{eq:4dconfigurationbis}
v = 
\begin{pmatrix} \upomega_{\text{p}}^1 \\ \upomega_{\text{p}}^2 \\ l_{11}\upomega_{\text{p}}^1+l_{12}\upomega_{\text{p}}^2 \\ 
l_{21}\upomega_{\text{p}}^1+l_{22}\upomega_{\text{p}}^2 \\ 0
\end{pmatrix}~,
\end{align}
where $\rho=l_{11}\tau-l_{12}$ and $\rho \tau =l_{21}\tau-l_{22}$. Recall that in the present context $v$ is a vector of $\Gamma_{2,10}$ rather than a vector of $\Gamma_{2,18}$, but it does not change 
the logic.

This configuration can be T-dualized by an $\GO(\Gamma_{2,18})$ transformation $\cT$ of the form~(\ref{eq:unwindingduality}), 
with $\bkappa_{1,2} \in \Gamma_8$ to a configuration describing a direct product $\text{K3}\times T^2$, endowed with an abelian bundle characterized by 
$\bUplambda '\in \Gamma_8 \otimes H^2(M,\Z)$ and quantized Wilson lines in the same $\GE_8$, i.e. with parameters $\ba_I '$ such that  $\ba_I '\cdot\bUplambda' \neq 0$.  A further duality---Wilson line integer shift of parameters $\bkappa'_{1,2} \in \Gamma_8$---allows to disentangle the Wilson lines from the abelian gauge bundle, see around~(\ref{eq:wlines_remov}), reaching the configuration 
specified by the vector 
\begin{align}
v''=\begin{pmatrix}
0\\0\\0\\0\\ \bUplambda''
\end{pmatrix}~,
\end{align}
with Wilson line parameters $\ba''_{1,2}$ given by~(\ref{eq:wilson_finalunwind}) and abelian gauge 
bundle specified by $\bUplambda''=-\bkappa_1\upomega^1_\text{p}-\bkappa_2\upomega^2_\text{p}$. For definiteness, and without loss of generality, let us consider that the non-zero 
entries of the 8-dimensional vector $-\bkappa_1\upomega^1_\text{p}-\bkappa_2\upomega^2_\text{p}$ are the first $s$ ones, such that the $\ba''_{1,2}$ Wilson line parameters 
have non-zero entries only for the remaining $8-s$ components.  

A better understanding of the dressed elliptic genus $Z_{\text{deg}} (\boldsymbol{q},\bar{\boldsymbol{q}} ; \boldsymbol{y}|v'')$ associated with the T-dual ``unwound'' configuration is obtained 
in the fermionic description, especially regarding the gauge bundle. 
The set of 8 chiral bosons with momentum lattice $\Gamma_8$ are then traded for a set of 8 left-moving fermions. For a given spin structure $[\alpha,\beta]$ the contribution 
to the path integral from the $8-s$ fermions supporting the Wilson lines of parameters $\ba_{I}^{''}$ is given by~\cite{Narain:1986am}
\begin{equation}
Z_w \left[ \begin{smallmatrix} \alpha \\ \beta \end{smallmatrix} \right] \left(\boldsymbol{q}\big|\ba_{I}^{''}\right) = \prod_{a=s+1}^8 \frac{1}{\eta(\boldsymbol{q})} 
\vartheta \left[ \begin{smallmatrix} \alpha + 2\ba_{I}^{'' a} \, \text{w}^I  
\\ \beta  + 2\ba_{I}^{'' a} \, \text{m}^I \end{smallmatrix} \right] \big(\boldsymbol{q}\big|0 \big)\, ,
\end{equation}
where we have used the orthonormal basis~(\ref{eq:CKbasis}). In this expression $\text{w}^I$ and $\text{m}^I$ are 
the winding numbers along the worldsheet space-like and time-like one-cycles respectively. After a Poisson resummation on 
$\text{m}^1$ and $\text{m}^2$, 
which are traded for $\text{n}_1$ and $\text{n}_2$, one obtains indeed left- and right-moving momenta of the expected form described in subsection~\ref{subsec:latticesetup}. 

As far as the abelian gauge bundle is concerned, the coupling of the $\Gamma_{2,10}$ lattice vectors with the background vector $v''$ given by~(\ref{eq:wlines_remov}) in the dressed elliptic 
genus~(\ref{eq:deg_fy}) takes the form 
\begin{equation}
e^{\, -\bp \cdot  v} = e^{\, \bp \cdot\big(\upomega^1_\text{p}\bkappa_1 +\upomega^2_\text{p}\bkappa_2 \big)} = 
\exp{\,\sum_{a=1}^s \ell^a \big(\upomega^1_\text{p}\bkappa_1^a +\upomega^2_\text{p}\bkappa_2^a \big)}~. 
\end{equation}
In the fermionic presentation, it gives a contribution to the dressed elliptic genus of the form: 
\begin{equation}
Z_l \left[ \begin{smallmatrix} \alpha \\ \beta \end{smallmatrix} \right] \big(\boldsymbol{q}\big|\bUplambda'\big) = \prod_{a=1}^s \frac{1}{\eta(\boldsymbol{q})} 
\vartheta \left[ \begin{smallmatrix} \alpha  \\ \beta \end{smallmatrix} \right] 
\Big( \boldsymbol{q}\Big|e^{\, \upomega^1_\text{p}\bkappa_1^a +\upomega^2_\text{p}\bkappa_2^a}\Big) \, .
\end{equation}
After summing over the spin structures, these terms will appear in the new supersymmetric index~(\ref{eq:new_susy_index}) exactly on the same footing as Fermi multiplets 
transforming as sections of abelian bundles over the base $M$, see the first term in the integrand of~(\ref{eq:deg_fy}). Thus we have demonstrated that, by the chain of Narain dualities  described in subsection~\ref{subsec:n=2duals}, the new supersymmetric index of flux compactifications with eight supercharges can be recast in the form of the new supersymmetric index for $\text{K3}\times T^2$ compactifications with appropriate Wilson lines and abelian gauge bundles.

\section{N=1 flux vacua in \texorpdfstring{$4$}{4} dimensions} \label{s:su3structure}
The observations in the previous sections show that four-dimensional heterotic vacua preserving $8$ supercharges can always be connected to flux-free $M\times T^2$ configurations by Narain T-duality. The N=2 requirements force the torus classes to be primitive $(1,1)$ forms, so these classes enter on the same footing as an abelian instanton in the gauge sector. It is then natural to expect that there exists a duality exchanging torus and gauge data.

Our goal in this section will be to repeat a similar T-duality analysis for geometries with only $4$ supercharges. As reviewed in section~\ref{s:fluxvacua}, anti-self-duality of the torus Chern classes is not required for heterotic geometries with minimal supersymmetry: the classes $\upomega^I$ can have non-zero $(2,0)\oplus(0,2)$ components as long as the complex combination $\upomega^1+\tau \upomega^2$ stays orthogonal to $\Omega$. Turning on such components leads to drastic consequences for the heterotic geometry. In particular, a crucial modification is the necessity of a formal $\alpha'$ expansion of the internal fields to solve the leading order equations of motion: the curvature $R_+$ of the $H$-twisted connection is no longer horizontal, and the Bianchi identity can only be solved at first order in $\alpha'$. The same $\alpha'$ expansion is necessary for a derivation of the flux quantization conditions as in~\cite{Melnikov:2012cv}. Another difference that will play an important role in the following is the quantization of Wilson line parameters. As described in section~\ref{s:fluxvacua}, there are no continuous deformations of the parameters $\ba_I$, and it is always possible to set $\ba_I=0$ by a large gauge transformation.

\subsection{Quantization of the K3 periods}
Supersymmetry and flux quantization impose strong conditions on the K3 surface $M$ used to build the heterotic geometry. A particularly simple example corresponds to configurations for which the torus Chern classes only have $(2,0)\oplus(0,2)$ components. To ensure supersymmetry, the complexified $T^2$ curvature must be proportional to the holomorphic $2$-form of $M$:
\begin{align}
\label{eq:attractiveK3}
\gamma\Omega=\upomega^1+\tau \upomega^2~,
\end{align}
where $\gamma$ is a non-zero complex constant. Consequently, supersymmetry fixes the periods of the K3 manifold. In particular, its Picard number is $\varrho(M)=20$: in other words, $M$ is an attractive K3 surface.\footnote{We refer to~\cite{Aspinwall:2005ad} for a pedagogical description of attractive K3 manifolds.} By the Shioda--Inose theorem~\cite{Shioda:1977}, this attractive K3 surface is completely characterized by the $\SL(2,\Z)$ equivalence class of the positive-definite even matrix $D$, where we denote by $D^{IJ}=\upomega^I.\upomega^J$ the intersection numbers of the torus classes. In addition, for the definition~(\ref{eq:attractiveK3}) of the holomorphic $2$-form to be consistent with the condition $\Omega.\Omega=0$, the torus complex structure is fixed to the value
\begin{equation}
\tau=\frac{-D^{12}+i\sqrt{\det D}}{D^{22}}~.
\end{equation}
After gauging away the Wilson line parameters, the quantization conditions~(\ref{eq:ModuliQuantization}) take the simple form
\begin{align}
\upnu_2-\tau\upnu_1=\rho(\upomega^1+\tau \upomega^2)~.
\end{align}
For this equation to admit solutions, the torus parameters $(\tau,\rho)$ are constrained to belong to the same quadratic field, corresponding to the conditions on the 
moduli of a rational $T^2$ CFT, as in the N=2 case.

Configurations with a purely holomorphic complexified $T^2$ curvature are quite special: in a generic N=1 geometry, the classes $\upomega^I$ also have $(1,1)$ components. When this is the case, the complexified $T^2$ class decomposes as
\begin{align}
\label{eq:mixedT2curvature}
\upomega^1+\tau \upomega^2 = \xi+\gamma\Omega~,
\end{align}
in terms of a complex primitive $(1,1)$ form $\xi$ and a complex constant $\gamma$. The quantization conditions~(\ref{eq:ModuliQuantization}) relate the Hodge duals of the torus classes to a linear combination of the four integral classes $\upomega^I$ and $\upnu_I$. Since the Hodge operator acts with a definite sign on $\xi$ and on $\Omega$, the quantization conditions can be used to extract their expansion in integral cohomology:
\begin{subequations}
\label{eq:mixedT2curvaturedecomposition}
\begin{align}
\gamma\Omega & =\frac{1}{2i\rho_{2}}\left(-\bar{\rho}(\upomega^{1}+\tau \upomega^{2})-\tau\upnu_{1}+\upnu_{2}\right)~, \\ \xi & =\frac{1}{2i\rho_{2}}\left(\rho(\upomega^{1}+\tau \upomega^{2})+\tau\upnu_{1}-\upnu_{2}\right)~.
\end{align}
\end{subequations}
In particular, the holomorphic $2$-form is defined by four integral classes, and the Picard number of the base satisfies $\varrho(M)\geq 18$. Note that K3 surfaces of high Picard rank are quite special in the K3 moduli space. In particular, it has been conjectured in~\cite{Oda:1980} that an analogue of the Shioda--Inose theorem exists when $\varrho(M)\geq18$, and the case $\varrho(M)=19$ was proved in~\cite{Morrison:1984}.

The decomposition~(\ref{eq:mixedT2curvaturedecomposition}) of $\xi$ and $\Omega$ in integral cohomology yields a solution to the flux quantization conditions. However, not every choice of integral classes leads to a well-defined K3 geometry. The forms $\xi$ and $\Omega$ must obey 
\begin{align}
\Omega.\Omega&=0~,&\Omega.\xi&=0~,&\Omega.\bar{\xi}&=0~,
\end{align}
as well as the inequalities $\Omega.\bar{\Omega}>0$ and $\xi.\bar{\xi}\leq0$. These equations fix the value of the torus moduli $(\tau,\rho)$ and constrain the allowed intersection numbers of the four integral classes. Indeed, the above equalities can be rewritten as polynomial equations for $\tau$, and the intersection numbers must be chosen such that these polynomials admit a root in the upper half-plane. When this holds, the complexified Chern class~(\ref{eq:mixedT2curvature}) defines a principal $T^2$ bundle over the K3 surface $M$ with quantized $\tau$ and $\rho$ parameters, which preserves minimal supersymmetry. Note that the geometry might be subject to additional constraints due to smoothness requirements for the K3 surface. Let us also emphasize that, while the complex constant $\gamma$ superficially looks like a continuous parameter that can be used to deform off the N=2 locus where the torus Chern classes are ASD, this is not the case. As already pointed out in~\cite{Melnikov:2014ywa}, such deformations of a N=2 geometry induce complex structure deformations of $X$ that are incompatible with the supersymmetry conditions on the flux. In the present analysis, we see another crucial difference: flux quantization fixes the $\SU(2)$ structure of the K3 base to a specific integral locus which cannot be reached by a smooth deformation of a generic N=2 geometry.

\subsection{Obstructions to unwinding}
The N=1 flux geometries admit an action of the T-duality group $\GO(\Gamma_{2,18})$. However, we will see that T-duality transformations act within this class, and these configurations cannot be connected by duality to a direct product $M\times T^2$. This is, of course, not surprising: any T-dual theory should preserve exactly $4$ supercharges, whereas a compactification on $M\times T^2$ preserves twice more supersymmetry, unless orbifolded. 

To study the $\GO(\Gamma_{2,18})$ action in detail, we can split the $2$-forms specifying the fibration structure of $X$ into their self-dual and anti-self-dual parts. For the Chern classes of the $T^2$ fibration, we can extract from (\ref{eq:mixedT2curvature}) the self-dual $(2,0)\oplus(0,2)$ components:
\begin{align}
\upomega^1_{+}&=\frac{i\bar{\tau}}{2\tau_2}\gamma\Omega-\frac{i\tau}{2\tau_2}\bar{\gamma}\bar{\Omega}~, & \upomega^2_{+}&=-\frac{i}{2\tau_2}\gamma\Omega+\frac{i}{2\tau_2}\bar{\gamma}\bar{\Omega}~.
\end{align}
Abelian gauge instantons are characterized by the vector of classes $\bUplambda=[\frac{1}{2\pi}\cF^s\bsigma_s]$ and, up to a large gauge transformation, the curvatures $\cF^s$ must be $(1,1)$ primitive forms. The vector $\bUplambda$ can be split as
\begin{align}
\bUplambda=\bUplambda_0-\ba_I \upomega^I~,
\end{align}
where $\bUplambda_0$ encodes anti-self-dual classes corresponding to a HYM connection pulled back from the base $M$. The Wilson line parameters $\ba_I$ are quantized and could be shifted away by a large gauge transformation, however we keep them in the present discussion as they are generically produced by T-duality. When the parameters $\ba_I$ are non-trivial, $\bUplambda$ has non-vanishing $(2,0)\oplus(0,2)$ components. Turning on those self-dual components introduces new terms in the expansion~(\ref{eq:mixedT2curvaturedecomposition}) of $\Omega$ in integral cohomology. The self-dual components of the classes $\upnu_I$ can be obtained from this expansion.

To summarize, supersymmetry imposes strong constraints on the self-dual part $v_+$ of $v$, which has to be of the form
\begin{align}
\label{eq:selfdualv}
v_+=\frac{i}{2\tau_2}\begin{pmatrix} \bar{\tau}\\ -1\\ \rho+\tfrac{1}{2}\ba_1\cdot(\ba_2-\bar{\tau}\ba_1)\\ \rho\bar{\tau}+\tfrac{1}{2}\ba_2\cdot(\ba_2-\bar{\tau}\ba_1)\\ \ba_2-\bar{\tau}\ba_1 \end{pmatrix} \gamma\Omega+\text{c.c.}~.
\end{align}
Note that $v_+$ is completely determined by the moduli and the complex constant $\gamma$. In addition, it is not too hard to see that the first four components of $v_+$ are non-zero, no matter what values the Wilson line parameters take. This is a first sign that the presence of $(2,0)\oplus(0,2)$ components for $v$ obstructs unwinding dualities: the analysis of section~\ref{s:4dimflux8supercharge} relied on dualizing $v$ using Wilson line shifts to set the classes $\upnu_1$ and $\upnu_2$ to zero, and this is no longer possible.

To make this argument more precise, we need more information on the T-duality orbit of $v_+$. Let us introduce the quantity
\begin{align}
h(\tau,\rho,\ba_1,\ba_2) =\frac{i}{2\tau_2}\begin{pmatrix} \bar{\tau}\\ -1\\ \rho+\tfrac{1}{2}\ba_1\cdot(\ba_2-\bar{\tau}\ba_1)\\ \rho\bar{\tau}+\tfrac{1}{2}\ba_2\cdot(\ba_2-\bar{\tau}\ba_1)\\ \ba_2-\bar{\tau}\ba_1 \end{pmatrix}
\end{align}
that appears in the self-dual part~(\ref{eq:selfdualv}) of $v$. While this $h$ is written as a vector, it is not a proper $\GO(2,18,\R)$ multiplet but has a more intricate transformation law under T-duality. The vector $v$ admits the decomposition 
\begin{align}
\label{eq:vSD-ASDdecomposition}
v=v_{(1,1)}+h\gamma\Omega+\bar{h}\bar{\gamma}\bar{\Omega}~,
\end{align}
and its contribution to the Bianchi identity can be written
\begin{equation}
-\tfrac{1}{2}v\bullet v=-\tfrac{1}{2}v_{(1,1)}\bullet v_{(1,1)}+\frac{\rho_2}{\tau_2}|\gamma|^2\Omega.\bar{\Omega}~,
\end{equation}
using $h\cdot h=0$ and $\bar{h}\cdot h=-\rho_2/\tau_2$. Both $h$ and $\gamma$ transform under T-duality. Their general transformation law can be derived from the action of $\GO(\Gamma_{2,18})$ generators. There is essentially one possibility for a duality $g\in \GO(\Gamma_{2,18})$ to preserve the form~(\ref{eq:vSD-ASDdecomposition}) of $v$: $g$ must act on $h$ as
\begin{align}
\label{eq:N=1Tduality}
g h(\tau,\rho,\ba_1,\ba_2)=f_g(\tau,\rho,\ba_1,\ba_2) h(\mu_g^{-1}(\tau,\rho,\ba_1,\ba_2))~,
\end{align}
with some complex moduli-dependent constant $f_g(\tau,\rho,\ba_1,\ba_2)$. When this is the case, the corresponding T-dual configuration $X_{v'}=X_{gv}$ respects the decomposition~(\ref{eq:vSD-ASDdecomposition}), with new parameters $(\tau',\rho',\ba'_1,\ba'_2)=\mu_g^{-1}(\tau,\rho,\ba_1,\ba_2)$ and a complex constant $\gamma'=f_g(\tau,\rho,\ba_1,\ba_2)\gamma$. We can easily see that the $\GO(\Gamma_{2,18})$ generators $g_{\text{t}}[\cR]$, $g_{\text{b}}[m]$, $g_{\text{s},I}[\bkappa]$ and $g_{\text{g}}[R]$ yield a transformation rule of this sort.\footnote{The $\SL(2,\Z)$ transformations $g_{\text{t}}[\cR]$ have a non-trivial factor $f_{g_\text{t}[\cR]}=(\cR^2{}_2-\cR^2{}_1\tau)^{-1}$, while the other generators have $f_g=1$.} However, the generators $g_{\text{i},I}$ and $g_{\text{ref},I}$ satisfy a different identity:
\begin{align}
\label{eq:N=1Tdualityconjugation}
g h(\tau,\rho,\ba_1,\ba_2)=\overline{f_g(\tau,\rho,\ba_1,\ba_2) h(\mu_g^{-1}(\tau,\rho,\ba_1,\ba_2))}~.
\end{align}
For those T-duality transformations, the dual geometries are characterized by a vector $v'=gv$ which splits in self-dual and anti-self-dual parts as
\begin{align}
v'=v'_{(1,1)}+h'\bar{\gamma}'\bar{\Omega}+\bar{h}'\gamma'\Omega~,
\end{align}
where $h'=h(\mu_g^{-1}(\tau,\rho,\ba_1,\ba_2))$ and $\gamma'=f_g(\tau,\rho,\ba_1,\ba_2)\gamma$. At first, this T-dual geometry seems incompatible with supersymmetry: it explicitly breaks the form~(\ref{eq:vSD-ASDdecomposition}) required by the supersymmetry equations. In particular, the complexified curvature of the $T^2$ fibration of $X_{g'}$ is given by
\begin{align}
\upomega'^1+\tau'\upomega'^2=\xi'+\bar{\gamma}'\bar{\Omega}~,
\end{align}
and has $(1,1)$ and $(0,2)$ components, which is incompatible with the conditions in~(\ref{eq:TorusHYM}). This apparent inconsistency is resolved by looking at the action of the transformation $g$ on the holomorphic $(1,0)$ form $\Theta$, which transforms as
\begin{align}
g~:~\Theta~\to~\Theta'=f_g(\tau,\rho,\ba_1,\ba_2)\,\bar{\Theta}~.
\end{align}
In particular, any T-duality under which $h$ transforms by~(\ref{eq:N=1Tdualityconjugation}) must act non-trivially on the $\SU(3)$ structure. The holomorphic $3$-form $\Omega_X$ transforms as
\begin{align}
g~:~\Omega_X~\to~\Omega_{X'}=e^{\text{i}\alpha'}e^{2\phi}\sqrt{\frac{\alpha'\rho'_{2}}{\tau'_{2}}}\;\Omega\wedge\bar{\Theta}~,
\end{align}
where the overall phase factor is defined by $e^{\text{i}\alpha'}\sqrt{\rho'_{2}/\tau'_{2}}=f_g(\tau,\rho,\ba_1,\ba_2)\sqrt{\rho_{2}/\tau_{2}}$ and could be absorbed in a redefinition of $\Omega_{X'}$. In the T-dual picture, $\Theta$ is now a $(0,1)$ form, and supersymmetry requires
\begin{align}
J\wedge\boldsymbol{F}&=0~, & \bar{\Omega}\wedge\boldsymbol{F}&=0~,
\end{align}
instead of~(\ref{eq:TorusHYM}). The complexified curvature $\boldsymbol{F}$ is constrained to have $(1,1)$ and $(0,2)$ components only, and this is the case for the T-dual geometry.

Examining the action of T-duality generators in detail, we find that every generator obeys one of the two conditions~(\ref{eq:N=1Tduality}) and~(\ref{eq:N=1Tdualityconjugation}). Consequently, the same holds for a generic $\GO(\Gamma_{2,18})$ element. As a result, all T-dual configurations of an N=1 geometry are non-trivially fibered: their complexified first Chern class has a non-trivial component proportional to $\Omega$ or to $\bar{\Omega}$. As expected, we find that Narain T-duality cannot undo the torus fibration of this class of backgrounds. Torsional geometries with minimal supersymmetry are disconnected from flux-free $M\times T^2$ compactifications.

\section{Outlook} \label{s:outlook}
In this work we examined a large class of heterotic flux vacua distinguished by the topologically non-K\"ahler target space as the underlying geometry.  In vacua that preserve $8$ supercharges we argued that all such vacua can be deformed to a locus in the moduli space where a T-duality relates the vacuum to a more familiar compactification on a topologically K\"ahler space.  By taking quotients of these dual pairs we uncovered surprising relations between K\"ahler and non-K\"ahler compactifications, and studying the implications of these relations may well provide some general lessons for heterotic flux vacua, and perhaps flux vacua in general.  

For instance, the $\Z_3$ quotient flux geometry described in section~(\ref{ss:n1quotient}) is singular.  While the existence of a torsional linear sigma model realization suggests that this is indeed a well-defined CFT, it is not clear whether the orbifold singularities can be resolved by a marginal deformation.  The existence of the conventional K\"ahler T-dual geometry may shed light on the nature of these singularities.  If there is indeed a mechanism to resolve these non-K\"ahler singularities, it may well generalize to other non-K\"ahler flux geometries.  

As we mentioned in the introduction, there are flux configurations that preserve $8$ supercharges and which are not covered by our analysis.  These are geometries introduced in~\cite{Fino:2019mvp}, where the base of the torus fibration is allowed to degenerate in such a way that the local geometry is of the form  $(\C^2 \times T^2 )/\Gamma$, where $\C^2/\Gamma$ is a standard K3 orbifold singularity, but there is an added shift on the torus fiber such that the action on the total space is free.\footnote{The simplest example is a $\Z_2$ action on $\C^2 \times S^1$ which acts on the local coordinates as $(z_1,z_2,\theta) \to (-z_1 ,-z_2,\theta+\pi)$.}  Since our analysis is ultimately based on compactifying an $8$-dimensional heterotic string theory on a K3 manifold $M$, it does not immediately extend to the case when $M$ has orbifold singularities.  

Nevertheless, we can still ask what is the action of T-duality on these configurations---after all the total space still has two commuting isometries.  A possible resolution comes from~\cite{Cheng:2022nso}, where just such singularities were studied in toroidal $\Z_n$ orbifolds of type II string, and it was argued that the T-dual of the local singularity of this form is a singular geometry $\C^2/\Z_n \times S^1$ equipped with a flat but topologically non-trivial $B$-field gerbe.  It would be interesting to see if such structures arise in the T-duals of the geometries studied in~\cite{Fino:2019mvp} and to precisely identify the singular K\"ahler geometries and the gerbe structures.

Another clear line of investigation is to return to the M/F-theoretic origins of heterotic flux vacua~\cite{Dasgupta:1999ss}.  For instance, we might consider the duals of $3$-dimensional heterotic flux vacua realized as M-theory compactification on K3$\times$K3 = $M\times M_f$, where we think of $M$ as the base K3 geometry, while identifying M-theory compactified on $M_f$ with the heterotic string on $T^3$.  The data describing the abelian gauge bundle and $T^3$ fibration on the heterotic side of the duality is encoded in the choice of the M-theory $G$-flux.  Taking $M_f$ to be elliptically fibered we can identify the corresponding $\Gamma_{2,18} = \Gamma_{2,2} + \Gamma_8+\Gamma_8$ lattice in $H^2(M_f,\Z)$ with the $T^2$ Narain lattice of the heterotic theory.  Then, as in~\cite{Dasgupta:1999ss}, we have a clear picture of the significance of various components of $G \in H^2(M,\Z) \times \Gamma_{2,18}$:  
\begin{enumerate}
\item activating components in the negative-definite part of $\Gamma_{2,2}$ will lead to backgrounds with $4$ supercharges;
\item a component in the positive-definite part of $\Gamma_{2,2}$ is dual to a non-trivial $T^2$ fibration;
\item a component in the $\Gamma_8 + \Gamma_8$ lattice corresponds to an abelian instanton.
\end{enumerate}
Finally, T-duality on the heterotic side can be understood as automorphisms of $M_f$ that act on the $\Gamma_{2,18}$ lattice.   This clear set-up seems ideally suited to investigating the action of various quotient constructions across string duality with reduced supersymmetry, and it should be useful to investigate which heterotic orbifold actions dualize to ``non-geometric'' actions on the M-theory side along the lines of~\cite{Gautier:2019qiq}.  

We can also consider adding non-perturbative ingredients to the discussion: for example, it seems reasonably clear that we can introduce $5$-branes in the $8$--dimensional compactification while preserving supersymmetry.  How does this addition affect the effective theory of the flux vacua?  

These and other questions indicate that while heterotic flux vacua may indeed be ``unwound'' to more familiar configurations, they provide a useful duality frame for a subset of the string landscape, and we suspect there are general lessons to be learned that may be applied to backgrounds where we do not have as much control.

\section*{Appendix}

\appendix

\section{Cocycle details} \label{app:cocyclerep}
In this appendix we give the explicit form for the cocycle factor $\varepsilon(\bp_1,\bp_2)$ used in our computations, as well as details of the phase calculations for the generators.

\subsection{Form of the cocycle}
As discussed in~\cite{Green:1987mn,Polchinski:1998rq,Tan:2015nja}, this is far from unique, but physical quantities are independent of the choice.  

For each $\Gamma_8$ factor, we define an antisymmetric bilinear form
\begin{align}
\Omega(\bell_1,\bell_2) = \frac{1}{2} \sum_{i>j} \left(\ell_1^i \ell_2^j - \ell_2^i \ell_1^j\right) \balpha_{i}\cdot\balpha_{j}~,
\end{align}
where the $\balpha_{i}$ denote the simple roots.  Decomposing
\begin{align}
\Gamma_{d,d+16} = \Gamma_{d,d} + \Gamma_8 + \Gamma_8~,
\end{align}
we then take
\begin{align}
\varepsilon(\bp_1,\bp_2) = \exp\left[ \frac{i\pi}{2} \left( \text{n}_{1i} \text{w}^{i}_2 -\text{w}_1^i \text{n}_{2i}\right) + i\pi \Omega(\bell_1,\bell_2) + i\pi \Omega(\bell'_1 ,\bell'_2) \right]
\end{align}
It is a nice exercise to check that this satisfies the conditions
\begin{align}
\varepsilon(\bp_2,\bp_1) &= e^{i\pi \bp_1\cdot\bp_2} \varepsilon(\bp_1,\bp_2)~, &
\varepsilon(\bp_1,\bp_3) \varepsilon(\bp_2,\bp_3)  &=
\varepsilon(\bp_1+\bp_2,\bp_3)~,&
\end{align}
and
\begin{align}
\varepsilon(\bp_1,\bp_2)\varepsilon(\bp_1+\bp_2,\bp_3) & =\varepsilon(\bp_1,\bp_2+\bp_3)\varepsilon(\bp_2,\bp_3)~.
\end{align}
The first two conditions are necessary so that the operators commute and the OPE closes.  The last condition follows from associativity of the OPE.

\subsection{Phases for generators of \texorpdfstring{$\GO(\Gamma_{1,17})$}{O(1,17)}}
In this appendix we provide some details for the phases in the actions on the Narain CFT quoted in the text.  We fix the form
\begin{align}
\bp &= \text{w} \be + \text{n} \be^\ast + \bL
\end{align}
as the initial form of the lattice vector.

We begin with the factorized duality $g_{\text{i}}$, for which
\begin{align}
\varphi_{g_\text{i}} (\bp) &= \text{n} \be + \text{w} \be^\ast + \bL~,
\end{align}
and therefore
\begin{align}
\frac{ \varepsilon(\varphi_{g_{\text{i}}} (\bp_1),\varphi_{g_{\text{i}}}(\bp_2))}{\varepsilon(\bp_1,\bp_2)} = e^{i\pi(\text{n}_1 \text{w}_2 + \text{w}_1 \text{n}_2)}~.
\end{align}
On the other hand, with $U(g_{\text{i}},\bp) = e^{i\pi\text{nw}}$, we calculate
\begin{align}
\frac{ U(g_{\text{i}},\bp_1 + \bp_2)}{U(g_{\text{i}},\bp_1) U(g_{\text{i}},\bp_2)} = e^{i\pi(\text{n}_1 \text{w}_2 + \text{w}_1 \text{n}_2)}~,
\end{align}
and we can also check that $U(g_{\text{i}},\varphi_{g_{\text{i}}}(\bp))U(g_{\text{i}},\bp) = 1$.

Next, for the reflection $g_{\text{ref}}$ 
\begin{align}
\varphi_{g_\text{ref}} (\bp) &= -\text{n} \be - \text{w} \be^\ast + \bL~,
\end{align}
\begin{align}
\frac{ \varepsilon(\varphi_{g_{\text{ref}}} (\bp_1),\varphi_{g_{\text{r}}}(\bp_2))}{\varepsilon(\bp_1,\bp_2)} =1~,
\end{align}
so we can set $U(g_{\text{ref}},\bp) = 1$.

The Wilson line shift is more complicated.  Using the explicit form of the cocycle and the action
\begin{align}
\bp' = \varphi_{g_\text{s}}[{\boldsymbol{\kappa}}] (\bp) = \text{w} \be + \left(\text{n} + {\boldsymbol{\kappa}}\cdot\bL -\ff{1}{2} {\boldsymbol{\kappa}}\cdot{\boldsymbol{\kappa}}\right) \be^{\ast} + \bL - {\boldsymbol{\kappa}} \text{w}
\end{align}
we find
\begin{align}
\frac{\varepsilon(\bp'_1,\bp'_2)}{\varepsilon(\bp_1,\bp_2)} = \exp\left[ i\pi \left(\text{w}_1 S_{{\boldsymbol{\kappa}}}(\bLambda_2) + \text{w}_2 S_{{\boldsymbol{\kappa}}}(\bLambda_1) \right)\right]~,
\end{align}
where
\begin{align}
S_{{\boldsymbol{\kappa}}}(\bLambda) = -\sum_{i}\boldsymbol{\kappa}^i \bL^i  - \sum_{i>j} \boldsymbol{\kappa}^i \bL^j  \balpha_i \cdot \balpha_j
= \Omega({\boldsymbol{\kappa}},\bL) -\ff{1}{2} {\boldsymbol{\kappa}}\cdot\bL~,
\end{align}
where $i,j \in \{1,\ldots, 16\}$, and we slightly abused our previous notation by combining the simple roots $\balpha$ and $\balpha'$ into a single set of $16$ vectors.  We compare this to the ratio of phases obtained with
\begin{align}
U(g_{\text{s}}[{\boldsymbol{\kappa}}],\bp) = e^{i\pi(\text{w}+1) S_{{\boldsymbol{\kappa}}}(\bL)}~,
\end{align}
which leads to
\begin{align}
\frac{ U(g_{\text{s}}[{\boldsymbol{\kappa}}],\bp_1 + \bp_2)}{U(g_{\text{s}}[{\boldsymbol{\kappa}}],\bp_1) U(g_{\text{s}}[{\boldsymbol{\kappa}}],\bp_2)} = 
\exp\left[ i\pi \left(\text{w}_1 S_{{\boldsymbol{\kappa}}}(\bL_2) + \text{w}_2 S_{{\boldsymbol{\kappa}}}(\bL_1) \right)\right]~.
\end{align}
We also have the composition
\begin{align}
U(g_{\text{s}}[{\boldsymbol{\kappa}}_2],\varphi_{g_{\text{s}}[{\boldsymbol{\kappa}}_1]}(\bp))
U(g_{\text{s}}[{\boldsymbol{\kappa}}_1],\bp) &= U(g_{\text{s}}[\boldsymbol{\kappa}_2+\boldsymbol{\kappa}_1],\bp) e^{-i\pi \text{w}(\text{w}+1) S_{{\boldsymbol{\kappa}}_2} ({\boldsymbol{\kappa}}_1)} \nonumber\\
&=U(g_{\text{s}}[\boldsymbol{\kappa}_2+\boldsymbol{\kappa}_1],\bp)~,
\end{align}
where the last equality follows because $S_{{\boldsymbol{\kappa}}_{2}} ({\boldsymbol{\kappa}}_1) \in\Z$.

Next we consider the phases associated to a rotation of $\Gamma_8+\Gamma_8$.  Since any such rotation can be obtained as a product of Weyl reflections (see, for example~\cite{Humphreys:1990rg}), we can restrict attention to $R_{\bbeta}$, where $\bbeta$ is a root, and the action on the simple roots is
\begin{align}
R_{\bbeta} (\balpha) = \balpha - (\bbeta\cdot\balpha) \bbeta~.
\end{align}
Denoting this reflection by $g_{\text{g}}[\bbeta]$, and setting $\bp' = \varphi_{g_{\text{g}}[\bbeta]}(\bp)$, we then compute
\begin{align}
\frac{\varepsilon(\bp'_1,\bp'_2)}{\varepsilon(\bp_1,\bp_2)} = \exp\left[ 2\pi i \left(S_{\bbeta}(\bL_1) \Omega(\bbeta,\bL_2) +S_{\bbeta}(\bL_2) \Omega(\bbeta,\bL_1)\right)\right]~.
\end{align}
We set
\begin{align}
U(g_{\text{g}}[\bbeta],\bp) = e^{2\pi i S_{\bbeta} (\bL) \Omega(\bbeta,\bL) + i \pi \Omega(\bbeta,\bL)}~,
\end{align}
and this satisfies
\begin{align}
\frac{\varepsilon(\bp'_1,\bp'_2)}{\varepsilon(\bp_1,\bp_2)} & =
\frac{U(g_{\text{g}}[\bbeta],\bp_1+\bp_2)}{U(g_{\text{g}}[\bbeta],\bp_1)U(g_{\text{g}}[\bbeta],\bp_2)}~,
\end{align}
as well as $U(g_{\text{g}}[\bbeta],\bp') U(g_{\text{g}}[\bbeta],\bp) = 1$.  The induced map on the moduli is
\begin{align}
\mu_{g_{\text{g}}[\bbeta]} (r,\ba)  = (r,\ba - (\ba \cdot\bbeta) \bbeta)~.
\end{align}

\subsection{Phases for generators of \texorpdfstring{$\GO(\Gamma_{2,18})$}{O(2,18)}}
In the following section, we collect the relevant details of the action of $\GO(\Gamma_{2,18})$ on vertex operators and on the associated moduli. The $T^2$ parameters are encoded in two constants $\tau$ and $\rho$ valued in the upper half-plane.

Factorized dualities act with a phase
\begin{align}
U(g_{\text{i},1},\bp)&=e^{i\pi\text{n}_1\text{w}^1}~,&U(g_{\text{i},2},\bp)&=e^{i\pi\text{n}_2\text{w}^2}~.
\end{align}
Their action on moduli is
\begin{align}
&\mu_{g_{\text{i},1}}(\tau,\rho,\ba_1,\ba_2)=\nonumber\\
&\left(\rho+\tfrac{1}{2}\ba_1\cdot(\ba_2-\bar{\tau} \ba_1),\frac{\rho_2\tau+\tfrac{1}{2}\tau_2\ba_1\cdot\ba_2}{\rho_2+\tfrac{1}{2}\tau_2\ba_1\cdot\ba_1},
\frac{-\tau_2}{\rho_2+\tfrac{1}{2}\tau_2\ba_1\cdot\ba_1}\ba_1,-\frac{\tau_1\rho_2+\tau_2\rho_1+\tfrac{1}{2}\tau_2\ba_1\cdot\ba_2}{\rho_2+\tfrac{1}{2}\tau_2\ba_1\cdot\ba_1}\ba_1+\ba_2\right)~,\nonumber\\
&\mu_{g_{\text{i},2}}(\tau,\rho,\ba_1,\ba_2)=\nonumber\\
&\left(
\frac{-\bar{\tau}}{\rho\bar{\tau}+\tfrac{1}{2}\ba_2\cdot(\ba_2-\bar{\tau}\ba_1)},
\frac{-\rho_2\bar{\tau}-\tfrac{1}{2}\tau_2\ba_1\cdot\ba_2}{\rho_2|\tau|^2+\tfrac{1}{2}\tau_2\ba_2\cdot\ba_2},\ba_1-\frac{\tau_1\rho_2-\tau_2\rho_1+\tfrac{1}{2}\tau_2\ba_1\cdot\ba_2}{\rho_2|\tau|^2+\tfrac{1}{2}\tau_2\ba_2\cdot\ba_2}\ba_2,\frac{-\tau_2}{\rho_2|\tau|^2+\tfrac{1}{2}\tau_2\ba_2\cdot\ba_2}\ba_2
\right)~.
\end{align}
Let us also quote the action $\mu_{g_{\text{i}}}$ of the product $g_{\text{i}}=g_{\text{i},1}g_{\text{i},2}$ of the two factorized dualities:
\begin{align}
&\mu_{g_{\text{i}}}(\tau,\rho,\ba_1,\ba_2)=\nonumber\\
&\Bigg(\frac{-\bar{\rho}-\tfrac{1}{2}\ba_1\cdot(\ba_2-\tau\ba_1)}{\tau\bar{\rho}+\tfrac{1}{2}\ba_2\cdot(\ba_2-\tau\ba_1)},~\frac{-\tau_2\bar{\rho}}{\tau_2|\rho|^2+\frac{1}{2}\rho_2|\ba_2-\tau\ba_1|^2+\tfrac{1}{4}\tau_2(\ba_1^2 \ba_2^2-(\ba_1\cdot\ba_2)^2)},\nonumber\\
&\frac{-(\rho_2|\tau|^2+\tfrac{1}{2}\tau_2\ba_2^2)\ba_1+(\tau_1\rho_2-\tau_2\rho_1+\tfrac{1}{2}\tau_2\ba_1\cdot\ba_2)\ba_2}{\tau_2|\rho|^2+\frac{1}{2}\rho_2|\ba_2-\tau\ba_1|^2+\tfrac{1}{4}\tau_2(\ba_1^2 \ba_2^2-(\ba_1\cdot\ba_2)^2)}
,~\frac{(\tau_1\rho_2+\tau_2\rho_1+\frac{1}{2}\tau_2\ba_1\cdot\ba_2)\ba_1-(\rho_2+\tfrac{1}{2}\tau_2\ba_1^2)\ba_2}{\tau_2|\rho|^2+\frac{1}{2}\rho_2|\ba_2-\tau\ba_1|^2+\tfrac{1}{4}\tau_2(\ba_1^2 \ba_2^2-(\ba_1\cdot\ba_2)^2)}
\Bigg)~,
\end{align}
where we denoted for brevity $\ba^2=\ba\cdot\ba$ and $|\ba|^2=\ba\cdot\bar{\ba}$.

For the $\SL(2,\Z)$ transformations $g_{\text{t}}[\cR]$ corresponding to $T^2$ isometries, the cocycle condition is compatible with $U(g_{\text{t}}[\cR],\bp)=1$, and the moduli action takes the following form:
\begin{align}
\mu_{g_{\text{t}}}[\cR](\tau,\rho,\ba_I)=\left(\frac{\cR^1{}_2+\cR^2{}_2\tau}{\cR^1{}_1+\cR^2{}_1\tau},\rho,\cR^J{}_I\ba_J\right)~,
\end{align}
with the usual $\PSL(2,\Z)$ action on $\tau$. We can also set $U(g_{\text{ref},I},\bp)=1$ for the $T^2$ reflections. The action of $\mu_{g_{\text{ref},I}}$ reads
\begin{align}
\mu_{g_{\text{ref},1}}(\tau,\rho,\ba_1,\ba_2)&=\left(-\bar{\tau},-\bar{\rho},-\ba_1,\ba_2\right)~,&\mu_{g_{\text{ref},2}}(\tau,\rho,\ba_1,\ba_2)&=\left(-\bar{\tau},-\bar{\rho},\ba_1,-\ba_2\right)~,
\end{align}
with a complex conjugation of the $T^2$ parameters.

Both $B$-field shifts and Wilson line shifts require a non-trivial phase. We take
\begin{align}
U(g_{\text{b}}[m],\bp)&=e^{i\pi m\text{w}^1\text{w}^2}~,\nonumber\\
U(g_{\text{s}}[\bkappa_I],\bp)&=e^{i\pi\text{w}^1S_{\bkappa_1}(\bL+\frac{1}{2}\bkappa_1)+i\pi\text{w}^2S_{\bkappa_2}(\bL+\frac{1}{2}\bkappa_2)+\frac{i\pi}{2}\text{w}^1\text{w}^2S_{\bkappa_1}(\bkappa_2)-S_{\bkappa_2}(\bkappa_1))}~.
\end{align}
The corresponding transformation law of the moduli is
\begin{align}
\mu_{g_{\text{b}}}[m](\tau,\rho,\ba_1,\ba_2)&=(\tau,\rho+m,\ba_1,\ba_2)~,\nonumber\\
\mu_{g_{\text{s}}}[\bkappa_I](\tau,\rho,\ba_1,\ba_2)&=\left(\tau,\rho+\tfrac{1}{2}\epsilon^{IJ}\ba_I\cdot\bkappa_{J},\ba_1-\bkappa_1,\ba_2-\bkappa_2\right)~.
\end{align}
Finally, the action of $\Gamma_8+\Gamma_8$ rotations is identical to the one already discussed for $\GO(\Gamma_{1,17})$ dualities.

\subsection{Phases in the T-dual description} \label{app:Tdualcocycles}
In this appendix, we make some comments on the action of the T-dual group $G'=\cT G\cT^{-1}$ on the vertex operators $\cV_\bp$ , given an action of the group $G$. Every element $g'$ of $G'$ is obtained by conjugating some $g\in G$ by the transformation $\cT$. Since $\cT$ has an action $\varphi_\cT$ on the $\Gamma_{d,d+16}$ lattice, we can obtain the lattice action of $g'=\cT g\cT^{-1}$ by
\begin{align}
\varphi_{g'}=\varphi_\cT\varphi_g\varphi_\cT^{-1}~.
\end{align}
Whenever $\varphi_\cT$ acts non-trivially on $\Gamma_{d,d+16}$, we must also define a phase $U(\cT,\bp)$ in order for the action of $\cT$ on vertex operators to be compatible with the operator product expansion. The phase associated to the inverse transformation $\cT^{-1}$ is
\begin{align}
U(\cT^{-1}\!,\bp)=U(\cT,\varphi_{\cT^{-1}}(\bp))^{-1}~,
\end{align}
which ensures that the product law $\cT^{-1}\circ(\cT\circ\cV_\bp)=\cV_\bp$ is satisfied. Every element $g$ of the group $G$ also has a corresponding phase $U(g,\bp)$. This phase obeys the cocycle condition~(\ref{eq:cocyclecondition}) and the group law~(\ref{eq:cocycleproduct}), since we assume a well-defined action of $G$.

Consider now an element $g'=\cT g\cT^{-1}$ of $G'$. We define the phase $U(g',\bp)$ by
\begin{align}
\label{eq:cocycleconjugate}
U(g',\bp)=U(\cT,\varphi_{g\cT^{-1}}(\bp))~U(g,\varphi_{\cT^{-1}}(\bp))U(\cT^{-1},\bp)~.
\end{align}
We can easily check that this phase obeys the cocycle condition~(\ref{eq:cocyclecondition}). 
Moreover, for two elements $g'_1$ and $g'_2$ of $G'$, we can check that $U(g'_2,\varphi_{g'_1}(\bp))U(g'_1,\bp)$ and $U(g'_2g'_1,\bp)$ agree. In other words, the product law~(\ref{eq:cocycleproduct}) for the phases associated to $G$ implies the product law for the phases associated to $G'$. We emphasize that this construction does not necessitate an embedding $G\subseteq H$ of the group $G$ in a bigger group $H$, with $\cT\in H$. The phases $U(g',\bp)$ define a consistent $G'$ action without having to specify such embedding in $H$ or the corresponding phases $U(h,\bp)$ for every $h\in H$.

As a particular case, notice that if an element $g\in G$ has a trivial lattice action $\varphi_g(\bp)=\bp$, then the conjugated element $g'$ also has $\varphi_{g'}(\bp)=\bp$ and the corresponding phase, defined in~(\ref{eq:cocycleconjugate}), takes the simple form
\begin{align}
U(g',\bp)=U(g,\varphi_{\cT}^{-1}(\bp))~.
\end{align}

Let us also describe in more details the case of a cyclic group $G\simeq\Z_k$. The action of $G$ is fully specified by the action of a generator $g$, with $\varphi_g^k(\bp)=\bp$ and
\begin{align}
U(g,\varphi_g^{k-1}(\bp))\dots U(g,\varphi_g^2(\bp))U(g,\varphi_g(\bp))U(g,\bp)=1~.
\end{align}
As an example, consider the product of Weyl reflections $R=R_{\bbeta_{2}}R_{\bbeta_{1}}$ introduced in section~\ref{s:4dimflux8supercharge}, where the two roots $\bbeta_1$ and $\bbeta_2$ are at a $\frac{2\pi}{3}$ angle, i.e. $\bbeta_{1}\cdot\bbeta_{2}=-1$. This transformation generates a $\Z_3$ subgroup of $\GO(\Gamma_8+\Gamma_8)$. Since $\bbeta_1\cdot\bbeta_2=-S_{\bbeta_1}(\bbeta_2)-S_{\bbeta_2}(\bbeta_1)$, we can---without loss of generality---order the two roots such that $S_{\bbeta_1}(\bbeta_2)$ is even and $S_{\bbeta_2}(\bbeta_1)$ is odd. We define the following cocycle\footnote{Here we slightly abuse notation by writing $U(R,\bp)$ instead of $U(g_{\text{g}}[R],\bp)$ and $\varphi_R$ instead of $\varphi_{g_{\text{g}}[R]}$.} associated to $R$:
\begin{align}
U(R,\bp)&=e^{i\pi\left(S_{\bbeta_{1}}(\bL)\bbeta_{1}\cdot\bL+S_{\bbeta_{2}}(R_{\bbeta_{1}}(\bL))\bbeta_{2}\cdot R_{\bbeta_{1}}(\bL)\right)}\nonumber\\
&=e^{i\pi\left(S_{\bbeta_{1}}(\bL)\bbeta_{1}\cdot\bL+S_{\bbeta_{2}}(\bL)(\bbeta_1+\bbeta_2)\cdot \bL+(\bbeta_1\cdot\bL)(\bbeta_1+\bbeta_2)\cdot\bL\right)}~.
\end{align}
We can easily check that $U(R,\bp)$ satisfies the cocycle property~(\ref{eq:cocyclecondition}). Moreover, we can compute
\begin{align}
U(R,\varphi_R(\bp))&=e^{i\pi\left(S_{\bbeta_1}(\bL)\bbeta_2\cdot\bL+S_{\bbeta_2}(\bL)\bbeta_1\cdot \bL+(\bbeta_1\cdot\bL)\bbeta_2\cdot\bL\right)}~,\nonumber\\
U(R,\varphi_{R^2}(\bp))&=e^{i\pi\left(S_{\bbeta_1}(\bL)(\bbeta_1+\bbeta_2)\cdot\bL+S_{\bbeta_2}(\bL)\bbeta_2\cdot \bL+(\bbeta_1\cdot\bL)\bbeta_1\cdot\bL\right)}~,
\end{align}
and we see that $U(R,\bp)$ satisfies the $\Z_3$ condition
\begin{align}
U(R,\bp)U(R,\varphi_R(\bp))U(R,\varphi_{R^2}(\bp))=1~.
\end{align}
As a result, any action obtained by conjugation of this $\Z_3$ group by some transformation $\cT$ will satisfy the same $\Z_3$ product law. For the unwinding T-duality $\cT=g_{\text{i}}g_{\text{s},1}[2\bbeta_1]g_{\text{s},2}[2\bbeta_2]$ discussed in section~\ref{s:4dimflux8supercharge}, the cocycle condition for $\cT$ is solved by the choice $U(\cT,\bp)=e^{i\pi\text{w}^I\text{n}_I}$. Computing the explicit form of the orbifold phase using~(\ref{eq:cocycleconjugate}) yields $U(\cT g_3\cT^{-1},\bp)=U(g_3,\bp)$.

\section{Some details on the cohomology of \texorpdfstring{$X$}{X}} \label{app:LeraySerre}
Because the base $M$ is simply connected, it is easy to use the Leray--Serre spectral sequence to calculate the integral cohomology of $X$.\footnote{Since $M$ is simply connected,  we can avoid the complications of having to work with local coefficients.  See, for example, chapter 5 of~\cite{Davis:2001at} for an introduction.}  This spectral sequence has $E_2^{p,q} = H^p(M,H^q(F))$ as its second stage, where $F=T^2$ denotes the torus fiber, and the spectral sequence converges to $H^{k}(X)$.  When working over a general module, as opposed to a vector space, the $E^{p,q}_{\infty}$ only determine a filtration of $H^{p+q}(X)$, and determining the precise $H^{p+q}(X)$ requires data beyond the spectral sequence~\cite{Davis:2001at}.   In our example $E^{p,q}_{\infty} = E^{p,q}_{3}$, and the extension problem turns out to be trivial for all groups except $H^4(X)$.  Fortunately, the torsion subgroup of $H^4(X)$ is isomorphic to the torsion subgroup in $H^3(X)$ by Poincar{\'e} duality and universal coefficients, and this shows that all extensions are trivial.

The universal coefficients theorem also determines the cohomology groups $H^p(M,H^q(F))$.  In general~\cite{Hatcher:2002at} these are given by the exact sequence
\begin{equation}
\begin{tikzcd}
0 \ar[r] &
\text{Ext}(H_{p-1}(M),H^q(F)) \ar[r] &
H^p(M, H^q(F)) \ar[r] &
\text{Hom}(H_p(M), H^q(F)) \ar[r] &
0
\end{tikzcd}
\end{equation}
But, this simplifies because $\text{Ext}(H,G) =0$ whenever $H$ is a freely generated.  Thus, we find
\begin{align}
H^p(M,H^q(F)) = \begin{cases}  H^p(M) \\  H^p(M) \times H^p(M) \\ H^p(M)~, \end{cases}
\end{align}
and the second page of the spectral sequence has the form
\begin{equation}
\begin{tikzcd}
H^0(M)   \ar[rrd,"d_{22}"]&	 0	&H^2(M) \ar[rrd,"d_{24}"]	&	0	&H^4(M) \\
H^0(M,H^1(F))	\ar[rrd,"d_{21}"]& 0 		&H^2(M,H^1(F)) \ar[rrd,"d_{23}"]& 0	& H^4(M,H^1(F))  \\
H^0(M)   &	 0	&H^2(M)	&	0	&H^4(M) 
\end{tikzcd}
\end{equation}
We indicated the $d_2$ map, and it is clear that $d_3 = 0$, so that $E^{p,q}_3 = E^{p,q}_{\infty}$.   All of the maps are fixed in terms of the cup product and the classes $\upomega^I$: 
\begin{equation}
\begin{tikzcd}
H^0(M, H^1(F)) \ar[r,"d_{21}"] & H^2(M) \\
(x_1, x_2) \ar[r,"d_{21}"] & (x_1 \upomega^1+x_2 \upomega^2)~,
\end{tikzcd}
\end{equation}
\begin{equation}
\begin{tikzcd}
H^0(M) \ar[r,"d_{22}"] & H^2(M, H^1(F) ) \ar[r,"d_{23}"] & H^4(M) \\
(x) \ar[r,"d_{22}"] & (x \upomega^1,x \upomega^2) & \\
	     & (C_1, C_2) \ar[r,"d_{23}"]  &(C_1 \cup \upomega^2 - C_2 \cup \upomega^1)~,
\end{tikzcd}
\end{equation}
\begin{equation}
\begin{tikzcd}
H^2(M) \ar[r,"d_{24}"] & H^4(M, H^1(F)) \\
(C) \ar[r,"d_{24}"] & (C \cup \upomega^1, C \cup \upomega^2)~.
\end{tikzcd}
\end{equation}
So, we find $H^0(X) = \Z$ and $H^6(X) = \Z$, while
\begin{align}
H^1(X) &= \ker d_{21}~,&
H^2(X) & = \ker d_{22} \oplus H^2(M)/\im d_{21}~,\nonumber\\
H^3(X) & = \ker d_{23} / \im d_{22}~, \nonumber\\
H^4(X) & = H^4(M)/\im d_{23} \oplus \ker d_{24}~,&
H^5(X) & =  \left(H^4(M) \oplus H^4(M) \right)/\im d_{24}~.
\end{align}
To study the groups in detail, we identify $H^4(M)\simeq \Z$, so that $C\cup \upomega^1 = C. \upomega^1 \in \Z$, and then we use the basis $\{\upomega_{\text{p}}^1, \upomega_{\text{p}}^2\,\ldots,\}$ and its dual basis.

We have three distinct situations to consider.  First, there is the trivial fibration with $m_1 = 0$.  In this case the cohomology is determined by the K\"unneth formula, and of course every group is free.  In what follows we will leave this case out and work with $m_1 >0$.

\begin{enumerate}
\item $H^1(X)$.  There are two non-trivial possibilities.
\begin{enumerate}
\item $m_1 >0$, $m_2 = 0$.  In this case just one circle is fibered non-trivially and $H^1(X) = \Z$.   We may as well factor out the trivial circle from $X =  Y \times S^1$ and work with $Y$ directly.
\item $m_1>0$, $m_2 >0$.  Both circles are non-trivially fibered, and $H^1(X) = 0$.
\end{enumerate}
\item $H^2(X)$.  If a circle is non-trivially fibered, $\ker d_{22} = 0$, so we just need to consider the quotient 
$H^2(M)/\im d_{21}$.
\begin{enumerate}
\item When $m_1 >0$ and $m_2 = 0$, then
$\im d_{21} = \text{Span}\{m_1 \upomega_{\text{p}}^1\}$, 
and therefore $H^2(X) = \Z^{21} \times \Z_{m_1}$.
\item When $m_1>0$ and $m_2 =0$, then $\im d_{21} = \text{Span}\{m_1 \upomega_{\text{p}}^1,m_1 m_2\upomega_{\text{p}}^1\}$, and therefore $H^2(X) = \Z^{20} \times\Z_{m_1} \times \Z_{m_1 m_2}$. 
\end{enumerate}
Notice that in either case the torsion subgroup reproduces the torsion subgroup of $H_1(X) =  \pi_1(X)/\CO{\pi_1(X)}{\pi_1(X)}$.
\item $H^3(X)$.   To study this group it is convenient to expand $C_2$ in the dual basis:
\begin{align}
C_2 = a \upomega_{\text{p} 1}^\ast + b \upomega_{\text{p} 2}^\ast +\Ch_2~.
\end{align}
Then
\begin{align}
d_{23} (C_1, C_2) =C_1. \upomega^2 - C_2 . \upomega^1 =  m_1\left( m_2 C_1 .\upomega^2_{\text{p}}- \Ch_2 . \upomega_{\text{p}}^1 - a\right)
= m_1\left( m_2 C_1 .\upomega^2_{\text{p}} - a\right)
~,
\end{align}
where we used $\Ch_2 . \upomega_{\text{p}}^1 = 0$.
We see therefore, that $a$ is determined, and $\ker d_{23} = \Z^{22} \times \Z^{21}$:
\begin{align}
\ker d_{23} = \{ (C_1,  (m_2 C_1. \upomega^2_{\text{p}} ) \upomega^\ast_{\text{p}1} + b \upomega^\ast_{\text{p}2} + \Ch_2) \}~.
\end{align}

The quotient by $\im d_{22}$ gives rise to possible torsion in $H^3(X)$.  This is because
\begin{align}
\im d_{22} = \text{Span} \{ (m_1 \upomega^1_{\text{p}}, m_1 m_2 \upomega^2_{\text{p}})\}~,
\end{align}
and therefore the element $(\upomega^1_{\text{p}}, m_2 \upomega^2_{\text{p}})$ generates a $\Z_{m_1}$ subgroup in $H^3(X)$.  This is the only source of torsion, so we conclude that whenever $m_1 >0$ 
\begin{align}
H^3(X) = \Z^{2\times 22 - 2} \times \Z_{m_1}~.
\end{align}
Note that when $m_2 = 0$, so that $X = Y \times S^1$, then the K\"unneth formula determines
\begin{align}
H^2(X) &= H^2(Y)~,&
H^3(X) &= H^3(Y) \times H^2(Y)~,
\end{align}
so that
\begin{align}
H^3(Y) = \Z^{21}~.
\end{align}

\item $H^4(X)$.  If $m_1>0$, then $\im d_{23} = m_1\Z$: 
\begin{align}
d_{23} ( 0, - \upomega^\ast_{\text{p}1}) = m_1~.
\end{align}
Thus, 
\begin{align}
H^4(M) / \im d_{23} = \Z_{m_1}~.
\end{align}
The free part of the group comes from $\ker d_{24}$, and this depends on whether $m_2 = 0$ or not:
\begin{enumerate}
\item $m_1 >0, m_2>0$.  Then $\ker d_{24} = \text{Span}\{\Ch\} = \Z^{20}$.
\item $m_1 >0, m_2 = 0$.  Then $\ker d_{24} = \text{Span}\{ \upomega_{\text{p}2}^\ast, \Ch\} = \Z^{21}$.
\end{enumerate}

\item $H^5(X)$.  Here we obtain
\begin{align}
\im d_{24} = \text{Span} \{ (m_1, 0) , (0, m_1 m_2)\}~.
\end{align}
So, the free part is the isomorphic to the free part of $H^1(X)$, while the torsion part is the Pontryagin dual to the torsion in $H^2(X)$.

\end{enumerate}

\bibliographystyle{utphys}
\bibliography{./newref}

\providecommand{\href}[2]{#2}\begingroup\raggedright\begin{thebibliography}{10}

\bibitem{Hull:1986kz}
C.~M. Hull, ``{Compactifications of the Heterotic Superstring},''
  \href{http://dx.doi.org/10.1016/0370-2693(86)91393-6}{{\em Phys. Lett. B}
  {\bfseries 178} (1986) 357--364}.

\bibitem{Strominger:1986uh}
A.~Strominger, ``{Superstrings with torsion},''
{\em Nucl. Phys.} {\bfseries B274} (1986) 253.

\bibitem{Goldstein:2002pg}
E.~Goldstein and S.~Prokushkin, ``{Geometric model for complex non-Kaehler
  manifolds with SU(3) structure},'' {\em Commun. Math. Phys.} {\bfseries 251}
  (2004) 65--78,
\href{http://arxiv.org/abs/hep-th/0212307}{{\ttfamily arXiv:hep-th/0212307}}.

\bibitem{Fu:2006vj}
J.-X. Fu and S.-T. Yau, ``{The theory of superstring with flux on non-Kaehler
  manifolds and the complex Monge-Ampere equation},'' {\em J. Diff. Geom.}
  {\bfseries 78} (2009) 369--428,
\href{http://arxiv.org/abs/hep-th/0604063}{{\ttfamily arXiv:hep-th/0604063}}.

\bibitem{Becker:2006et}
K.~Becker, M.~Becker, J.-X. Fu, L.-S. Tseng, and S.-T. Yau, ``{Anomaly
  cancellation and smooth non-Kaehler solutions in heterotic string theory},''
  {\em Nucl. Phys.} {\bfseries B751} (2006) 108--128,
\href{http://arxiv.org/abs/hep-th/0604137}{{\ttfamily arXiv:hep-th/0604137}}.

\bibitem{Dasgupta:1999ss}
K.~Dasgupta, G.~Rajesh, and S.~Sethi, ``{M theory, orientifolds and G-flux},''
  {\em JHEP} {\bfseries 08} (1999) 023,
\href{http://arxiv.org/abs/hep-th/9908088}{{\ttfamily arXiv:hep-th/9908088}}.

\bibitem{Banks:1988yz}
T.~Banks and L.~J. Dixon, ``{Constraints on string vacua with space-time
  supersymmetry},'' {\em Nucl. Phys.} {\bfseries B307} (1988) 93--108.

\bibitem{Melnikov:2010pq}
I.~V. Melnikov and R.~Minasian, ``{Heterotic sigma models with N=2 space-time
  supersymmetry},'' \href{http://dx.doi.org/10.1007/JHEP09(2011)065}{{\em JHEP}
  {\bfseries 1109} (2011) 065},
  \href{http://arxiv.org/abs/1010.5365}{{\ttfamily arXiv:1010.5365 [hep-th]}}.

\bibitem{Melnikov:2012cv}
I.~V. Melnikov, R.~Minasian, and S.~Theisen, ``{Heterotic flux backgrounds and
  their IIA duals},'' \href{http://dx.doi.org/10.1007/JHEP07(2014)023}{{\em
  JHEP} {\bfseries 07} (2014) 023},
\href{http://arxiv.org/abs/1206.1417}{{\ttfamily arXiv:1206.1417 [hep-th]}}.

\bibitem{Fino:2019mvp}
A.~Fino, G.~Grantcharov, and L.~Vezzoni, ``{Solutions to the
  Hull\textendash{}Strominger System with Torus Symmetry},''
  \href{http://dx.doi.org/10.1007/s00220-021-04223-7}{{\em Commun. Math. Phys.}
  {\bfseries 388} no.~2, (2021) 947--967},
  \href{http://arxiv.org/abs/1901.10322}{{\ttfamily arXiv:1901.10322
  [math.DG]}}.

\bibitem{Adams:2006kb}
A.~Adams, M.~Ernebjerg, and J.~M. Lapan, ``{Linear models for flux vacua},''
\href{http://arxiv.org/abs/hep-th/0611084}{{\ttfamily arXiv:hep-th/0611084}}.

\bibitem{Adams:2009zg}
A.~Adams and J.~M. Lapan, ``{Computing the Spectrum of a Heterotic Flux
  Vacuum},''
\href{http://arxiv.org/abs/0908.4294}{{\ttfamily arXiv:0908.4294 [hep-th]}}.

\bibitem{Israel:2015aea}
D.~Israel and M.~Sarkis, ``{New supersymmetric index of heterotic
  compactifications with torsion},''
  \href{http://dx.doi.org/10.1007/JHEP12(2015)069}{{\em JHEP} {\bfseries 12}
  (2015) 069},
\href{http://arxiv.org/abs/1509.05704}{{\ttfamily arXiv:1509.05704 [hep-th]}}.

\bibitem{Israel:2016xfu}
D.~Israel and M.~Sarkis, ``{Dressed elliptic genus of heterotic
  compactifications with torsion and general bundles},''
  \href{http://dx.doi.org/10.1007/JHEP08(2016)176}{{\em JHEP} {\bfseries 08}
  (2016) 176},
\href{http://arxiv.org/abs/1606.08982}{{\ttfamily arXiv:1606.08982 [hep-th]}}.

\bibitem{Angelantonj:2016gkz}
C.~Angelantonj, D.~Israel, and M.~Sarkis, ``{Threshold corrections in heterotic
  flux compactifications},''
  \href{http://dx.doi.org/10.1007/JHEP08(2017)032}{{\em JHEP} {\bfseries 08}
  (2017) 032},
\href{http://arxiv.org/abs/1611.09442}{{\ttfamily arXiv:1611.09442 [hep-th]}}.

\bibitem{Becker:2007ea}
M.~Becker, L.-S. Tseng, and S.-T. Yau, ``{Heterotic Kahler/non-Kahler
  Transitions},'' {\em Adv.Theor.Math.Phys.} {\bfseries 12} (2008) 1147--1162,
\href{http://arxiv.org/abs/0706.4290}{{\ttfamily arXiv:0706.4290 [hep-th]}}.

\bibitem{Sethi:2007bw}
S.~Sethi, ``{A Note on heterotic dualities via M-theory},''
  \href{http://dx.doi.org/10.1016/j.physletb.2007.10.043}{{\em Phys. Lett. B}
  {\bfseries 659} (2008) 385--387},
  \href{http://arxiv.org/abs/0707.0295}{{\ttfamily arXiv:0707.0295 [hep-th]}}.

\bibitem{Evslin:2008zm}
J.~Evslin and R.~Minasian, ``{Topology change from (heterotic) Narain
  T-duality},'' {\em Nucl. Phys.} {\bfseries B820} (2009) 213--236,
\href{http://arxiv.org/abs/0811.3866}{{\ttfamily arXiv:0811.3866 [hep-th]}}.

\bibitem{Andriot:2009fp}
D.~Andriot, R.~Minasian, and M.~Petrini, ``{Flux backgrounds from Twists},''
  \href{http://dx.doi.org/10.1088/1126-6708/2009/12/028}{{\em JHEP} {\bfseries
  12} (2009) 028}, \href{http://arxiv.org/abs/0903.0633}{{\ttfamily
  arXiv:0903.0633 [hep-th]}}.

\bibitem{Martelli:2010jx}
D.~Martelli and J.~Sparks, ``{Non-Kahler heterotic rotations},''
  \href{http://dx.doi.org/10.4310/ATMP.2011.v15.n1.a4}{{\em Adv. Theor. Math.
  Phys.} {\bfseries 15} no.~1, (2011) 131--174},
  \href{http://arxiv.org/abs/1010.4031}{{\ttfamily arXiv:1010.4031 [hep-th]}}.

\bibitem{Andriot:2011iw}
D.~Andriot, ``{Heterotic string from a higher dimensional perspective},''
  \href{http://dx.doi.org/10.1016/j.nuclphysb.2011.10.007}{{\em Nucl. Phys. B}
  {\bfseries 855} (2012) 222--267},
  \href{http://arxiv.org/abs/1102.1434}{{\ttfamily arXiv:1102.1434 [hep-th]}}.

\bibitem{Israel:2013hna}
D.~Isra\"el, ``{T-Duality in Gauged Linear Sigma-Models with Torsion},''
  \href{http://dx.doi.org/10.1007/JHEP11(2013)093}{{\em JHEP} {\bfseries 11}
  (2013) 093}, \href{http://arxiv.org/abs/1306.6609}{{\ttfamily arXiv:1306.6609
  [hep-th]}}.

\bibitem{Barth:2004ne}
W.~P. Barth, K.~Hulek, C.~A.~M. Peters, and A.~Van~de Ven, {\em Compact complex
  surfaces}, vol.~4.
\newblock Springer-Verlag, Berlin, second~ed., 2004.

\bibitem{Aspinwall:1996mn}
P.~S. Aspinwall, ``{K3 surfaces and string duality},''
\href{http://arxiv.org/abs/hep-th/9611137}{{\ttfamily arXiv:hep-th/9611137}}.

\bibitem{Chaudhuri:1995bf}
S.~Chaudhuri and J.~Polchinski, ``{Moduli space of CHL strings},''
  \href{http://dx.doi.org/10.1103/PhysRevD.52.7168}{{\em Phys. Rev. D}
  {\bfseries 52} (1995) 7168--7173},
  \href{http://arxiv.org/abs/hep-th/9506048}{{\ttfamily arXiv:hep-th/9506048}}.

\bibitem{Kachru:1995wm}
S.~Kachru and C.~Vafa, ``{Exact results for N=2 compactifications of heterotic
  strings},'' \href{http://dx.doi.org/10.1016/0550-3213(95)00307-E}{{\em
  Nucl.Phys.} {\bfseries B450} (1995) 69--89},
  \href{http://arxiv.org/abs/hep-th/9505105}{{\ttfamily arXiv:hep-th/9505105
  [hep-th]}}.

\bibitem{Ferrara:1995yx}
S.~Ferrara, J.~A. Harvey, A.~Strominger, and C.~Vafa, ``{Second quantized
  mirror symmetry},''
  \href{http://dx.doi.org/10.1016/0370-2693(95)01074-Z}{{\em Phys.Lett.}
  {\bfseries B361} (1995) 59--65},
  \href{http://arxiv.org/abs/hep-th/9505162}{{\ttfamily arXiv:hep-th/9505162
  [hep-th]}}.

\bibitem{Israel:2023itj}
D.~Israel and Y.~Proto, ``{A Worldsheet Approach to $N=1$ Heterotic Flux
  Backgrounds},'' \href{http://arxiv.org/abs/2302.01889}{{\ttfamily
  arXiv:2302.01889 [hep-th]}}.

\bibitem{Becker:2009df}
K.~Becker and S.~Sethi, ``{Torsional heterotic geometries},'' {\em Nucl. Phys.}
  {\bfseries B820} (2009) 1--31,
\href{http://arxiv.org/abs/0903.3769}{{\ttfamily arXiv:0903.3769 [hep-th]}}.

\bibitem{Melnikov:2014ywa}
I.~V. Melnikov, R.~Minasian, and S.~Sethi, ``{Heterotic fluxes and
  supersymmetry},'' \href{http://dx.doi.org/10.1007/JHEP06(2014)174}{{\em JHEP}
  {\bfseries 06} (2014) 174},
\href{http://arxiv.org/abs/1403.4298}{{\ttfamily arXiv:1403.4298 [hep-th]}}.

\bibitem{Oda:1980}
T.~Oda, ``{A note on the Tate conjecture for $K3$ surfaces},''
  \href{http://dx.doi.org/10.3792/pjaa.56.296}{{\em Proceedings of The Japan
  Academy Series A} {\bfseries 56} (01, 1980) 296--300}.

\bibitem{Morrison:1984}
D.~R. Morrison, ``On {K3} surfaces with large {Picard} number,''
  \href{http://dx.doi.org/10.1007/BF01403093}{{\em Inventiones mathematicae}
  {\bfseries 75} no.~1, (Feb., 1984) 105--121}.
  \url{https://doi.org/10.1007/BF01403093}.

\bibitem{Giveon:1994fu}
A.~Giveon, M.~Porrati, and E.~Rabinovici, ``{Target space duality in string
  theory},'' \href{http://dx.doi.org/10.1016/0370-1573(94)90070-1}{{\em
  Phys.Rept.} {\bfseries 244} (1994) 77--202},
\href{http://arxiv.org/abs/hep-th/9401139}{{\ttfamily arXiv:hep-th/9401139
  [hep-th]}}.

\bibitem{Font:2020rsk}
A.~Font, B.~Fraiman, M.~Gra\~na, C.~A. N\'u\~nez, and H.~P. De~Freitas,
  ``{Exploring the landscape of heterotic strings on $T^d$},''
  \href{http://dx.doi.org/10.1007/JHEP10(2020)194}{{\em JHEP} {\bfseries 10}
  (2020) 194}, \href{http://arxiv.org/abs/2007.10358}{{\ttfamily
  arXiv:2007.10358 [hep-th]}}.

\bibitem{Cheng:2022nso}
P.~Cheng, I.~V. Melnikov, and R.~Minasian, ``{Flat equivariant gerbes:
  holonomies and dualities},''
  \href{http://dx.doi.org/10.1007/JHEP04(2023)074}{{\em JHEP} {\bfseries 04}
  (2023) 074}, \href{http://arxiv.org/abs/2207.06885}{{\ttfamily
  arXiv:2207.06885 [hep-th]}}.

\bibitem{Ginsparg:1986bx}
P.~H. Ginsparg, ``{Comment on Toroidal Compactification of Heterotic
  Superstrings},''
\href{http://dx.doi.org/10.1103/PhysRevD.35.648}{{\em Phys.Rev.} {\bfseries
  D35} (1987) 648}.

\bibitem{Tan:2015nja}
H.~S. Tan, ``{T-duality Twists and Asymmetric Orbifolds},''
  \href{http://dx.doi.org/10.1007/JHEP11(2015)141}{{\em JHEP} {\bfseries 11}
  (2015) 141}, \href{http://arxiv.org/abs/1508.04807}{{\ttfamily
  arXiv:1508.04807 [hep-th]}}.

\bibitem{Polchinski:1998rq}
J.~Polchinski, {\em {String theory. Vol. 1: An introduction to the bosonic
  string}}.
\newblock Cambridge University Press,
2007.
\newblock

\bibitem{Green:1987mn}
M.~Green, J.~Schwarz, and E.~Witten, {\em Superstring Theory, Volume 1}.
\newblock Cambridge University Press, 1987.

\bibitem{Fraiman:2018ebo}
B.~Fraiman, M.~Gra\~na, and C.~A. N\'u\~nez, ``{A new twist on heterotic string
  compactifications},'' \href{http://dx.doi.org/10.1007/JHEP09(2018)078}{{\em
  JHEP} {\bfseries 09} (2018) 078},
  \href{http://arxiv.org/abs/1805.11128}{{\ttfamily arXiv:1805.11128
  [hep-th]}}.

\bibitem{Davis:2001at}
J.~F. Davis and P.~Kirk, {\em Lecture notes in algebraic topology}, vol.~35 of
  {\em Graduate Studies in Mathematics}.
\newblock Amer. Math. Soc., 2001.

\bibitem{Hofer:1993tpb}
T.~H{\"o}fer, ``{Remarks on torus principal bundles},''
  \href{http://dx.doi.org/10.1215/kjm/1250519346}{{\em Journal of Mathematics
  of Kyoto University} {\bfseries 33} no.~1, (1993) 227 -- 259}.
  \url{https://doi.org/10.1215/kjm/1250519346}.

\bibitem{Honecker:2006dt}
G.~Honecker, ``{Massive U(1)s and heterotic five-branes on K3},'' {\em
  Nucl.Phys.} {\bfseries B748} (2006) 126--148,
  \href{http://arxiv.org/abs/hep-th/0602101}{{\ttfamily arXiv:hep-th/0602101
  [hep-th]}}.

\bibitem{Honecker:2006qz}
G.~Honecker and M.~Trapletti, ``{Merging Heterotic Orbifolds and K3
  Compactifications with Line Bundles},''
  \href{http://dx.doi.org/10.1088/1126-6708/2007/01/051}{{\em JHEP} {\bfseries
  0701} (2007) 051},
\href{http://arxiv.org/abs/hep-th/0612030}{{\ttfamily arXiv:hep-th/0612030
  [hep-th]}}.

\bibitem{Louis:2011hp}
J.~Louis, M.~Schasny, and R.~Valandro, ``{6D Effective Action of Heterotic
  Compactification on K3 with Nontrivial Gauge Bundles},''
  \href{http://dx.doi.org/10.1007/JHEP04(2012)028}{{\em JHEP} {\bfseries 04}
  (2012) 028}, \href{http://arxiv.org/abs/1112.5106}{{\ttfamily arXiv:1112.5106
  [hep-th]}}.

\bibitem{Kumar:2009zc}
V.~Kumar and W.~Taylor, ``{Freedom and Constraints in the K3 Landscape},''
  \href{http://dx.doi.org/10.1088/1126-6708/2009/05/066}{{\em JHEP} {\bfseries
  0905} (2009) 066},
\href{http://arxiv.org/abs/0903.0386}{{\ttfamily arXiv:0903.0386 [hep-th]}}.

\bibitem{MR658473}
C.~H. Taubes, ``Self-dual {Y}ang-{M}ills connections on non-self-dual
  {$4$}-manifolds,'' {\em J. Differential Geom.} {\bfseries 17} no.~1, (1982)
  139--170.

\bibitem{Witten:1985ga}
E.~Witten, ``Global anomalies in string theory,'' in {\em Argonne symposium on
  geometry, anomalies and topology}, W.~A. Bardeen, ed., Argonne.
\newblock 1985.

\bibitem{Freed:1986zx}
D.~Freed, ``Determinants, torsion, and strings,''
\href{http://dx.doi.org/10.1007/BF01221001}{{\em Commun.Math.Phys.} {\bfseries
  107} (1986) 483--513}.

\bibitem{Gukov:2002wpj}
S.~Gukov and C.~Vafa, ``{Rational Conformal Field Theories and Complex
  Multiplication},'' in {\em {25th International Workshop on Fundamental
  Problems of High-Energy Physics and Field Theory}}, pp.~93--123.
\newblock 2002.

\bibitem{Sharpe:2003cs}
E.~Sharpe, ``{Discrete torsion and shift orbifolds},''
  \href{http://dx.doi.org/10.1016/S0550-3213(03)00412-7}{{\em Nucl. Phys. B}
  {\bfseries 664} (2003) 21--44},
  \href{http://arxiv.org/abs/hep-th/0302152}{{\ttfamily arXiv:hep-th/0302152}}.

\bibitem{Sharpe:2000ki}
E.~R. Sharpe, ``{Discrete torsion},''
  \href{http://dx.doi.org/10.1103/PhysRevD.68.126003}{{\em Phys. Rev. D}
  {\bfseries 68} (2003) 126003},
  \href{http://arxiv.org/abs/hep-th/0008154}{{\ttfamily arXiv:hep-th/0008154}}.

\bibitem{Vafa:1986wx}
C.~Vafa, ``{Modular Invariance and Discrete Torsion on Orbifolds},''
  \href{http://dx.doi.org/10.1016/0550-3213(86)90379-2}{{\em Nucl. Phys. B}
  {\bfseries 273} (1986) 592--606}.

\bibitem{MR2128387}
Y.~Ruan, ``{Discrete Torsion and Twisted Orbifold Cohomology},'' {\em Journal
  of Symplectic Geometry} {\bfseries 2} no.~1, (2003) 001 -- 24.

\bibitem{Becker:2008rc}
M.~Becker, L.-S. Tseng, and S.-T. Yau, ``{New Heterotic Non-Kahler
  Geometries},''
\href{http://arxiv.org/abs/0807.0827}{{\ttfamily arXiv:0807.0827 [hep-th]}}.

\bibitem{Giusti:2023mqz}
F.~Giusti and C.~Spotti, ``{A K\"ummer construction for Chern-Ricci flat
  balanced manifolds},'' \href{http://arxiv.org/abs/2309.12909}{{\ttfamily
  arXiv:2309.12909 [math.DG]}}.

\bibitem{Nikulin:1980}
V.~V. Nikulin, ``Finite automorphism groups of {K{\"a}hler} {K3} surfaces,''
  {\em Trans. Moscow Math. Soc.} {\bfseries 38} (1980) 71--135.

\bibitem{Sterk:1985}
H.~Sterk, ``Finiteness results for algebraic {K3} surfaces,''
  \href{http://dx.doi.org/10.1007/BF01168156}{{\em Mathematische Zeitschrift}
  {\bfseries 189} no.~4, (Dec., 1985) 507--513}.
  \url{https://doi.org/10.1007/BF01168156}.

\bibitem{YANG2014230}
H.~Yang, ``Equivariant cohomology and sheaves,''
  \href{http://dx.doi.org/https://doi.org/10.1016/j.jalgebra.2014.05.009}{{\em
  Journal of Algebra} {\bfseries 412} (2014) 230--254}.
  \url{https://www.sciencedirect.com/science/article/pii/S0021869314002622}.

\bibitem{Artebani:2011}
M.~Artebani, A.~Sarti, and S.~Taki, ``K3 surfaces with non-symplectic
  automorphisms of prime order,''
  \href{http://dx.doi.org/10.1007/s00209-010-0681-x}{{\em Mathematische
  Zeitschrift} {\bfseries 268} no.~1, (June, 2011) 507--533}.
  \url{https://doi.org/10.1007/s00209-010-0681-x}.

\bibitem{Cheng:2023owv}
P.~Cheng, I.~V. Melnikov, and R.~Minasian, ``{Flat F-theory and friends},''
  \href{http://arxiv.org/abs/2306.00865}{{\ttfamily arXiv:2306.00865
  [hep-th]}}.

\bibitem{Artebani:2008}
M.~Artebani and A.~Sarti, ``Non-symplectic automorphisms of order 3 on {K3}
  surfaces,'' \href{http://dx.doi.org/10.1007/s00208-008-0260-1}{{\em
  Mathematische Annalen} {\bfseries 342} no.~4, (Dec., 2008) 903--921}.
  \url{https://doi.org/10.1007/s00208-008-0260-1}.

\bibitem{Borcea:1996mxz}
C.~Borcea, ``{K3 surfaces with involution and mirror pairs of Calabi-Yau
  manifolds},'' {\em AMS/IP Stud. Adv. Math.} {\bfseries 1} (1996) 717--743.

\bibitem{Voisin:1993mir}
C.~Voisin, ``{Miroirs et involutions sur les surfaces K3},'' {\em Ast\'erisque}
  {\bfseries 218} (1993) 273.

\bibitem{Cecotti:1992qh}
S.~Cecotti, P.~Fendley, K.~A. Intriligator, and C.~Vafa, ``{A New
  supersymmetric index},''
  \href{http://dx.doi.org/10.1016/0550-3213(92)90572-S}{{\em Nucl. Phys. B}
  {\bfseries 386} (1992) 405--452},
  \href{http://arxiv.org/abs/hep-th/9204102}{{\ttfamily arXiv:hep-th/9204102}}.

\bibitem{Benini:2013nda}
F.~Benini, R.~Eager, K.~Hori, and Y.~Tachikawa, ``{Elliptic genera of
  two-dimensional N=2 gauge theories with rank-one gauge groups},''
\href{http://arxiv.org/abs/1305.0533}{{\ttfamily arXiv:1305.0533 [hep-th]}}.

\bibitem{Benini:2013xpa}
F.~Benini, R.~Eager, K.~Hori, and Y.~Tachikawa, ``{Elliptic genera of 2d N=2
  gauge theories},''
\href{http://arxiv.org/abs/1308.4896}{{\ttfamily arXiv:1308.4896 [hep-th]}}.

\bibitem{Witten:1986bf}
E.~Witten, ``{Elliptic genera and quantum field theory},''
\href{http://dx.doi.org/10.1007/BF01208956}{{\em Commun.Math.Phys.} {\bfseries
  109} (1987) 525}.

\bibitem{Narain:1986am}
K.~S. Narain, M.~H. Sarmadi, and E.~Witten, ``{A Note on Toroidal
  Compactification of Heterotic String Theory},''
  \href{http://dx.doi.org/10.1016/0550-3213(87)90001-0}{{\em Nucl. Phys. B}
  {\bfseries 279} (1987) 369--379}.

\bibitem{Aspinwall:2005ad}
P.~S. Aspinwall and R.~Kallosh, ``{Fixing all moduli for M-theory on K3xK3},''
  \href{http://dx.doi.org/10.1088/1126-6708/2005/10/001}{{\em JHEP} {\bfseries
  0510} (2005) 001},
\href{http://arxiv.org/abs/hep-th/0506014}{{\ttfamily arXiv:hep-th/0506014
  [hep-th]}}.

\bibitem{Shioda:1977}
T.~Shioda and H.~Inose, {\em On Singular K3 Surfaces},
  \href{http://dx.doi.org/10.1017/CBO9780511569197.010}{pp.~119--136}.
\newblock Cambridge University Press, 1977.

\bibitem{Gautier:2019qiq}
Y.~Gautier, C.~M. Hull, and D.~Isra\"el, ``{Heterotic/type II Duality and
  Non-Geometric Compactifications},''
  \href{http://dx.doi.org/10.1007/JHEP10(2019)214}{{\em JHEP} {\bfseries 10}
  (2019) 214}, \href{http://arxiv.org/abs/1906.02165}{{\ttfamily
  arXiv:1906.02165 [hep-th]}}.

\bibitem{Humphreys:1990rg}
J.~E. Humphreys, \href{http://dx.doi.org/10.1017/CBO9780511623646}{{\em
  Reflection groups and {C}oxeter groups}}, vol.~29 of {\em Cambridge Studies
  in Advanced Mathematics}.
\newblock Cambridge University Press, Cambridge, 1990.
\newblock \url{https://doi.org/10.1017/CBO9780511623646}.

\bibitem{Hatcher:2002at}
A.~Hatcher, {\em Algebraic topology}.
\newblock Cambridge University Press, 2002.

\end{thebibliography}\endgroup

\end{document}